\documentclass[manuscript]{aastex62}
\shorttitle{Molecular cloud cores in the 50MC}
\shortauthors{Uehara et al.}

\begin{document}
\title{Molecular Cloud Cores in the Galactic Center 50 $\rm km~s^{-1}$ Molecular Cloud}

\email{uehara@vsop.isas.jaxa.jp}

\author{Kenta Uehara}
\affil{Department of Astronomy, The University of Tokyo, Bunkyo, Tokyo 113-0033, Japan}

\author{Masato Tsuboi}
\affil{Department of Astronomy, The University of Tokyo, Bunkyo, Tokyo 113-0033, Japan}
\affiliation{Institute of Space and Astronautical Science(ISAS), \\
Japan Aerospace Exploration Agency,\\
3-1-1 Yoshinodai, Chuo-ku, Sagamihara, Kanagawa 252-5210, Japan}

\author{Yoshimi Kitamura}
\affiliation{Institute of Space and Astronautical Science(ISAS), \\
Japan Aerospace Exploration Agency,\\
3-1-1 Yoshinodai, Chuo-ku, Sagamihara, Kanagawa 252-5210, Japan}
\collaboration{}

\author{Ryosuke Miyawaki}
\affiliation{College of Arts and Sciences, J.F. Oberlin University, Machida, Tokyo 194-0294, Japan}

\author{Atsushi Miyazaki}
\affiliation{Japan Space Forum, Kandasurugadai, Chiyoda-ku,Tokyo,101-0062, Japan}



\begin{abstract}

The Galactic Center 50 km s$^{-1}$ Molecular Cloud (50MC) is the most remarkable molecular cloud in the Sagittarius A region. 
This cloud is a candidate for the massive star formation induced by cloud-cloud collision (CCC) with a collision velocity of $\sim30\rm~km~s^{-1}$ that is estimated from the velocity dispersion. 
We observed the whole of the 50MC with a high angular resolution ($\sim2\farcs0\times1\farcs4$) in ALMA cycle 1 in the H$^{13}$CO$^+~J=1-0$ and ${\rm C^{34}S}~J=2-1$ emission lines. 
We identified 241 and 129 bound cores with a virial parameter of less than 2, which are thought to be gravitationally bound, in the H$^{13}$CO$^+$ and ${\rm C^{34}S}$ maps using the clumpfind algorithm, respectively. 
In the CCC region, the bound ${\rm H^{13}CO^+}$ and ${\rm C^{34}S}$ cores are 119 and 82, whose masses are $68~\%$ and $76~\%$ of those in the whole 50MC, respectively. 
The distribution of the core number and column densities in the CCC are biased to larger densities than those in the non-CCC region. 
The distributions indicate that the CCC compresses the molecular gas and increases the number of the dense bound cores. 
Additionally, the massive bound cores with masses of $>3000~M_{\odot}$ exist only in the CCC region, although the slope of the core mass function (CMF) in the CCC region is not different from that in the non-CCC region. 
We conclude that the compression by the CCC efficiently formed massive bound cores even if the slope of the CMF is not changed so much by the CCC. 
\end{abstract}



\section{Introduction} \label{sec:intro}
The Central Molecular Zone (CMZ) in the Galactic Center (GC) region is a molecular cloud complex extending $\sim500\rm~pc$ along the Galactic plane \citep{morris1996}. 
The physical properties of the molecular gas in the CMZ are quite different from those in the Galactic disk region. 
The gas in the CMZ is much denser, warmer, and more turbulent than that in the disk region ($\ga1000\rm~cm^{-3}$, $\sim10-100$ K and $\sim15-50\rm~km~s^{-1}$ in the CMZ). 
In the CMZ, there are bright, young massive clusters that are hardly seen in the disk region, including the Arches cluster, Quintuplet cluster, and Central cluster \citep[e.g.][]{figer1999}. 
Thus, massive star formation must have occurred in such severe turbulent conditions in the CMZ. 
However, we cannot demonstrate what mechanism was responsible for the formation of the star clusters in the CMZ because the cradle molecular gas has already been dissipated from around these clusters. 
One of the promising mechanisms for the cluster formation is cloud-cloud collision (CCC) in the CMZ \citep{hasegawa1994,2015PASJ...67..109T}, because the CCC probably makes massive stars efficiently \citep[e.g.][]{habe1992,2009ApJ...696L.115F,2010ApJ...709..975O,inoue2013}. 

The $50\rm~km~s^{-1}$ molecular cloud (50MC) is one of the bright molecular clouds in molecular emission lines in the Sagittarius A (Sgr A) region. 
The cloud includes four compact $\rm H_{II}$ regions A-D which are conspicuous in the Paschen $\alpha$ recombination line \citep{2011ApJ...735...84M} 
and radio continuum emission \citep{goss1985,ekers1983}. 
Thus, this cloud is considered to be a young massive star formation site, which does not yet dissipate the molecular gas. 
The 50MC also interacts with the supernova remnant called Sgr A east \citep{1985ApJ...288..575H,tsuboi2009}. 
The kinetic temperature of the 50MC has been estimated to be $>100$ K from the H$_2$CO observations \citep{ao2013} and at $\sim400$ K from the NH$_3$ observations \citep{mills2013}. 

The 50MC was observed by the Nobeyama Radio Observatory 45m telescope (NRO45) and the Nobeyama Millimeter Array (NMA). 
From the NMA observation, 37 molecular cloud cores were identified from the CS $J=1-0$ emission line maps \citep{tsuboi2012}. 
A half-shell-like feature with the high brightness temperature ratio of the SiO $ J=2-1$ and $\rm H^{13}CO^+~J=1-0$ emission lines (up to 8) was found in the 50MC using NRO45 \citep{tsuboi2011}. 
This feature has been proposed to be result from the CCC between the 50MC and a smaller cloud \citep{2015PASJ...67..109T}. 
The molecular cloud cores identified in the CCC region have a top-heavy molecular cloud Core Mass Function (CMF) and are more massive than those in the non-CCC region \citep{2015PASJ...67..109T}. 
{  
According to recent simulations \citep[e.g.][]{inoue2013},  because the effective velocity width, $\sqrt{V_{\rm s}^{2}+\Delta V^{2}+V_{\rm Alf}^{2}}$, becomes large by compression of the magnetic field, where the $V_{\rm s}$, $\Delta V$, and $V_{\rm Alf}$ are the sound velocity, velocity dispersion, and Alfv\'en velocity, respectively, the effective Jeans mass of the molecular core becomes large in the CCC.
Consequently, massive stars can be formed in the region because molecular cloud cores would not fragment until their masses exceed the effective Jeans masses. 
}
Thus, because molecular cloud cores would {{ not fragment} until the core mass exceeds the effective Jeans masses, massive stars can be formed in the region.
The 50MC is thought to be a candidate of the massive star forming region (SFR) induced by the CCC \citep{2015PASJ...67..109T}. 
The collision velocity is estimated to be $\sim30\rm~km~s^{-1}$ from the velocity dispersion. 
However, the number of the molecular cloud cores was too small to obtain conclusive results because both the observation area and angular resolution were not sufficient \citep[e.g.][]{tsuboi2012}. 
Thus, we need a larger mapping area with high angular resolution and high sensitivity to understand the properties of the cores. 

We observed the whole of the 50MC using the Atacama Large Millimeter/submillimeter Array (ALMA). 
The observation details and the results will be presented in \S\ref{sec:obs} and \S\ref{sec:map}, respectively. 
We will identify the { core candidate}s using the data in \S\ref{sec:core}. 
In \S\ref{sec:property}, we clarify the statistical relations of the bound cores in the two different environments of the 50MC and the Orion A and in the two region of the CCC and non-CCC regions. 

\section{Observation}\label{sec:obs}
We observed the 50MC located at $(l,b)=(-0\farcs018,-0\farcs072)$ as an ALMA cycle 1 program using the 12m array, Atacama Compact Array (ACA) and Total Power telescope (TP) on May 31, 2013 to Jan 23, 2015 (2012.1.00080.S,PI M.Tsuboi). 
The center frequencies of the spectral window (SPW) 0, 1, 2 and 3 are 97.987, 96.655, 86.910 and 85.723 GHz, respectively. 
The bandwidth and frequency resolution of each SPW are 937.5 MHz and 244.141 kHz/1ch, respectively. 
These SPWs include some molecular emission lines (CS $J=2-1$, C$^{34}$S $J=2-1$, SiO $v=0~J=2-1$, $^{29}$SiO $v=0~J=2-1$, H$^{13}$CO$^+~J=1-0$, ${\rm CH_{3}OH}~J_{K_{a},K_{c}}=2_{1,1}-1_{1,0}$ and so on) and the H42$\alpha$ recombination line. 
This observation was performed with mosaic mapping of 137 pointings (12m array) and 52 pointings (ACA). 
Figures \ref{fig:guide}-A and B show the mapping region, which is a square area of $330''\times330''$ covering the whole of the 50MC. 
The obtained data of the 12m array and TP were reduced using the manual script and those of the ACA were reduced using the pipeline script in the CASA software \citep{2007ASPC..376..127M}. 
The data of the 12m array and ACA were concatenated with the task "concat" in the uv-plane. 
Furthermore, the interferometer map from the concatenated data was created using the "briggs" weighting with a robust parameter of 0.5 in the "clean" of the CASA. 
The interferometer maps and the TP maps were combined with the task "feathering" by making the sum of these maps in the uv-plane in order to restore the missing large-scale information because the resolved-out scale of the ACA is $\sim80''$. 
Finally, we created the channel maps of the H$^{13}$CO$^+~J=1-0~(\rm\nu=86.754~GHz)$ and ${\rm C^{34}S}~J=2-1~(\rm\nu=96.413~GHz)$ emission lines. 
The angular resolution of the H$^{13}$CO$^+~J=1-0$ map is $2\farcs04\times1\farcs41$ at $\rm PA=-39.40^\circ$ corresponding to $\rm0.083~pc\times0.058~pc$ at the distance to the Galactic Center ($8.5$ kpc), and that of the C$^{34}$S$~J=2-1$ map is $2\farcs00\times1\farcs35$ ($0.056~{\rm pc}\times0.082~{\rm pc}$) at $\rm PA=-30\fdg3$. 
These are about 5 times higher than those of the previous interferometer observations \citep[e.g.][]{tsuboi2009}. 
The physical resolution in our ALMA observation is equal to that in the Orion A cloud observed by current single dish telescopes \citep[e.g.][]{ikeda2007}. 
Thus, it becomes possible to directly compare and contrast massive star forming processes in the GC 50MC and the typical Galactic disk molecular cloud, the Orion A cloud. 
Although the original velocity resolution is $\rm\sim0.75~km~s^{-1}$, the velocity resolutions of the created maps are $2\rm~km~s^{-1}$ for improving the noise level. 
The rms noise levels of the ${\rm H^{13}CO^+}~J=1-0$ and ${\rm C^{34}S}~J=2-1$ channel maps are $\sigma_{\rm rms}=0.16\rm~K$ in brightness temperature.

\begin{figure}[ht!]
\begin{center}
\includegraphics[scale=0.42]{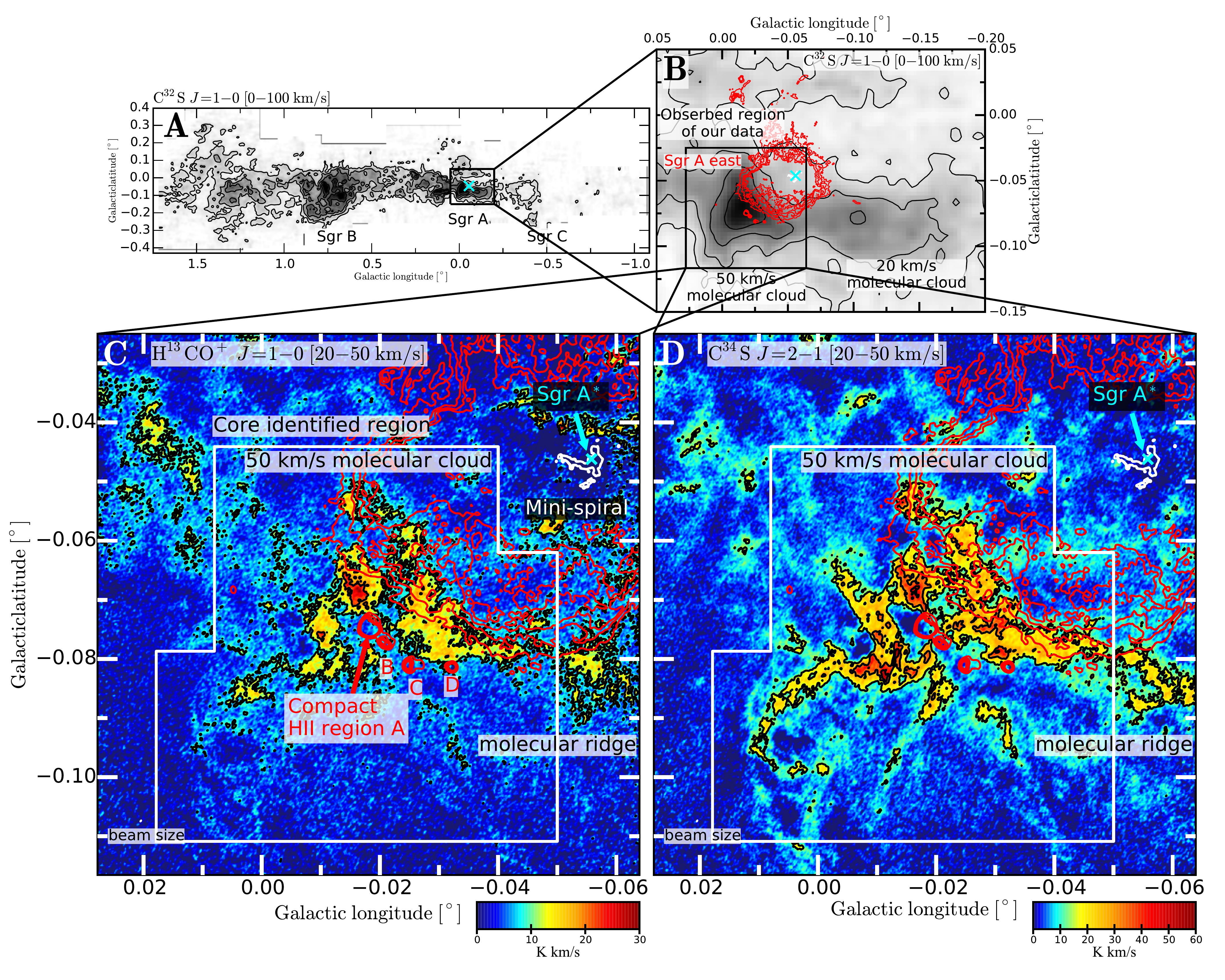}
\end{center}
\caption{{[A] Integrated intensity map of the CMZ in the ${\rm C^{32}S}~J=1-0$ emission line  with contours (black solid line) \citep{tsuboi1999}. 
The cyan cross shows the position of the Sgr A*. 
[B] Integrated intensity map of the ${\rm C^{32}S}~J=1-0$ emission line of the Sgr A region with contours (black solid line) \citep{tsuboi1999}. 
Red contours show the 6cm continuum emission map \citep{yusef-zadeh1987}. 
The contour levels are $0.02,  0.04,  0.06$ and $0.08\rm~Jy~beam^{-1}$. 
The black thick line shows the region observed with ALMA. 
[C] Integrated intensity map of ${\rm H^{13}CO^+}~J=1-0$ with contours (black solid line) is shown as a guide map. 
The integration range is $V_{\rm LSR}=20-50\rm~km~s^{-1}$. 
The color bar shows the intensity range of $0-30\rm~K~km~s^{-1}$, and the contour levels are $10$ and $20\rm~K~km~s^{-1}$. 
White contours show the continuum emission map in the 100 GHz band of the mini-spiral observed by ALMA \citep{2016PASJ...68L...7T}. 
	The contour levels are $0.01$ and $0.015\rm~Jy~beam^{-1}$. 
[D] Integrated intensity map of ${\rm C^{34}S}~J=2-1$ with contours (black solid line) is shown as a guide map. 
The integration range is $V_{\rm LSR}=20-50\rm~km~s^{-1}$. 
The color bar shows the intensity range of $0-60\rm~K~km~s^{-1}$, and the contour levels are $15,~30$ and $45\rm~K~km~s^{-1}$. 
}
	}\label{fig:guide}
\end{figure}

\begin{table}
	\caption{The parameters of the ALMA observation}\label{tab:pyspara}
	\begin{center}
 	\begin{tabular}{ccc}
		\hline\hline
		Source&G-0.02-0.07&\\
		Obs. region [$\rm arcsec^{2}$]&$330\times330$ &\\
		Mosaic pointing (12m array)&137 pointings &\\
		Mosaic pointing (7m array)&52 pointings &\\
		\hline\hline
		& ${\rm H^{13}CO^+}~J=1-0$ & ${\rm C^{34}S}~J=2-1$ \\
		\hline
		Rest frequency [GHz] & 86.754& 96.413\\
		Angular resolution [$\rm arcsec^{2}$]&$2.04\times1.41$&$2.00\times1.35$\\
		Physical resolution $[{\rm pc^{2}}]$ (@$8.5\rm~kpc$) &$\rm0.083\times0.058$ &$0.082\times0.056$\\
		Peak intensity\tablenotemark{a} [K]&2.1&3.9\\
		Conversion factor [$\rm K/(Jy~beam^{-1})$]&56.4&48.7\\
		Peak velocity\tablenotemark{a} $[\rm km~s^{-1}]$&52&49\\
		FWHM velocity width\tablenotemark{a} $[\rm km~s^{-1}]$& $34$&37\\
		\hline\hline
	\end{tabular}
	\tablenotetext{1}{We applied the Gaussian fitting to the core identified region shown in Figure \ref{fig:guide}. }
	\end{center}
\end{table}

\section{Channel maps of the ${\rm H^{13}CO^+}~J=1-0$ and ${\rm C^{34}S}~J=2-1$ emission lines}\label{sec:map}

First of all, we evaluated how much the feathering method restores the missing flux of structures extending larger than the observable scale by ACA. 
Figure \ref{fig:spectrum} shows the spectra of the ALMA feathering map, the ALMA interferometer map, and the NRO45 map. The flux density ratio of the ALMA interferometer map to the NRO45 map is $\sim0.15$, whereas the ratio of the ALMA feathering map to the NRO45 map is as high as $\sim0.8$. 
Considering the calibration error between the ALMA feathering map and NRO45 map, the feathering method restores the missing flux quite well. 
We use the missing flux restored maps by the feathering method hereafter in this paper. 

\begin{figure}[ht!]
\plotone{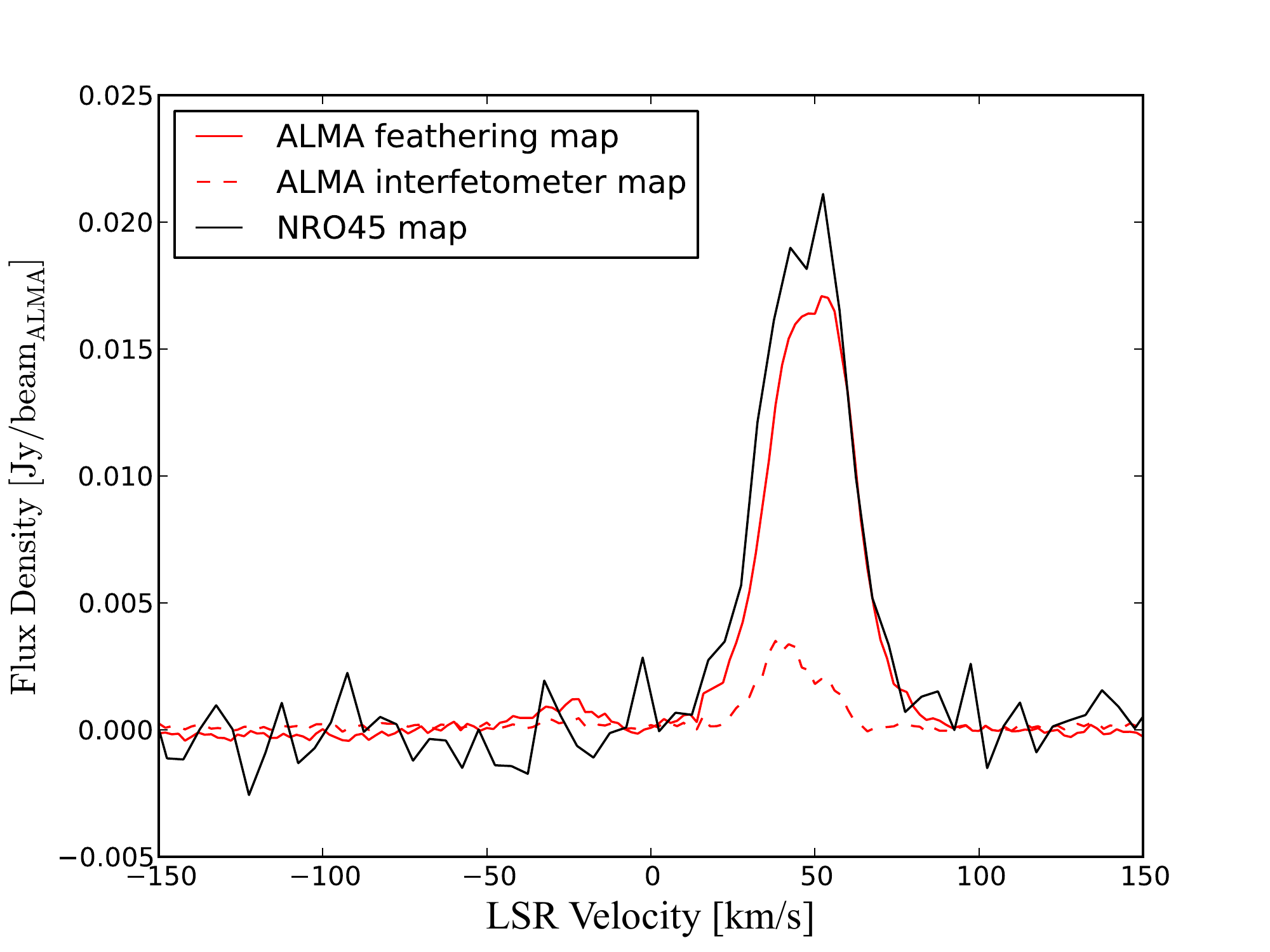}
\caption{Spectra of the ALMA feathering ${\rm H^{13}CO^+}$ map (red solid line), the ALMA interferometer map (red dashed line), and the NRO45 map (black solid line). These spectra were calculated inside a circle with a radius of 30 arcsec, centered on the 50MC center of $(l,b)=(-0\fdg02,-0\fdg07)$. The intensity scales of the spectra are shown in the units of $\rm Jy/(2\farcs04\times1\farcs41)$. }\label{fig:spectrum}
\end{figure}

The channel maps of the ${\rm H^{13}CO^+}~J=1-0$ emission line are shown in Figure \ref{fig:chanmap}. 
These maps show the dense region of the 50MC because the H$^{13}$CO$^+~J=1-0$ emission line has a high critical density of $\sim10^5\rm~cm^{-3}$. 
We show the velocity range of $0\rm~km~s^{-1}$ to $100\rm~km~s^{-1}$ in $V_{\rm LSR}$ because the 50MC is not detected out of this velocity range. 
In the same velocity range, the channel maps of the ${\rm C^{34}S}~J=2-1$ emission line also are shown in Figure \ref{fig:chanmap34}. 
{ The positions of Sgr A* (white star) and four HII regions (white cross) are shown on each channel map. }

The molecular ridge \citep[e.g.][]{coil2000,park2004} is seen in the velocity range of $V_{\rm LSR}=26-60\rm~km~s^{-1}$ to the west of the 50MC in both the ${\rm H^{13}CO^+}$ and ${\rm C^{34}S}$ maps. 
The bright components only in the ${\rm H^{13}CO^+}$ maps exist on $(l,b)=(-0\fdg046,-0\fdg056)$ with $V_{\rm LSR}=48-78\rm~km~s^{-1}$ which was observed in the ${\rm HCO^+}~J=1-0$ and HCN $J=1-0$ emission lines as the dense clumps \citep{2005ApJ...622..346C}. 
The other bright components exist on $(l,b)=(-0\fdg051,-0\fdg056)$ with $V_{\rm LSR}=18-28\rm~km~s^{-1}$. 
The compact HII region D is found at $(l,b)=(-0\fdg04,-0\fdg08)$ by absorption in the maps of $V_{\rm LSR}=32-48\rm~km~s^{-1}$ in the ${\rm H^{13}CO^+}$ maps. 
These features are summarized in Figure \ref{fig:guide}-C.

These channel maps with $0.1\rm~pc$ resolution revealed that the 50MC has clumpy and filamentary structures as shown in Figure \ref{fig:chanmap} and \ref{fig:chanmap34}. 
The filamentary structures are conspicuous in the range of $30-40\rm~km~s^{-1}$ in $V_{\rm LSR}$ and radially extended from around the center of the 50MC in the ${\rm H^{13}CO^+}$ maps. 
On the other hand, in the ${\rm C^{34}S}$ maps, the filamentary structures can be confirmed more clearly compared with those in the ${\rm H^{13}CO^+}$ maps. 
The filamentary structures are apparent in the range of $24-40\rm~km~s^{-1}$ in $V_{\rm LSR}$ and radially extended from around the center of the 50MC. 
The filamentary structures are also found distinctly in the CS $J=2-1$ channel maps \citep{2017IAUS..322..162U}. 
Filamentary structures in molecular clouds are found in the Galactic disk region \citep[e.g.][]{2010A&A...518L.102A} from the Herschel survey observations \citep{pilbratt2010}. 
It has been revealed that the molecular clouds ubiquitously exist as filamentary structures in the Galactic disk region. 
On the other hand, the filamentary structures in the CMZ have been detected in G0.253$+$0.016 with ALMA \citep{2015ApJ...802..125R}, but other molecular clouds with filamentary structures have not yet been found in the CMZ. 
The existence of a number of the filamentary structures in the 50MC strongly suggests that the filamentary structures are also ubiquitous in the molecular clouds in the CMZ. 

\begin{figure}[ht!]
\plotone{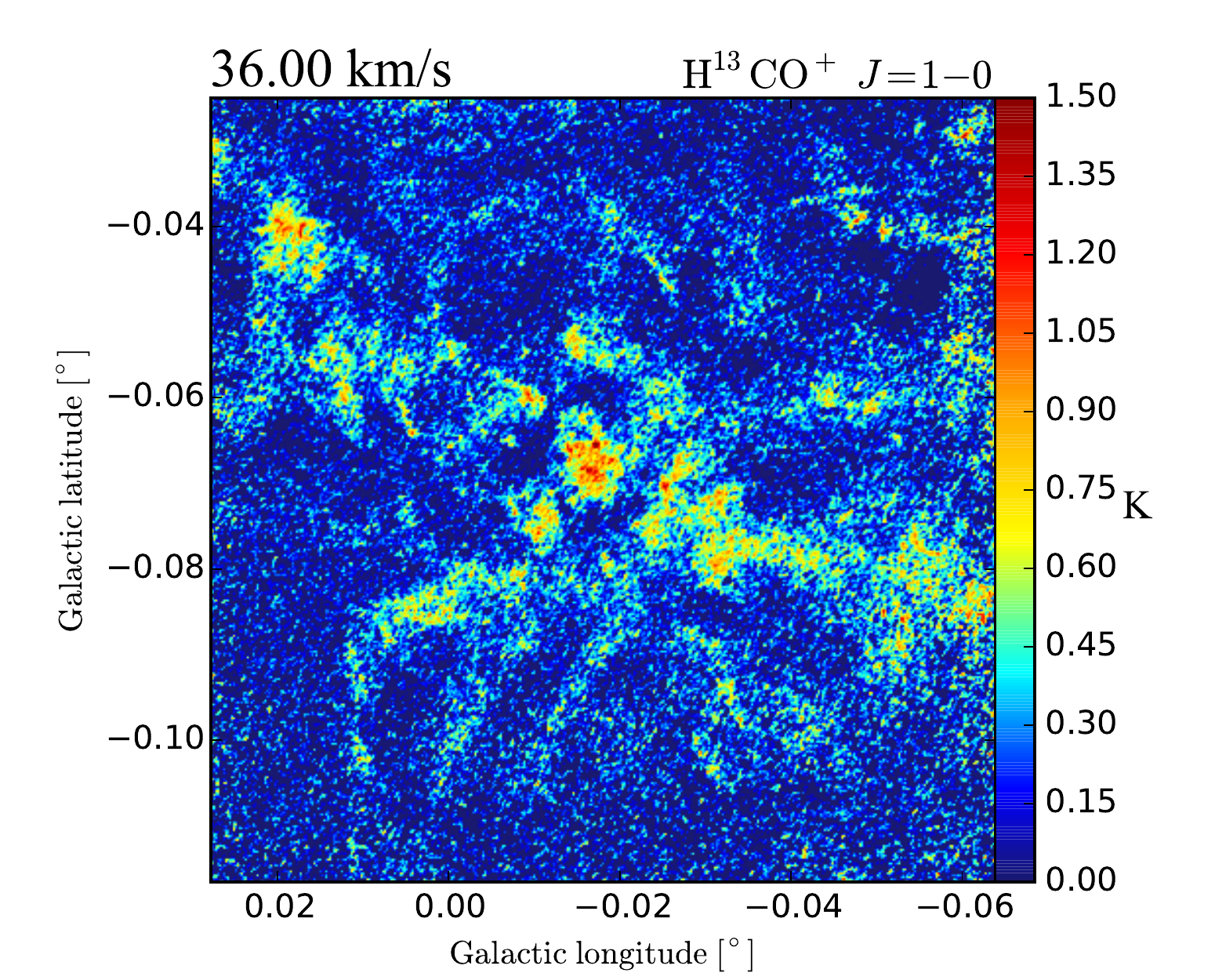}
\caption{Channel maps with velocity widths of $2\rm~km~s^{-1}$ of the ${\rm H^{13}CO^+}~J=1-0$ emission line. The central velocity of each channel is shown in the upper left corner of each panel, which ranges from 0 to 100 $\rm km~s^{-1}$ in $V_{\rm LSR}$. 
Figure \ref{fig:chanmap} is published in its entirety in \url{http://www.vsop.isas.jaxa.jp/~nakahara/figure_uehara/201812/fig3.zip}. 
The range of the brightness temperature $T_{\rm B}$ in the maps is $-28.4$ to $3.6\rm~K$. 
The color bars of the $T_{\rm B}$ are shown on the right side of the rightmost maps with the range of $0.0-1.5\rm~K$ because we do not consider the absorption at the Sgr A$^*$ and H$_{\rm II}$ region D. 
The white star and white crosses show the position of the Sgr A* and four HII regions, respectively. 
The inset figures of each panel show the beam size. 
Additionally, we supply the channel map movie (\url{http://www.vsop.isas.jaxa.jp/~nakahara/figure_uehara/201812/map_h13co+.mp4})
}
\label{fig:chanmap}
\end{figure}

\begin{figure}[ht!]
\plotone{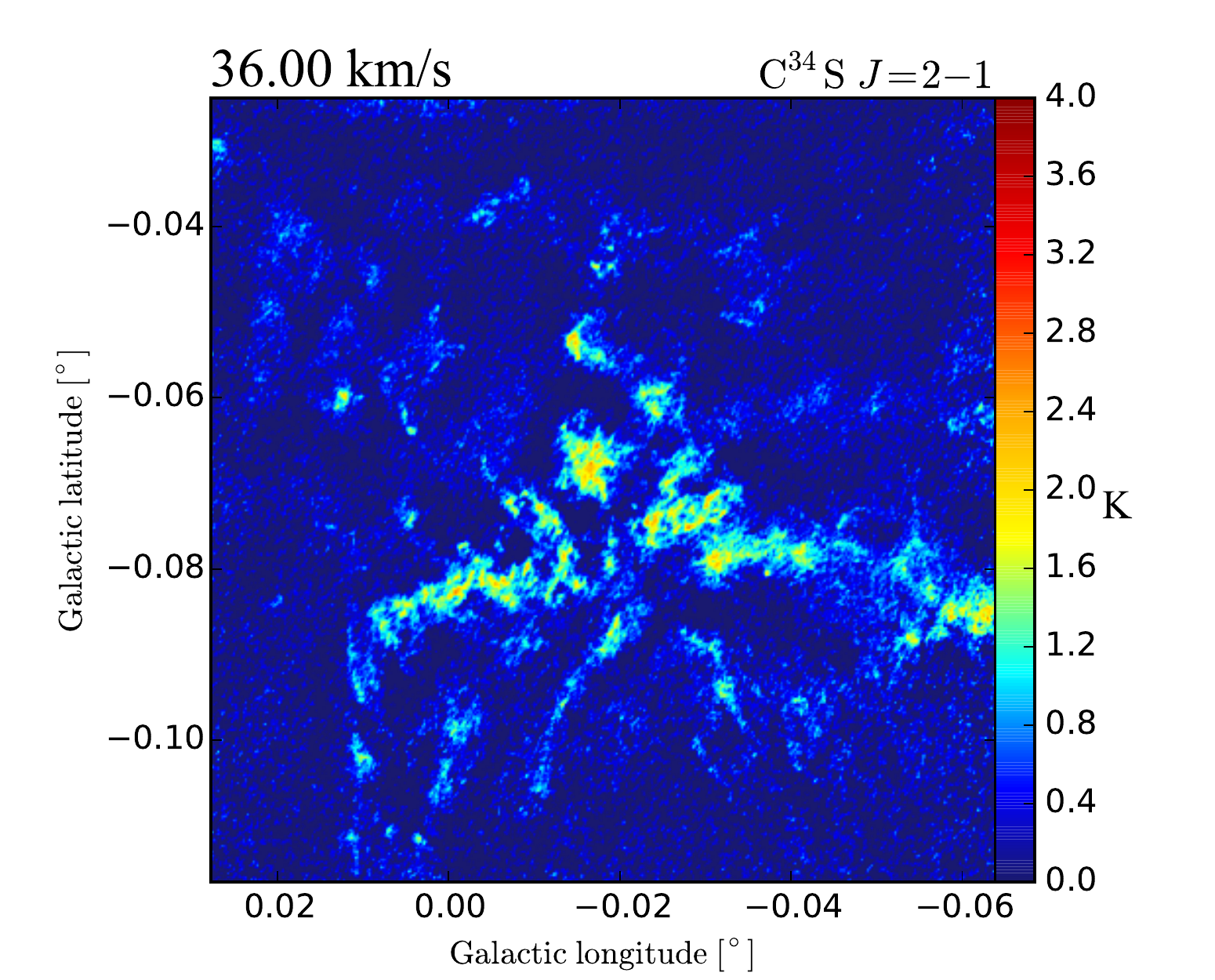}
\caption{Channel maps with velocity widths of $2\rm~km~s^{-1}$ of the $\rm C^{34}S$ $J=2-1$ emission line. The central velocity of each channel is shown on the upper left corner of each panel, which ranges from 0 to 100 $\rm km~s^{-1}$ in $V_{\rm LSR}$. 
Figure \ref{fig:chanmap34} is published in its entirety in {\url{http://www.vsop.isas.jaxa.jp/~nakahara/figure_uehara/201812/fig4.zip}}. 
The range of the brightness temperature $T_{\rm B}$ in the maps is $-4.5$ to $4.0\rm~K$. The color bars of the $T_{\rm B}$ are shown on the right side of the rightmost maps with the range of $0.0-4.0\rm~K$ because we do not consider the absorption at the Sgr A$^*$ and H$_{\rm II}$ region D. 
The white star and white crosses show the position of the Sgr A* and four HII regions, respectively. 
The inset figures of each panel show the beam size. 
{ We supply the channel map movie (\url{http://www.vsop.isas.jaxa.jp/~nakahara/figure_uehara/201812/map_c34s.mp4})}}\label{fig:chanmap34}
\end{figure}
\clearpage
%
\section{Molecular Cloud Core Identification}\label{sec:core}
\subsection{Identification of the molecular cloud { core candidate}s}\label{sec:idcore}

We identified molecular cloud cores in the 3D ($l-b-v_{\rm LSR}$) H$^{13}$CO$^+~J=1-0$ and ${\rm C^{34}S}~ J=2-1$ data with the velocity range of $0-100\rm~km~s^{-1}$ in $V_{\rm LSR}$ using the $clumpfind$ algorithm \citep{williams1994}. 
The clumpfind algorithm finds local peaks as clumps in the data cube and does not determine whether those clumps are bound or not. 
Therefore, we refer to the identified clumps as { core candidates} in this section. 
Before the clumpfind was applied, the maps were resampled using a sample interval of $1\farcs5\times1\farcs5$, which corresponds to the FWHM of the ALMA beam. 
We analyzed { core candidate}s only within the white boundary line in Figure \ref{fig:guide} in order to examine the { core candidate}s in the 50MC. 
In the clumpfind algorithm, the parameters, Lowest contour level and Contour increment, were set to $2\sigma$ and $2\sigma$ corresponding to 0.32 K and 0.32 K, respectively. 
The original FWHM velocity width is calculated as
\begin{eqnarray}
	\Delta v=2\sqrt{2\ln2}\left[\frac{\displaystyle\sum_{i} v_{i}^{2}T_{i}}{\displaystyle\sum_{i} T_{i}}-\left(\frac{\displaystyle\sum_{i} v_{i}T_{i}}{\displaystyle\sum_{i} T_{i}}\right)^{2}\right]^{1/2}, \label{eq:ori-v}
\end{eqnarray}
where $v_{i}$ and $T_{i}$ are the radial velocity and intensity of the $i$-th pixel in each { core candidate}, respectively. 
The FWHM velocity width $\Delta V_{\rm FWHM}$ is corrected for the velocity resolution by 
\begin{eqnarray}
	\Delta V_{\rm FWHM}=\left(\Delta v^{2}-2{[\rm km~s^{-1}]}^{2}\right)^{1/2}. 
\end{eqnarray}

The original radius of the { core candidate} is defined as the effective circular radius, 
\begin{eqnarray}
	r=\left(\frac{A}{\pi}\right)^{1/2}, \label{eq:ori-r}
\end{eqnarray}
where $A$ is the projected area of each { core candidate} derived by the clumpfind. 
The beam-deconvolved radius $R$ of the { core candidate} is calculated by 
\begin{eqnarray}
	R=\left[r^2-\left\{\frac{\sqrt{\theta_{\rm major}\times\theta_{\rm minor}}}{\sqrt{2\ln2}}\left(2\ln\frac{T_{\rm peak}}{\Delta T}\right)^{1/2}\right\}^{2}\right]^{1/2}, 
\end{eqnarray}
where $T_{\rm peak}$, $\Delta T$, $\theta_{\rm major}$ and $\theta_{\rm minor}$ are the peak temperature in the { core candidate}, the threshold level in the { core candidate} identification, the beam semi major axis and the beam semi minor axis, respectively \citep{williams1994}. 
The { core candidate}s whose deconvolved radii are smaller than 0.035pc were rejected because the deconvolved radius is less than the mean beam radius of $\sqrt{\theta_{\rm major}\times\theta_{\rm minor}}/2$, $\sim0.035\rm~pc$. 
Furthermore, we rejected the detected { core candidate}s that do not have three or more pixels with intensities of $\geq3\sigma$ or the { core candidate}s that do not have two or more velocity channels. 
Finally, we identified 3293 { core candidate}s in the ${\rm H^{13}CO^+}$ data and 3192 { core candidate}s in the ${\rm C^{34}S}$ data, respectively. 
The number of the identified { core candidate}s in these data is $\sim100$ times larger than that in the previous work using the CS $J=1-0$ emission line \citep{tsuboi2012}. 
This number is also $\sim15$ times larger than the number of the cores in the Orion A cloud identified by \cite{ikeda2007}. 
The large number seems enough for statistical analysis. 
Because our observation resolved the { core candidate}s nearly to the minimum spatial scale, 0.06 pc, observed by existing single dish telescopes in the Orion A molecular cloud \citep{ikeda2007}, it becomes possible to directly compare the { core candidate} properties in the 50MC and a typical Galactic disk molecular cloud, the Orion A cloud, as mentioned in \S\ref{sec:obs}. 
The { core candidate}s are distributed throughout the 50MC.

\subsection{Mass estimation of the { core candidate}}\label{sec:mass}
The physical parameters of the ${\rm H^{13}CO^+}$ and ${\rm C^{34}S}$ { core candidate}s are estimated in this section and are summarized in Table \ref{tab:core} and \ref{tab:core34}, respectively. 

Firstly, we calculate the column densities of the $\rm H^{13}CO^+~J=1-0$ and ${\rm C^{34}S}$ { core candidate}s.  
From the total intensities of the H$^{13}$CO$^+~J=1-0$ and ${\rm C^{34}S}$ emission lines of the { core candidate}s, $\int T_{\rm MB}dV[{\rm K~km/s}]$, the { core candidate} masses are estimated assuming the local thermodynamic equilibrium (LTE) condition. 
The $\rm H_2$ column densities of the ${\rm H^{13}CO^+}$ { core candidate}s are estimated from the equation given by, 
\begin{eqnarray}\label{eq:N}
	N_{\rm H_2}{\rm[cm^{-2}]}=
	\frac{2.53\times10^{11} }{1-\exp(-4.16/T_{\rm ex,H^{13}CO^+})}
	\frac{\displaystyle \int T_{\rm MB}dV[{\rm K~km/s}]}{X({\rm H^{13}CO^+})}. 
\end{eqnarray}
Here $T_{\rm ex,H^{13}CO^+}$ is the excitation temperature of the H$^{13}$CO$^+~J=1-0$ emission line; $X(\rm H^{13}CO^+)$ is the fractional abundance, $X({\rm H^{13}CO^+}) = N_{\rm H^{13}CO^+}/N_{\rm H_{2}}$, which is the relative abundance of H$^{13}$CO$^+$ molecules to total $\rm H_2$ molecules. 
The $X(\rm H^{13}CO^+)$ is assumed to be $6\times10^{-11}$ with uncertainty of a factor of 2 in the CMZ \citep{amo2011}. 
\cite{amo2011} estimated that the fractional abundances of other CMZ clouds are $X(\rm H^{13}CO^+)=(5-25)\times10^{-11}$, while those of the galactic disk clouds are $X(\rm H^{13}CO^+)=(5-18)\times10^{-11}$. 
It is similar to that in the Orion A of $4.8\times10^{-11}$ \citep{ikeda2007}. 
On the other hand, the $\rm H_2$ column densities of the ${\rm C^{34}S}$ { core candidate}s are estimated from the equation given by, 
\begin{eqnarray}\label{eq:Nc34s}
	N_{\rm H_2}{\rm[cm^{-2}]}=
	9.03\times10^{11} \frac{\exp(2.31/T_{\rm ex,C^{34}S})}{1-\exp(-4.63/T_{\rm ex,C^{34}S})}
	\frac{\displaystyle \int T_{\rm MB}dV[{\rm K~km/s}]}{X({\rm C^{34}S})}. 
\end{eqnarray}
Here $T_{\rm ex, C^{34}S}$ is the excitation temperature of the ${\rm C^{34}S}~J=2-1$ emission line; the fractional abundance $X(\rm {\rm C^{34}S})$ is assumed to be $4.87\times10^{-10}$ \citep{amo2011}. 
For the Orion A cloud, \cite{1997ApJ...482..245U} derived $[{\rm C^{32}S}]/[{\rm ^{12}CO}]=(0.5-2)\times10^{-4}$. 
Using $[{\rm ^{12}CO}]/[{\rm H_{2}}]=6.0\times10^{-5}$ \citep{1982ApJ...262..590F,2002ApJ...578..211S} and $\rm [^{32}S]/[^{34}S]=22.4$, a fractional abundance is estimated to be $X(\rm {\rm C^{34}S})=(1.3-5.3)\times10^{-10}$ in the Orion A. 
This value is consistent with that in the 50MC.

{ 
Using the RADEX LVG (Large Velocity Gradient) algorithm \citep{tak2007}, we estimated the $T_{\rm ex}$ from the ${\rm C^{32}S}~J=1-0$ \citep[NMA:][]{tsuboi2009} and $2-1$ (our ACA+TP data) emission line data within the circle with 78 arcsec radius centered on the 50MC center (the field of view of the NMA). 
The results of the LVG analysis are shown in Figure \ref{fig:lvg}. 
The black thick lines and the red thick lines show the brightness temperature, $T_{\rm B}({\rm C^{32}S}~J=2-1)$, and the brightness temperature ratio, $T_{\rm B}({\rm C^{32}S}~J=2-1)/T_{\rm B}({\rm C^{32}S}~J=1-0)$, respectively. 
The core candidates identified by the ${\rm H^{13}CO^+}$ (panel A) and ${\rm C^{34}S}$ (panel B) observations are plotted on the LVG diagrams as the colored filled circles. 
The colors of the plotted circles indicate the excitation temperature of the ${\rm C^{32}S}~J=2-1$ transition in each core candidate. 
Because the $T_{\rm ex}$ scatters in the wide range of $5-150\rm~K$ for both the ${\rm H^{13}CO^+}$ and ${\rm C^{34}S}$ core candidates, we used the $T_{\rm ex}$ obtained for each core candidate in the column density estimation.

}

\begin{figure}[ht!]
\plottwo{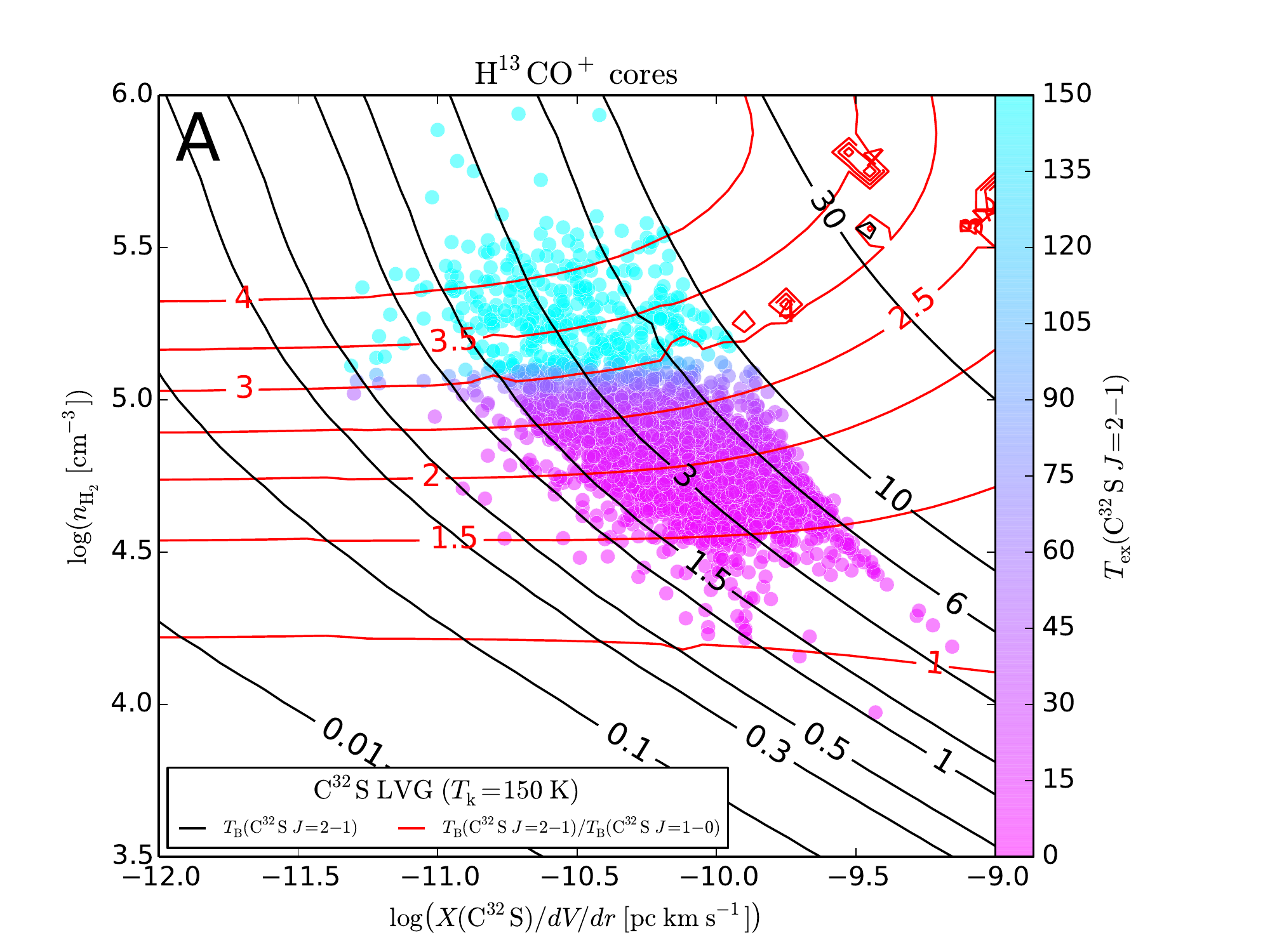}{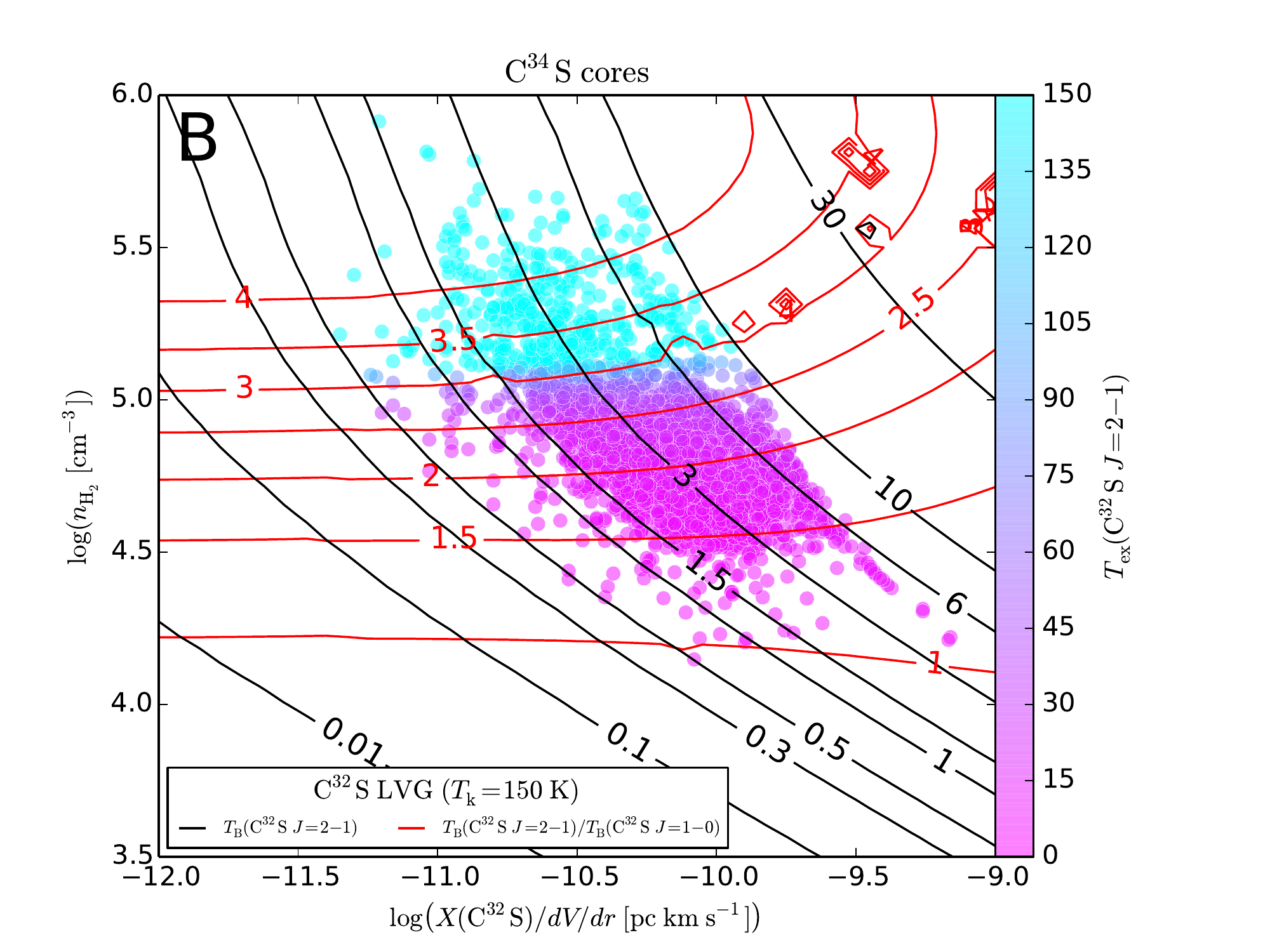}
\caption{Curves of the brightness temperature $T_{\rm B}({\rm C^{32}S}~J=2-1)$ (black lines) and the intensity ratio $T_{\rm B}({\rm C^{32}S}~J=2-1)/T_{\rm B}({\rm C^{32}S}~J=1-0)$ (red lines) are shown on the planes of $\rm H_{2}$ number density vs. CS fractional abundance per velocity gradient. 
[A] The colored filled circles show the observed data of the ${\rm H^{13}CO^+}$ { core candidate}s. 
The color bar shows the excitation temperature of each ${\rm H^{13}CO^+}$ { core candidate}s. 
[B] The colored filled circles show the observed data of the ${\rm C^{34}S}$ { core candidate}. 
The color bar shows the excitation temperature of each ${\rm C^{34}S}$ { core candidate}. 
    }\label{fig:lvg}
\end{figure}

Finally, we calculate the { core candidate} masses from the equation given by, 
\begin{eqnarray}\label{eq:M}
	M[M_{\odot}]=\mu[M_{\odot}]\sum_m\sum_n\left(N_{\rm H_{2}}(m,n){\rm [cm^{-2}]}\Omega[{\rm cm^2}]\right). 
\end{eqnarray}
The summation is done over each { core candidate} area. 
Here $\Omega$ is the physical area of a pixel of the map; $\mu$ is the mean mass of the molecular gas per $\rm H_2$ molecule. 
These values are $\Omega=3.64\times10^{34}\rm cm^2$ for the $1\farcs5$ pixel size at the 8.5 kpc distance and $\mu=2.35\times10^{-57}M_\odot$. 
{ 
We also estimate the mass detection limit in our identification. 
According to our identification, at least 3 pixels of $3\sigma=0.48\rm~K$ or more are included in a core. 
Since the $T_{\rm ex}$ obtained by the LVG analysis is $5-150\rm~K$, the mass detection limit is estimated to be $0.6-13~M_{\odot}$ using equation \ref{eq:N}-\ref{eq:M}. 
Thus, the detection limit of the ${\rm H^{13}CO^+}$ core mass comes to be $0.6~M_{\odot}$. 
Similarly, the detection limit of the ${\rm C^{34}S}$ core mass is estimated to be $0.4~M_{\odot}$. 

The average and range of the ${\rm H^{13}CO^+}$ { core candidate} masses are $(2.3\pm3.7)\times10^{2}M_\odot$ and $4.3 -4500M_\odot$, respectively (See the column 9 in Table \ref{tab:core}). 
The total { core candidate} mass is estimated to be $7.6\times10^5 M_\odot$. 
The total LTE mass of the 50MC is $1.3\times10^6 M_\odot$ from the H$^{13}$CO$^+~J=1-0$ channel maps with the velocity range of $0-100\rm~km~s^{-1}$ in $V_{\rm LSR}$. 
Thus, the ratio of the total core candidate mass to the mass of the whole of the 50MC is $58\%$ ($=7.6\times10^5 M_\odot/1.3\times10^6M_\odot$). 

The average and range of the ${\rm C^{34}S}$ { core candidate} masses also are $(1.9\pm3.6)\times10^{2}M_\odot$ and $2.4 -5500M_\odot$, respectively (See the column 9 in Table \ref{tab:core34}). 
The total mass of the { core candidate} is estimated to be $6.1\times10^5~M_\odot$. 
The total LTE mass of the 50MC is $7.3\times10^5~M_\odot$ from the ${\rm C^{34}S}~J=2-1$ channel maps with the velocity range of $0-100\rm~km~s^{-1}$ in $V_{\rm LSR}$. 
Thus, the mass ratio of the core candidates to the whole 50MC is $83\%$ ($=6.1\times10^5~M_\odot/7.3\times10^5~M_\odot$). 
The total LTE mass of the 50MC is smaller than that derived from the ${\rm H^{13}CO^+}~J=1-0$ maps as estimated above. 
Because the uncertainty of the abundance that is used to estimate the mass is up to a factor of 2 \citep{amo2011}, the two masses coincide within the uncertainties. 
}

Additionally, assuming a sphere shape with radius $R$, the mean number density of the { core candidate}, $\overline{n}$, is given by
\begin{eqnarray}
	\overline{n}~[{\rm cm^{-3}}]=\frac{3M_{\rm LTE}[M_\odot]}{4\pi\mu[M_\odot]R[{\rm cm}]^3}. \label{eq:n}
\end{eqnarray}
These values are summarized in the column 11 in Table \ref{tab:core} and \ref{tab:core34}. 
{ The average and range of the $\overline{n}$ are $(1.7 \pm2.0) \times10^5~{\rm cm^{-3}}$ and $(0.10-25) \times10^5~{\rm cm^{-3}}$ for the ${\rm H^{13}CO^+}$ { core candidate}, respectively. 
The $\overline{n}$ values are comparable to the critical number density of the H$^{13}$CO$^+~J=1-0$ emission line of $\sim10^5~\rm cm^{-3}$. 
For the ${\rm C^{34}S}$ { core candidate}s, the average and range of the densities $\overline{n}$ are $(1.0 \pm1.4) \times10^5~{\rm cm^{-3}}$ and $(0.05-15) \times10^5~{\rm cm^{-3}}$, respectively. 
The $\overline{n}$ values are also comparable to the critical number density of the ${\rm C^{34}S}~J=2-1$ emission line of $\sim10^5~\rm cm^{-3}$. 
}

We estimate the virial masses of the { core candidate}s assuming no external pressure and no magnetic field. 
The virial masses are calculated by the equation 
\begin{eqnarray}
	M_{\rm vir}[M_\odot]&=&5R[{\rm pc}]\sigma[{\rm km~s^{-1}}]^2/G\\
	&=&210\times R[{\rm pc}] \Delta V_{\rm FWHM}[\rm km~s^{-1}]^2, \label{eq:mvir}
\end{eqnarray}
where $\sigma$ and $\Delta V_{\rm FWHM}[\rm km~s^{-1}]$ are the velocity dispersion and the FWHM velocity width of the { core candidate}, respectively (see the column 6 and 8 in the Table \ref{tab:core} and \ref{tab:core34}). 
The average and range of the virial masses are $(17.9\pm15.3)\times10^{2}  M_\odot$ and $(0.83-228)\times10^{2}M_\odot$ for the ${\rm H^{13}CO^+}$ { core candidate}s, respectively. 
Those of the ${\rm C^{34}S}$ { core candidate}s are $(18.8\pm13.6)\times10^{2}  M_\odot$ and $(0.76-137)\times10^{2}M_\odot$, respectively.

In addition, the virial parameters defined by the ratio of the virial mass and the LTE mass ($\alpha=M_{\rm vir}/M_{\rm LTE}$) are calculated. 
{ The large $\alpha$ ($>1$) indicates that the { core candidate} is unbound by self-gravity, whereas the small $\alpha$ ($<1$) indicates that the { core candidate} is bound by self-gravity. }
These values are summarized in the column 10 in Table \ref{tab:core} and \ref{tab:core34}. 
{ For the ${\rm H^{13}CO^+}$ { core candidate}s, the average and range of the virial parameter are $19\pm25$ and $0.40-450$, respectively. 
These values of the ${\rm C^{34}S}$ { core candidate}s are $38\pm52$ and $0.8-810$, respectively. }
The virial parameter in the 50MC is two to three orders of magnitude larger than that in the Orion A \citep{ikeda2007} and is also larger than that of the whole of the 50MC ($\alpha\sim2$) \citep{tsuboi2011}. 
They suggest that the gas in the { core candidate}s is strongly turbulent and are often unbound by self-gravity.


\section{Discussion}\label{sec:property}
\subsection{Identification of Bound cores}\label{sec:r}
{ 
Because the range of the virial parameters of the cores in the Orion A is 0.2-4, almost all of the cores are likely to be bound by self-gravity \citep{ikeda2007}. 
On the other hand, the virial parameters in the 50MC scatter in the range of 0.4-810 (\S\ref{sec:mass}), indicating a mixture of bound and unbound cores. 
The criterion for the bound cores is nominally that the virial parameter is less than unity. 
Because the uncertainty of the fractional abundance of $X({\rm H^{13}CO^+})$ is as large as a factor of 2 \citep{amo2011}, the ${\rm H^{13}CO^+}$ core candidate mass may be underestimated down to a factor of 0.5. 
Therefore, we consider the ${\rm H^{13}CO^+}$ core candidates with the virial parameters of less than 2 as ''bound cores''. 
Additionally, in the previous works \citep[e.g.][]{ikeda2007,2009ApJ...691.1560I}, the cores identified by the clumpfind have the radii of $\sim0.1\rm~pc$. 
The radii of the bound ${\rm H^{13}CO^+}$ cores in the 50MC are comparable to those in the previous works. 
The 241 bound cores were identified. 
The number corresponds to  $7\%~(=241/3293)$ of all the identified core candidates. 
The positions of the bound cores are shown in Figure \ref{fig:core}-A. 
Thus, we use only the bound cores in the CMF analysis hereafter in order to compare the cores between the 50MC and the Orion A. 
Meanwhile, the bound ${\rm C^{34}S}$ cores with $\alpha<2$ ($\sim4\%=129/3192$ of all the identified ${\rm C^{34}S}$ core candidates) are plotted in Figure \ref{fig:core}-B (the uncertainty in $X({\rm C^{34}S})$ is also a factor of $\sim2$). 
The physical parameters of the bound ${\rm H^{13}CO^+}$ and ${\rm C^{34}S}$ cores are summarized in Table \ref{tab:bound}. 
On the other hand, we call the core candidates with $\alpha>2$ ''transient cores''.

\begin{figure}[ht!]
	\plottwo{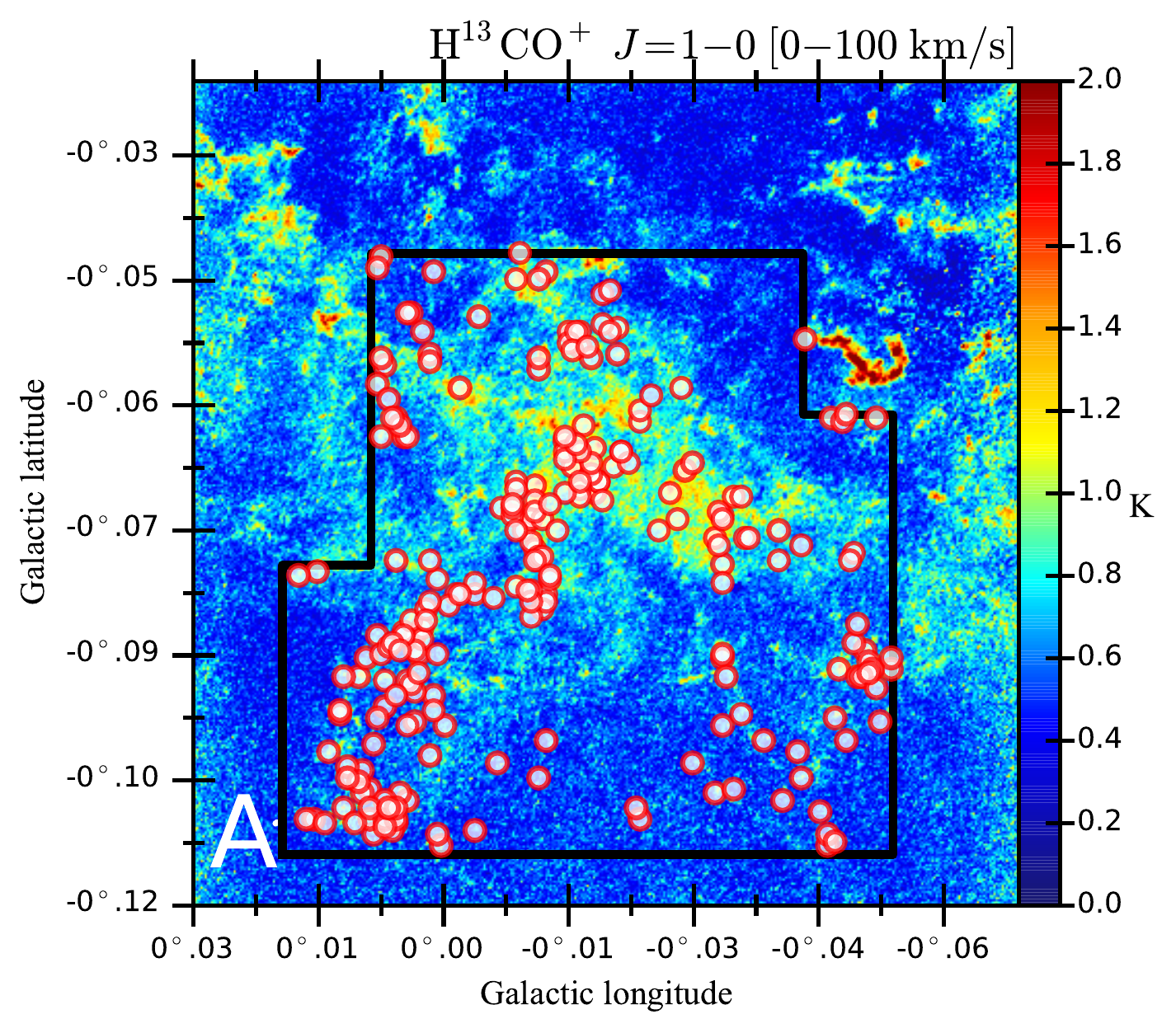}{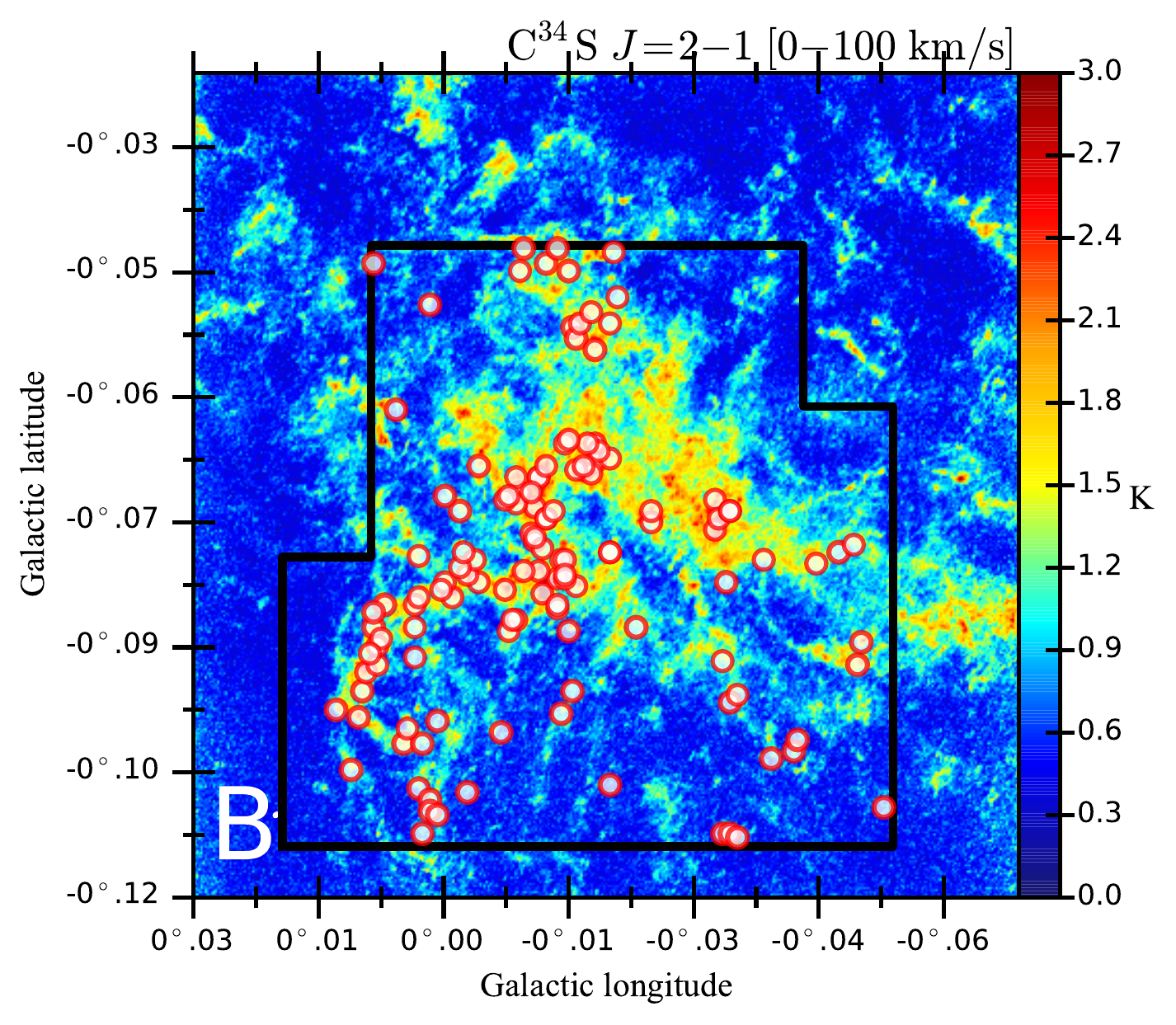}
	\caption{[A] Positions of the identified bound ${\rm H^{13}CO^+}$ cores with $\alpha<2$ on the $T_{\rm peak}$ map of the ${\rm H^{13}CO^+}~J=1-0$ emission line in the units of brightness temperature ($\rm K$) over the velocity range of $V_{\rm LSR}=0-100\rm~km~s^{-1}$. The cores outside the polygon with the black thick line are rejected. 
    [B] Positions of the identified bound ${\rm C^{34}S}$ cores with $\alpha<2$ on the $T_{\rm peak}$ map of the ${\rm C^{34}S} J=2-1$ emission line in the units of brightness temperature ($\rm K$) over the velocity range of $V_{\rm LSR}=0-100\rm~km~s^{-1}$. The cores outside the polygon with the black thick line are rejected. }\label{fig:core}
\end{figure}

Figure \ref{fig:h2c} shows the bound cores identified in both the ${\rm H^{13}CO^+}~J=1-0$ and ${\rm C^{34}S}~J=2-1$ emission lines. 
We regard the bound cores that satisfy the following two criteria as the identification by the both lines: 
\begin{enumerate}
\item
The distance between the centers of the $\rm C^{34}S $ and $\rm H^{13}CO^+ $ cores is smaller than the larger radius of the two cores. 
\item
The difference between the center LSR velocities of the $\rm C^{34}S $ and $\rm H^{13}CO^+ $ cores is smaller than the larger velocity width of the two cores. 
\end{enumerate}
Finally, $38~\%~(=49/129)$ of ${\rm C^{34}S}$ bound cores are found to have the ${\rm H^{13}CO^+}$ counterparts. }

\begin{figure}
        \plotone{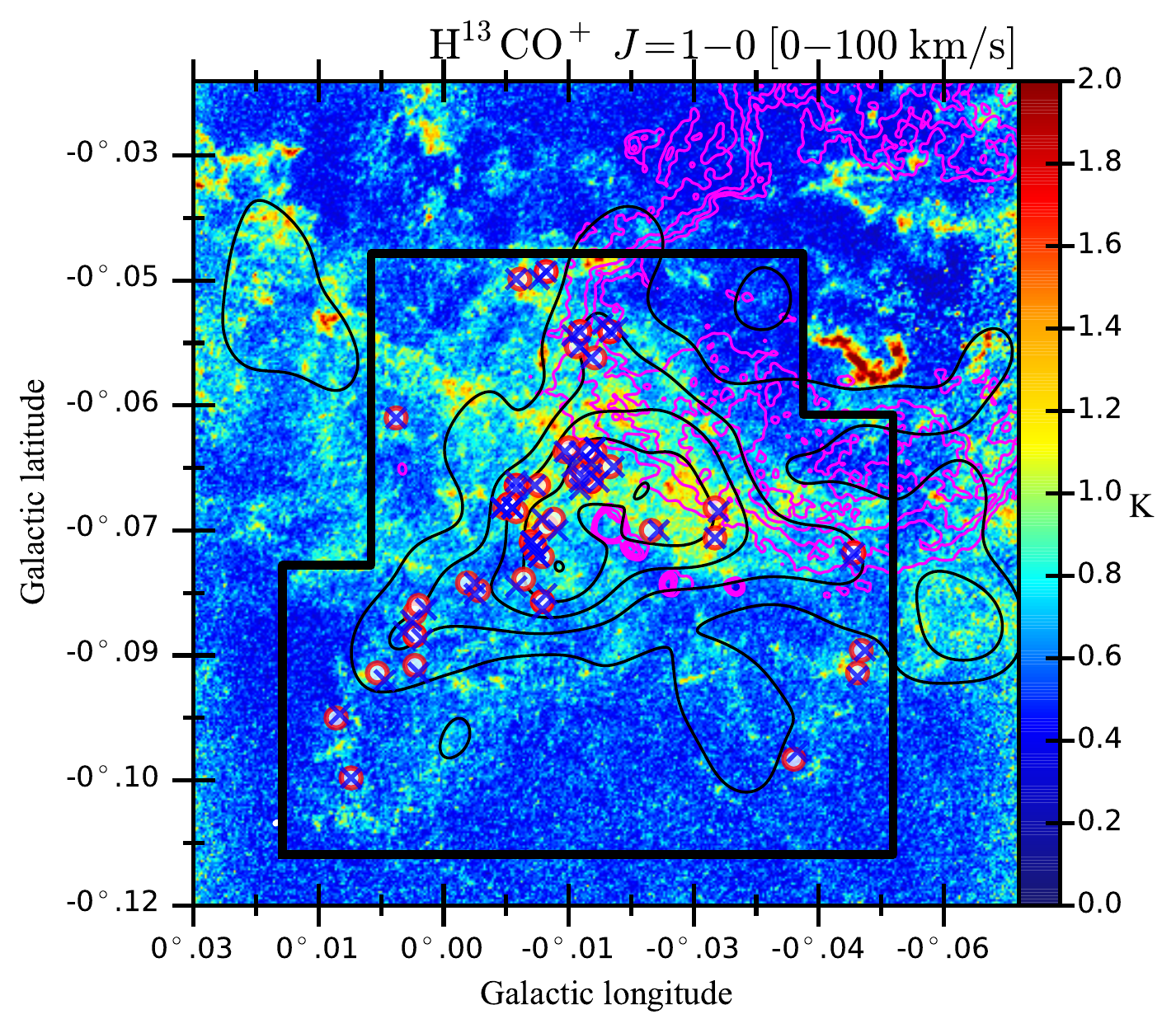}
    \caption{[A] Bound cores identified in both the ${\rm H^{13}CO^+}~J=1-0$ (blue cross) and ${\rm C^{34}S}~J=2-1$ (red open circle) emission lines plotted on the same ${\rm H^{13}CO^+}$ map as in Figure \ref{fig:core}. 
    Red contours show the H42$\alpha$~recombination line. 
    Black contours show the $R_T=T_{\rm B}({\rm SiO}~J=2-1)/T_{\rm B}({\rm H^{13}CO^+}~J=1-0)$ smoothed to $26''$ resolution. 
    The contour levels are 1.4,2.5,3.6,4.7 and 5.8. 
    }\label{fig:h2c}
\end{figure}

\subsection{Relation between the bound and transient cores}
We compare the physical parameters of the bound and transient cores to understand the formation process of the bound cores. 
Note that the massive bound cores with greater than $\sim1000~M_{\odot}$ might be precursors of stellar clusters. 
The average and range of the bound ${\rm H^{13}CO^+}$ core LTE masses, $M_{\rm LTE}$, are $960\pm850~M_{\odot}$ and $48-4500\rm~M_{\odot}$, respectively. 
The core with the smallest mass of $48~M_{\odot}$ is larger than the detection limit of the ${\rm H^{13}CO^+}$  core mass; the mass detection limit is $13~M_{\odot}$ at $T_{\rm ex, {\rm H^{13}CO^+} }=150\rm~K$. 
On the other hand, the average and range of the ${\rm H^{13}CO^+}$ { transient core} LTE masses are $170\pm210~M_{\odot}$ and $4.3-2000\rm~M_{\odot}$, respectively. 
The average and range of the bound core masses seem to be larger than those of the { transient core}s. 
Figure \ref{fig:mcum}-A shows the cumulative distribution functions (CDFs) for the LTE masses of the bound (red line) and transient (black line) ${\rm H^{13}CO^+}$ cores. 
The mass distribution of the bound ${\rm H^{13}CO^+}$ cores is also biased to a larger mass than that of the transient ${\rm H^{13}CO^+}$ cores. 
The mass ratio of the total bound core mass to the total gas mass is $18~\%(=2.3\times10^{5}~M_{\sun}/1.3\times10^{6}~M_{\sun})$. 
{ The low mass ratio is consistent with the lower star formation rate in the CMZ, $\sim0.1~M_{\odot}\rm~yr^{-1}$ \citep[e.g.][]{1989IAUS..136...89G,2009ApJ...702..178Y,2017MNRAS.469.2263B}, than that in the disk region, $\sim2.0~M_{\odot}\rm~yr^{-1}$ \citep[e.g.][]{2011AJ....142..197C}. }

Figure \ref{fig:mvircum}-A shows the CDFs for the virial masses of the bound (red line) and transient (black line) ${\rm H^{13}CO^+}$ cores. 
The average and range of the bound and { transient core} virial masses are indicated in Figure \ref{fig:mvircum}-A. 
The average and range of the bound cores are consistent with those of the { transient core}s within the uncertainties, respectively. 
However, the virial mass distribution of the bound ${\rm H^{13}CO^+}$ cores seems to be biased to a slightly smaller value than that of the transient ${\rm H^{13}CO^+}$ cores as shown in Figure \ref{fig:mvircum}-A.

The mean velocity widths of the bound and transient ${\rm H^{13}CO^+}$ cores are $5.4\pm1.8\rm~km~s^{-1}$ and $6.8\pm2.1\rm~km~s^{-1}$, respectively. 
In addition, the velocity width ranges of the bound and transient ${\rm H^{13}CO^+}$ cores are $2.1-11\rm~km~s^{-1}$ and $2.2-22\rm~km~s^{-1}$, respectively. 
The mean velocity width of the bound cores is consistent with that of the { transient core}s within the uncertainties. 
However, Figure \ref{fig:dvcum}-A shows the CDF for the velocity widths of the bound (red line) and transient (black line) ${\rm H^{13}CO^+}$ cores, indicating that the velocity width distribution of the bound cores is biased to a smaller velocity width than that of the { transient core}s. 

Figure \ref{fig:rcum}-A shows the CDFs for the radii of the bound (red line) and transient (black line) ${\rm H^{13}CO^+}$ cores. 
The average and range of the bound and transient ${\rm H^{13}CO^+}$ cores radii are shown in Figure \ref{fig:rcum}-A. 
The mean radius of the bound cores is consistent with that of the { transient core}s within the uncertainties. 
On the other hand, the mean and range of the radii of the cores in the Orion A cloud are $0.14\pm0.03\rm~pc$ and $0.06-0.23\rm~pc$, respectively, which are similar to those of the bound cores in 50MC.
The radius distribution of the bound cores seems to be slightly larger than that of the transient cores as shown in Figure \ref{fig:rcum}-A.

As shown above, the smaller velocity widths of the bound ${\rm H^{13}CO^+}$ cores make them bound by self-gravity, compared to the larger widths of the unbound cores. 
The mean radius ratio of the bound and transient ${\rm H^{13}CO^+}$ cores is $1.1$ and the velocity width ratio is $0.8$. 
From these ratios, the mean virial mass of the bound ${\rm H^{13}CO^+}$ cores are $1.1\times0.8^{2}=0.7$ times smaller than that of the transient ${\rm H^{13}CO^+}$ cores because the virial mass is proportional to the radius and the square of the velocity width. 
Additionally, the small virial parameters of the bound ${\rm H^{13}CO^+}$ cores also depend on the distribution of the LTE masses biased to the large-mass side.

On the other hand, the bound ${\rm C^{34}S}$ cores have the larger LTE masses than the transient ${\rm C^{34}S}$ cores (see Figure \ref{fig:mcum}-B) although the virial masses of the bound ${\rm C^{34}S}$ cores are consistent with those of the transient ${\rm C^{34}S}$ cores shown in Figure \ref{fig:mvircum}-B. 
The core with the smallest mass of $48~M_{\odot}$ is larger than the detection limit of the ${\rm C^{34}S}$ core mass; the mass detection limit is $5~M_{\odot}$ at $T_{\rm ex, {\rm C^{34}S} }=150\rm~K$. 
The mass ratio of the total bound core mass to the total gas mass is $23~\%(=1.7\times10^{5}~M_{\sun}/7.3\times10^{5}~M_{\sun})$. 
Additionally, the bound ${\rm C^{34}S}$ cores have the larger radii than the transient ${\rm C^{34}S}$ cores (see Figure \ref{fig:rcum}-B) although the velocity widths of the bound ${\rm C^{34}S}$ cores are consistent with those of the transient ${\rm C^{34}S}$ cores (see Figure \ref{fig:dvcum}-B). 
The mean radius ratio of the bound and unbound ${\rm C^{34}S}$ cores is 1.29, and the velocity width ratio is 0.84.
From these ratios, the mean virial masses of the bound and unbound ${\rm C^{34}S}$ cores are not different from each other because the mean virial mass of the bound ${\rm C^{34}S}$ cores is $1.29\times0.84^{2}=0.92$ times smaller than that of the unbound ${\rm C^{34}S}$ cores. 
Thus, because the virial masses distribution of the bound ${\rm C^{34}S}$ cores is consistent with that of the transient cores, the ${\rm C^{34}S}$ cores need to have large masses in order for the virial parameter to be smaller than 2.

\begin{figure}
    \plottwo{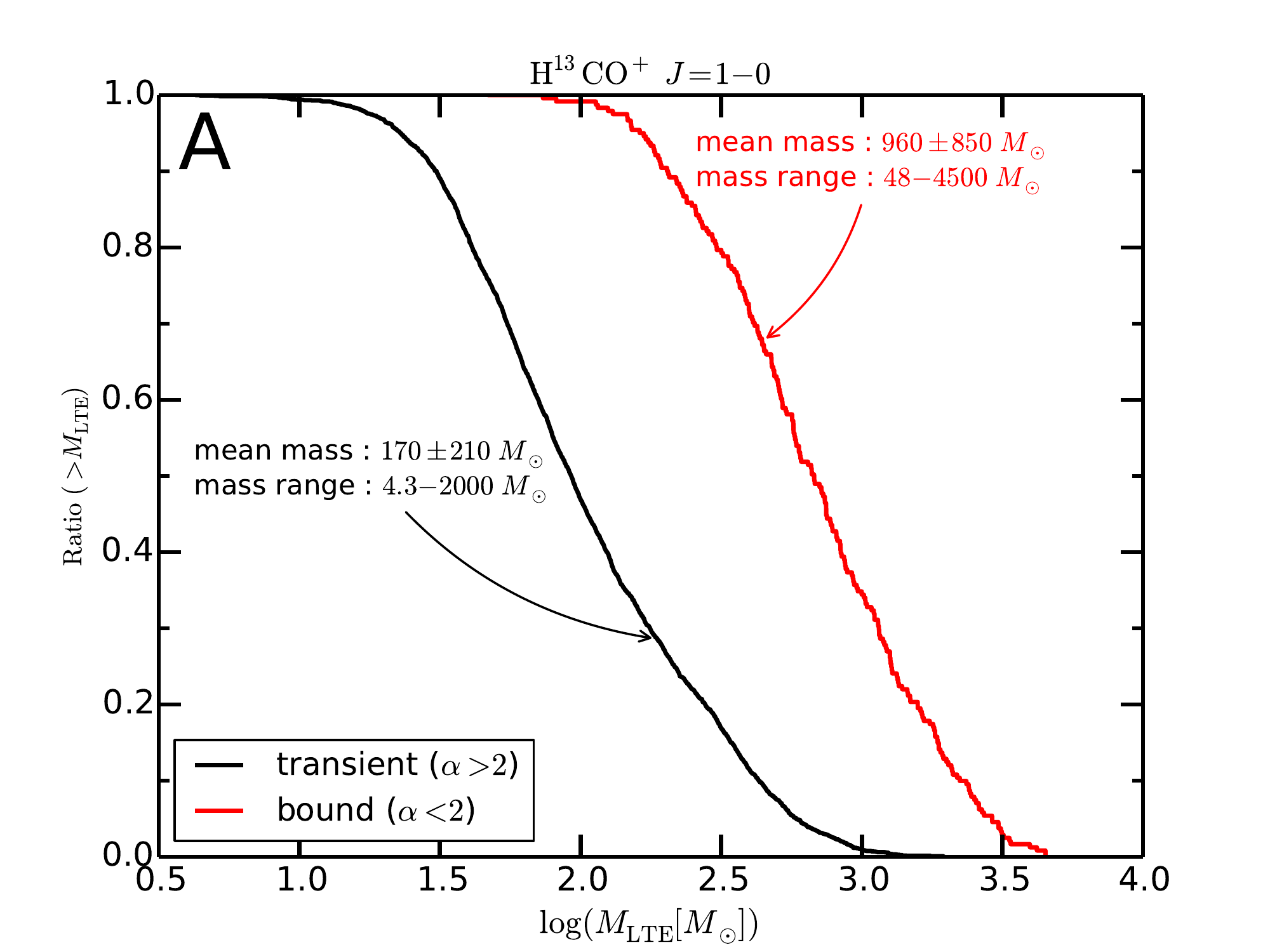}{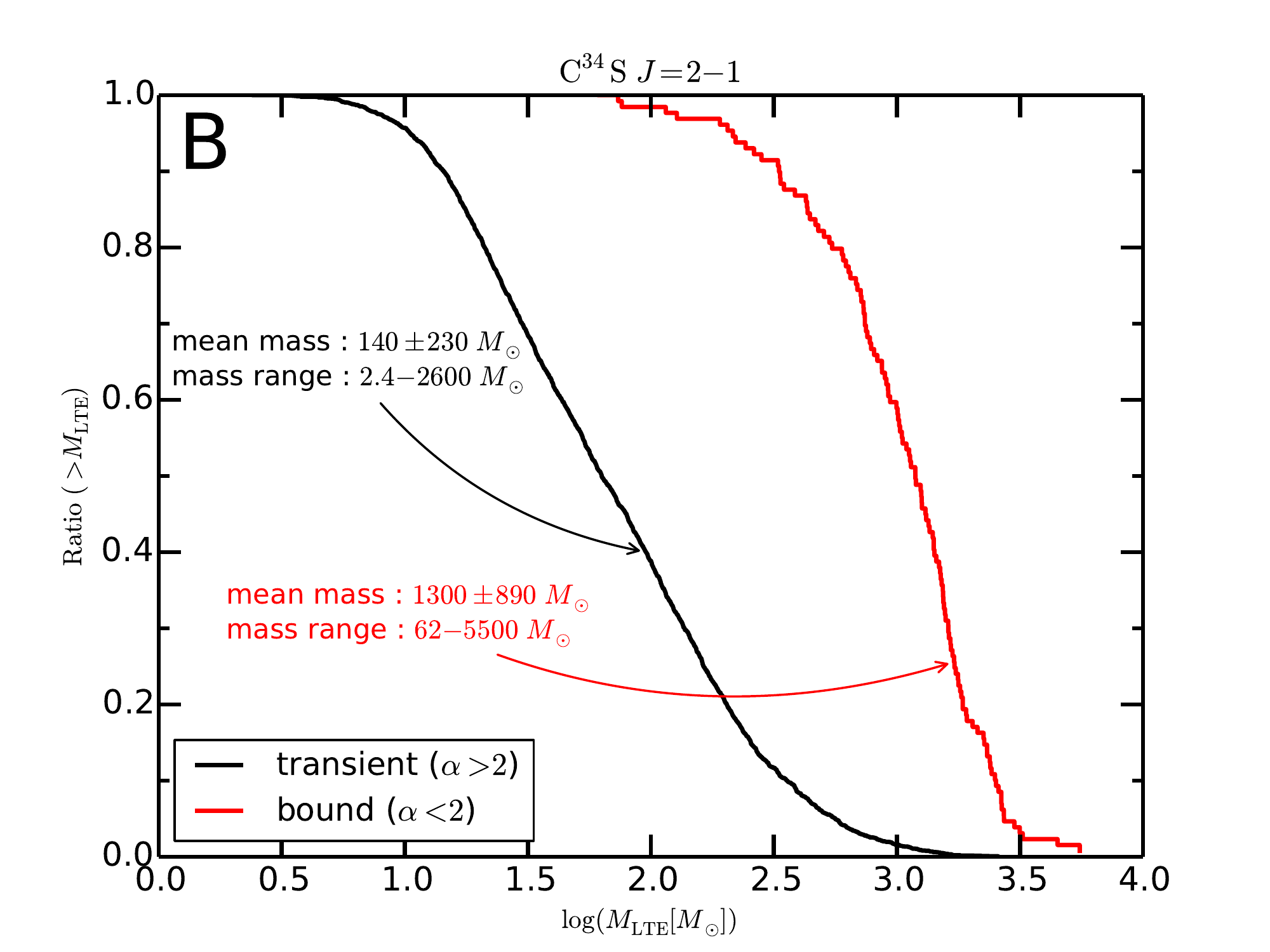}
    \caption{[A] CDF of the ${\rm H^{13}CO^+}~J=1-0$ core mass. The black thick line shows the distribution of the { transient core}s with $\alpha>2$, while the red thick line shows the distribution of the bound cores with $\alpha<2$. [B] CDF of the ${\rm C^{34}S}~J=2-1$ core mass. }\label{fig:mcum}
\end{figure}

\begin{figure}
    \plottwo{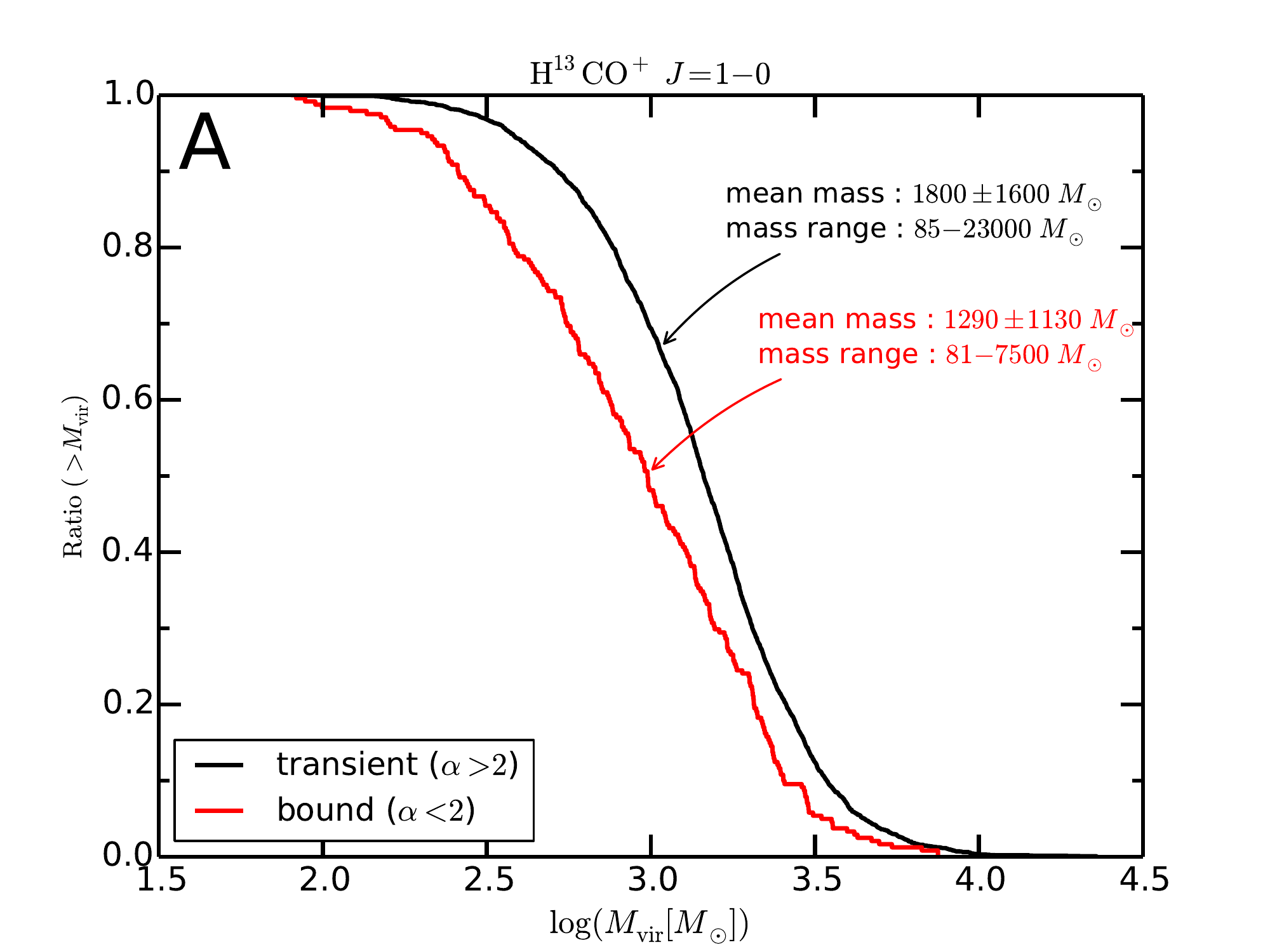}{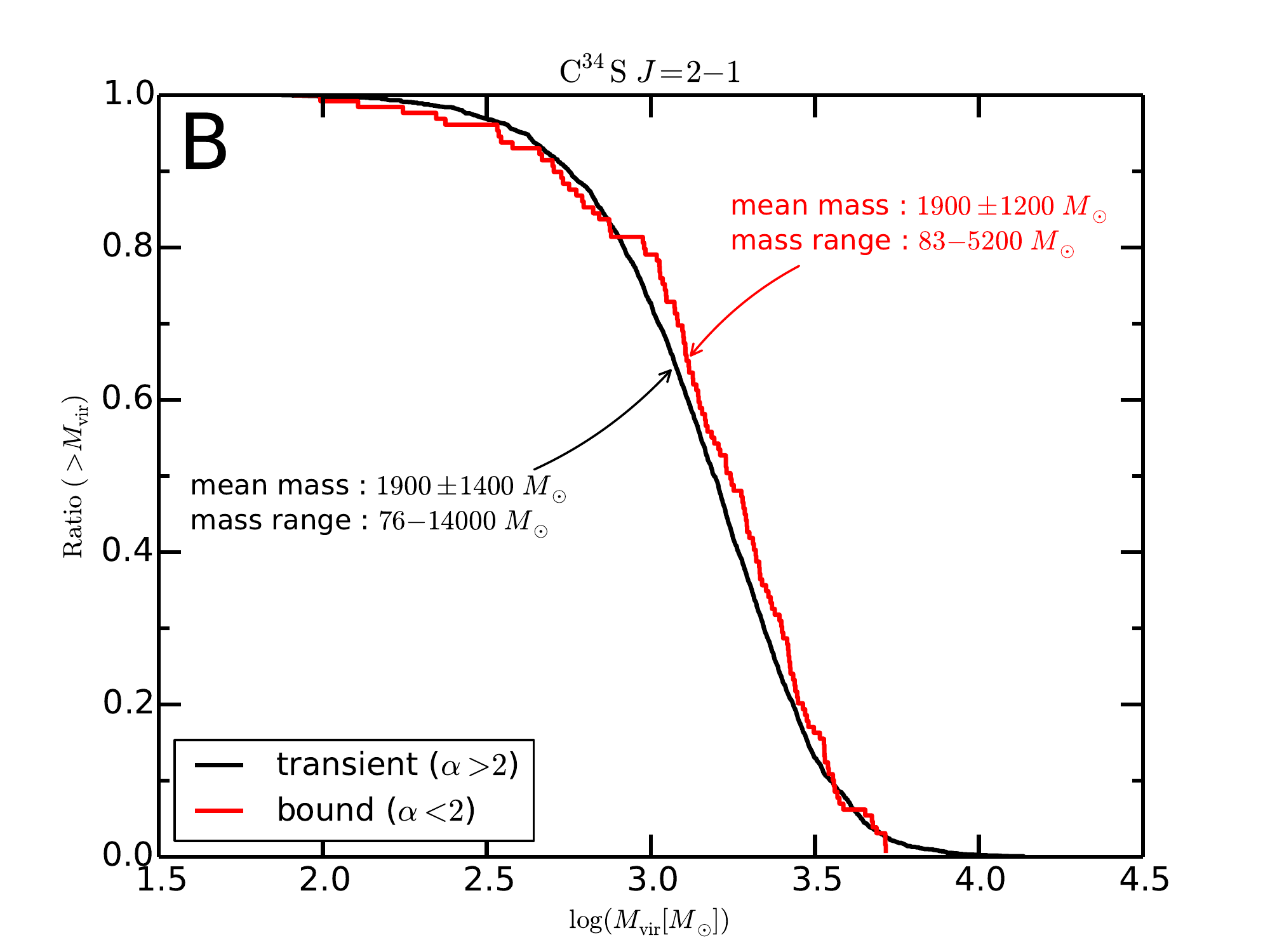}
    \caption{[A] CDF of the ${\rm H^{13}CO^+}~J=1-0$ core virial mass. The black thick line shows the distribution of the { transient core}s with $\alpha>2$, while the red thick line shows the distribution of the bound cores with $\alpha<2$. [B] CDF of the ${\rm C^{34}S}~J=2-1$ core virial mass. }\label{fig:mvircum}
\end{figure}

\begin{figure}
    \plottwo{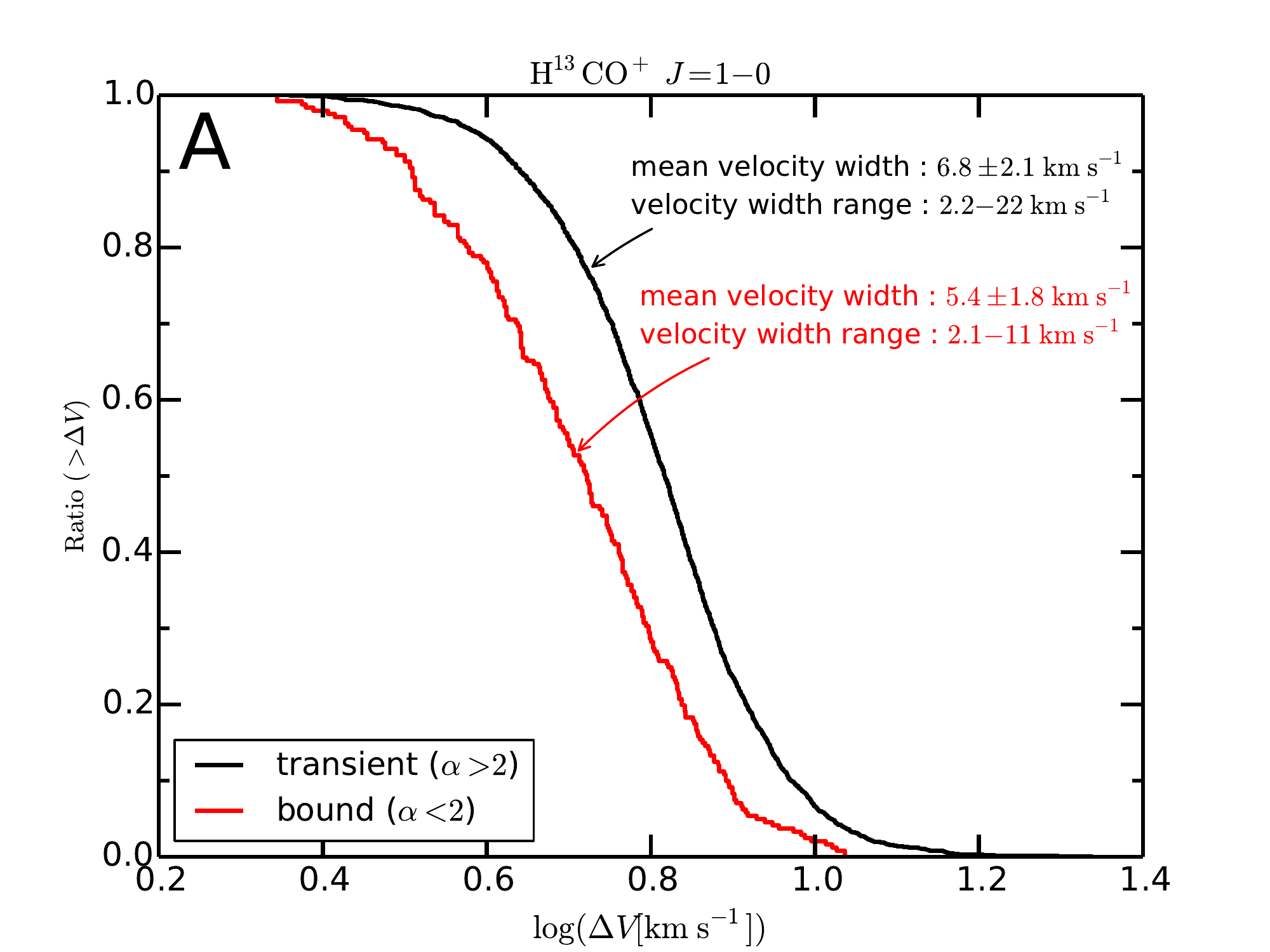}{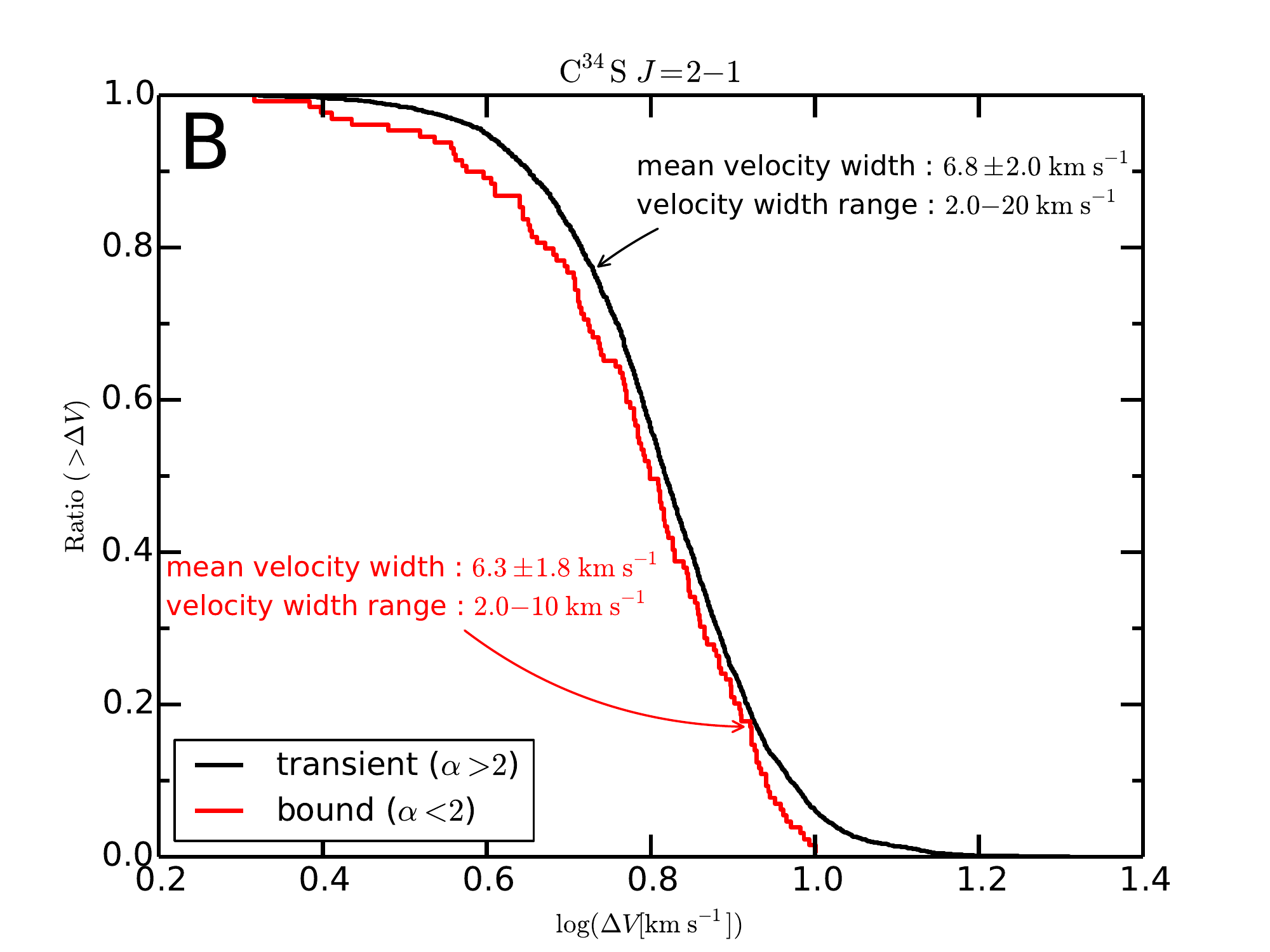}
    \caption{[A] CDF of the ${\rm H^{13}CO^+}~J=1-0$ core velocity width. The black thick line shows the distribution of the { transient core}s with $\alpha>2$, while the red thick line shows the distribution of the bound cores with $\alpha<2$. [B] CDF of the ${\rm C^{34}S}~J=2-1$ core velocity width. }\label{fig:dvcum}
\end{figure}

\begin{figure}
    \plottwo{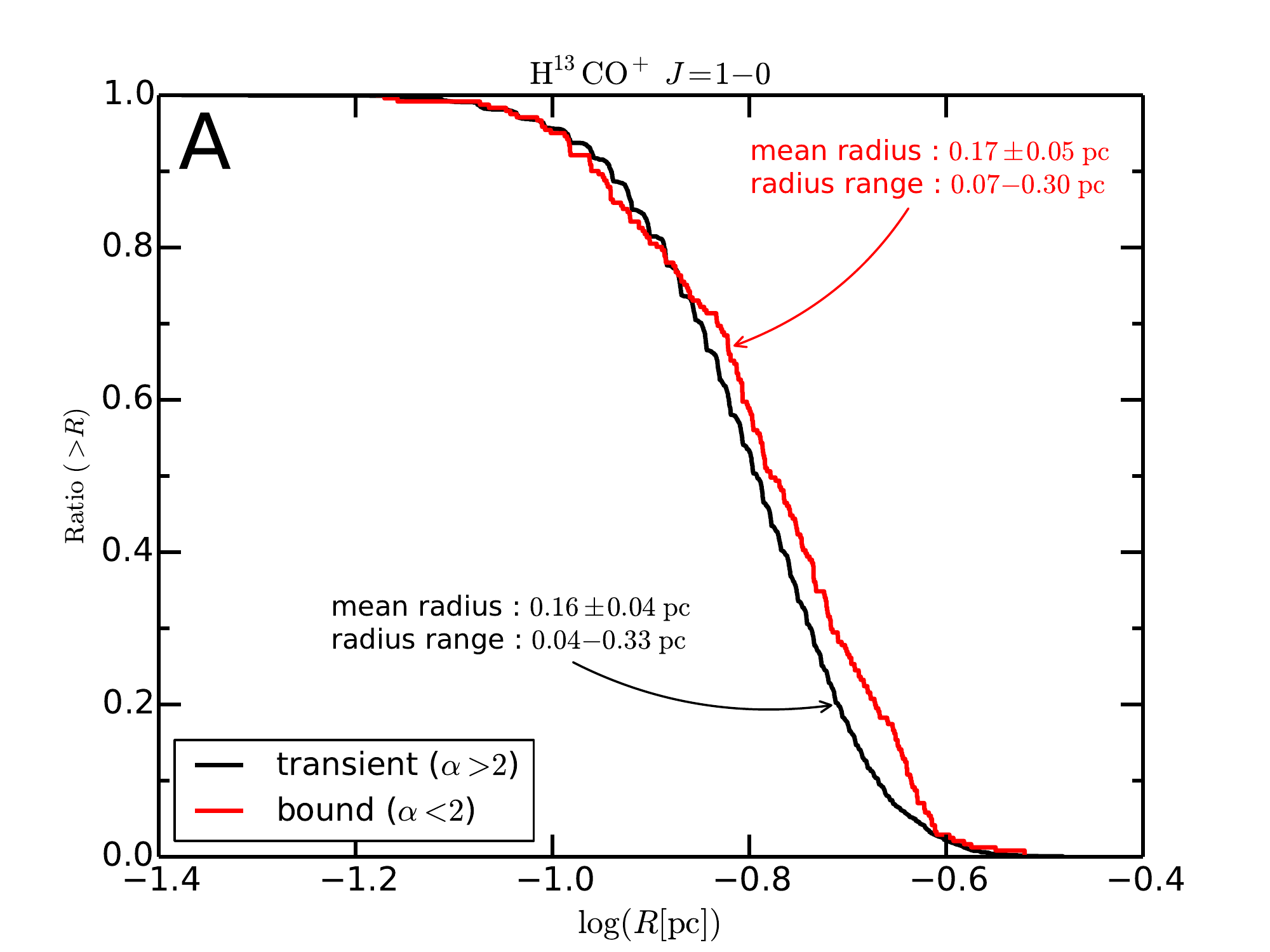}{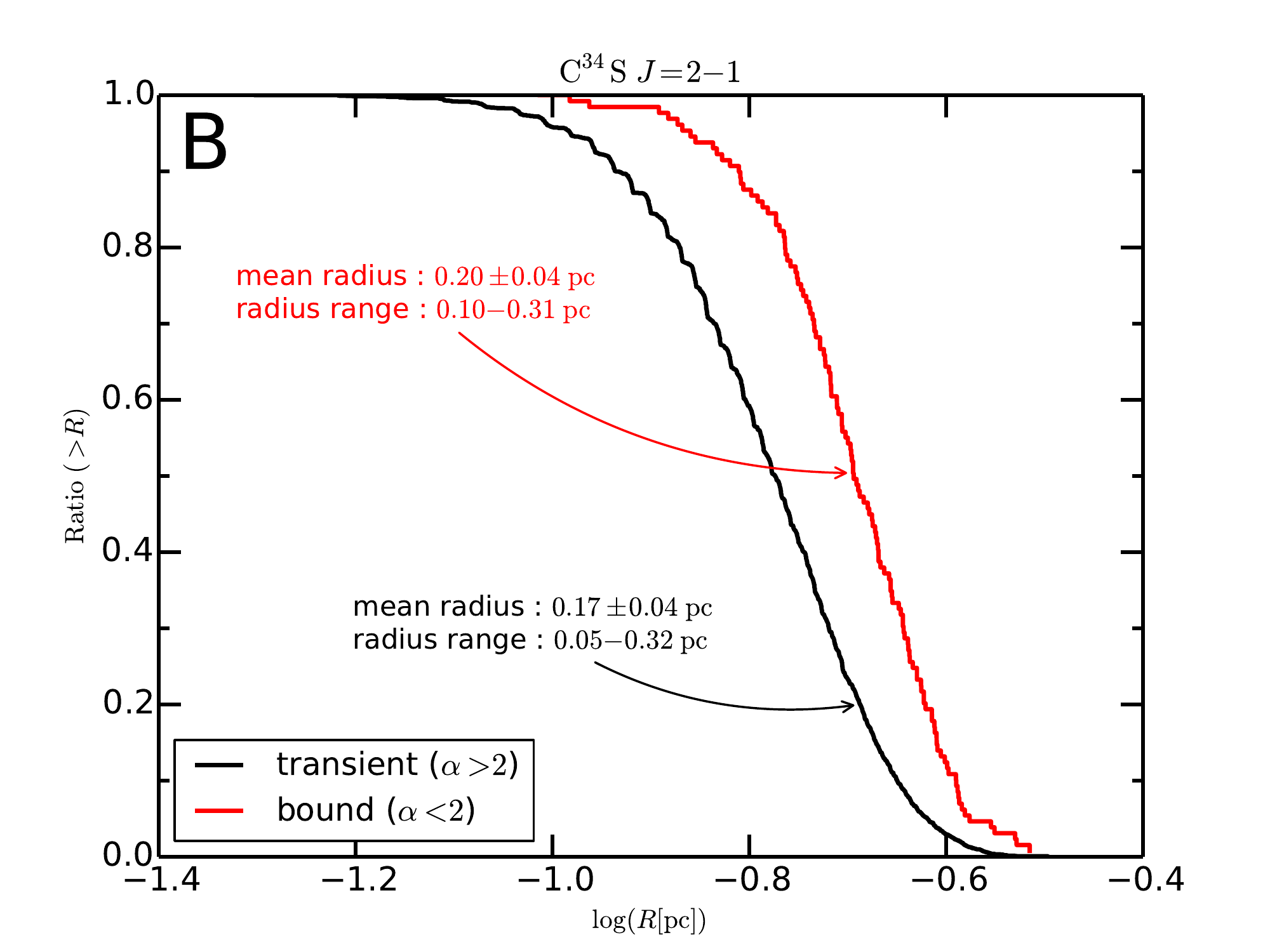}
    \caption{[A] CDF of the ${\rm H^{13}CO^+}~J=1-0$ core radii. The black thick line shows the distribution of the { transient core}s with $\alpha>2$, while the red thick line shows the distribution of the bound cores with $\alpha<2$. [B] CDF of the ${\rm C^{34}S}~J=2-1$ core radii. }\label{fig:rcum}
\end{figure}


\subsection{Comparison of the bound cores in the 50MC and Orion A}\label{sec:cmf}
The spatial resolution of our data, $\sim0.07\rm~pc$, is similar to that in the Orion A observed by the NRO45 \citep[e.g. $\sim0.05\rm~pc$; ][]{ikeda2007}. 
Therefore, we can compare directly massive star forming processes in the GC 50MC and the typical Galactic disk molecular cloud, the Orion A cloud. 

Firstly, most of the ${\rm H^{13}CO^+}$ { core candidate}s in the 50MC ($\sim93\%=3052/3293$) have $\sim10-100$ times larger virial parameters than those in the Orion A \citep{ikeda2007} and are unbound only by self-gravity. 
Similarly, most of the ${\rm C^{34}S}$ { core candidate}s ($\sim96\%=3063/3192$) have $\sim10-100$ times larger virial parameters than those in the Orion A. 
Thus, some external pressure is needed for confinement of the unbound { core candidate}s. 
The { core candidate}s are probably embedded in the ambient gas that is observed in lower critical density lines such as the ${\rm CO}~J=1-0$ and ${\rm CS}~J=1-0$ emission lines. 
If the ambient gas has low density but is highly turbulent, the { core candidate}s may be bound by the external pressure of the gas. 
In this paper, note that the { core candidate}s with $\alpha>2$ are treated as { transient core}s (probably pressure-confined cores) which we will discuss in a future paper. 

Figure \ref{fig:r-dv} shows the radius-velocity width relation of the dense cores ($R$-$\Delta V$ relation). 
{ The velocity widths of the bound cores in the 50MC are 10 times larger than those of the cores in the Orion A cloud. 
However, the radii of the bound cores detected in the 50MC are similar to those of the cores in the Orion A cloud. 

\begin{figure}[ht!]
\plottwo{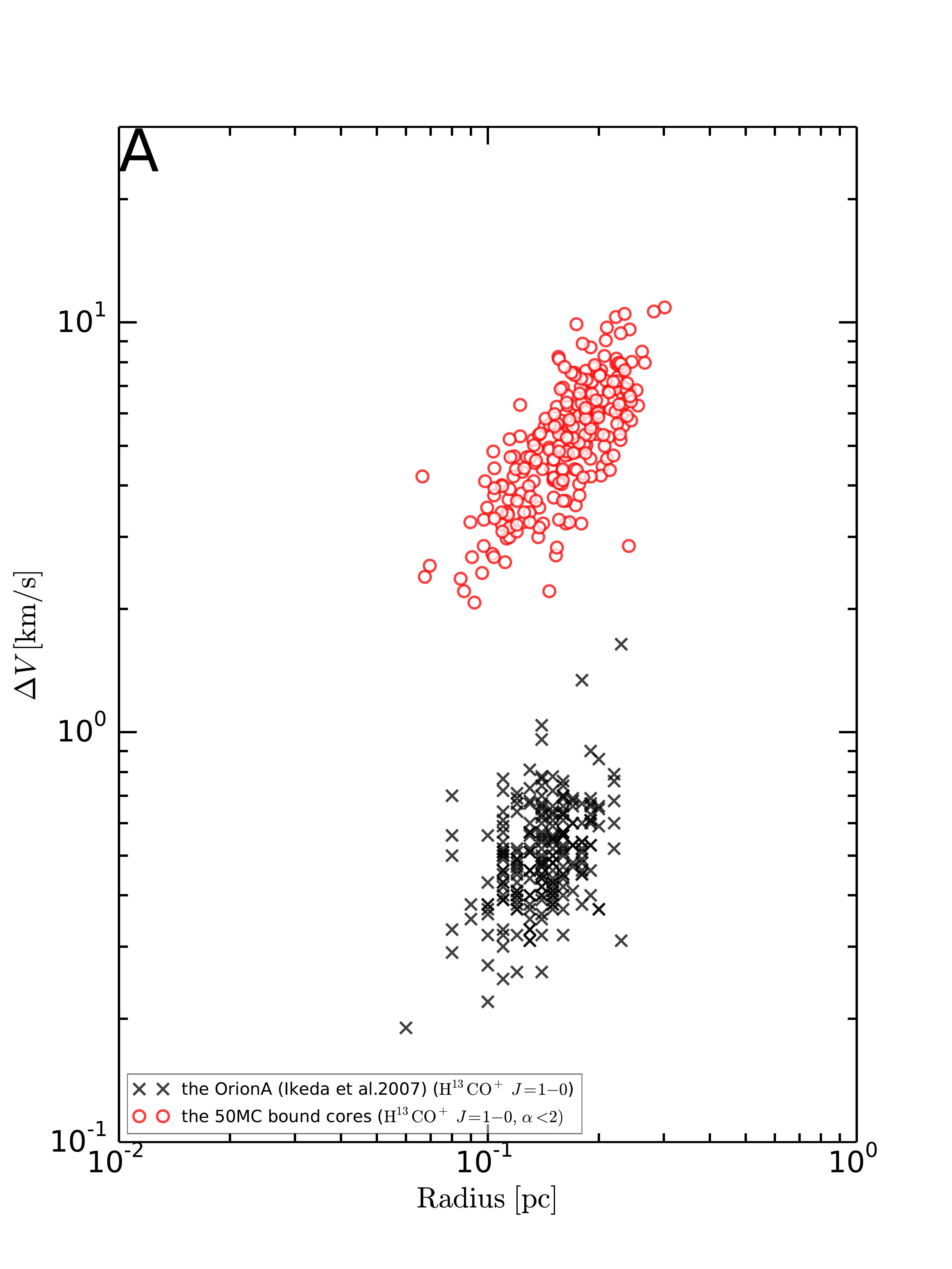}{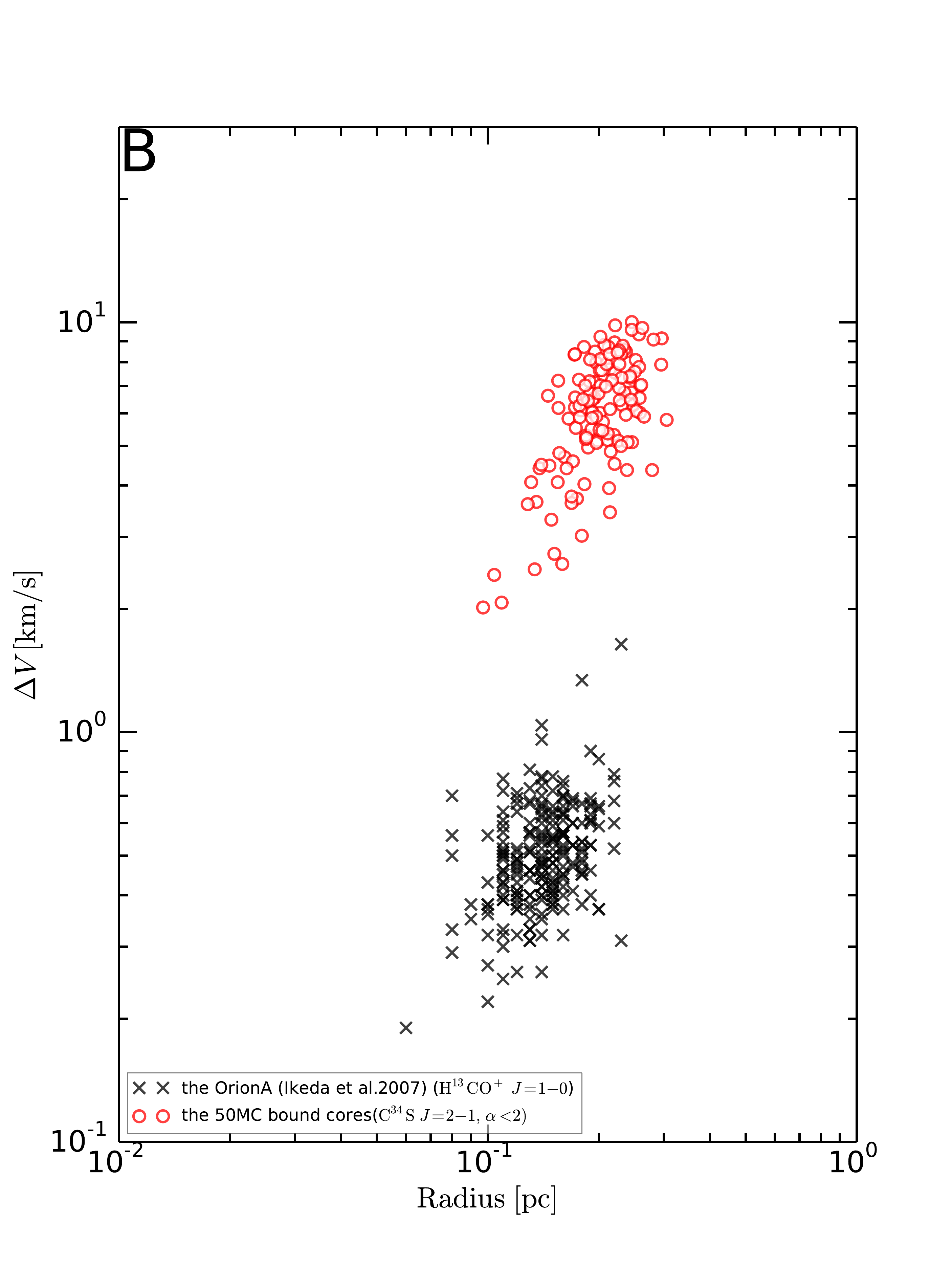}
\caption{[A] Radius-velocity width relation of the ${\rm H^{13}CO^+}$ cores. 
[B] Radius-velocity width relation of the ${\rm C^{34}S}$ cores. 
The red circles show the bound cores in the 50MC. 
The black squares show the cores in the Orion A \citep{ikeda2007}.  
}\label{fig:r-dv}
\end{figure}

}

%

Figure \ref{fig:cmf}-A shows the histograms of the LTE masses of the bound ${\rm H^{13}CO^+}$ cores in the 50MC (red bar) and the Orion A (black bar). 
The cores in the 50MC and the Orion A have different mass distributions. 
{ The mean mass in the Orion A is $12.3\pm12.0~M_{\odot}$, whereas the mean mass in the 50MC is $960 \pm850 M_\odot$ which is $\sim80$ times larger than that in the Orion A. }

The CMF of the bound ${\rm H^{13}CO^+}$ cores in the whole of the 50MC is shown in the Figure \ref{fig:cmf}-B (red circle). 
For comparison, Figure \ref{fig:cmf}-B also shows the CMF in the Orion A molecular cloud (black square) observed by the H$^{13}$CO$^+~J=1-0$ emission line \citep{ikeda2007}. 
{ 
The CMF distributions in the 50MC and the Orion A are quite different from each other (also see Figure \ref{fig:cmf}-A). 
Because the CMF of the 50MC becomes flat below $\sim450~M_\odot$ which is larger than the detection limit of the ${\rm H^{13}CO^+}$ core mass of $13~M_{\odot}$ at $T_{\rm ex, {\rm H^{13}CO^+} }=150\rm~K$, we analyze the CMF in the 50MC above $400~M_\odot$ by using a usual single power-law function given by 
\begin{eqnarray}
	\frac{dN}{dM}\propto M^{-\alpha_{\rm cmf}} \label{eq:spl}. 
\end{eqnarray}
Here $dN$ is the number of the cores whose masses are in the range of $M$ to $M+dM$; 
$\alpha_{\rm cmf}$ is the power-law index. 
The best-fit $\alpha_{\rm cmf}$ value is $1.48\pm0.14$, which is smaller than that of $2.3\pm0.1$ in the Orion A \citep{ikeda2007}. 
Therefore, the CMFs of the bound ${\rm H^{13}CO^+}$ in the 50MC have a top-heavy distribution compared with those in the Orion A and in the previous work \citep{2015PASJ...67..109T}. 

\begin{figure}
        \plottwo{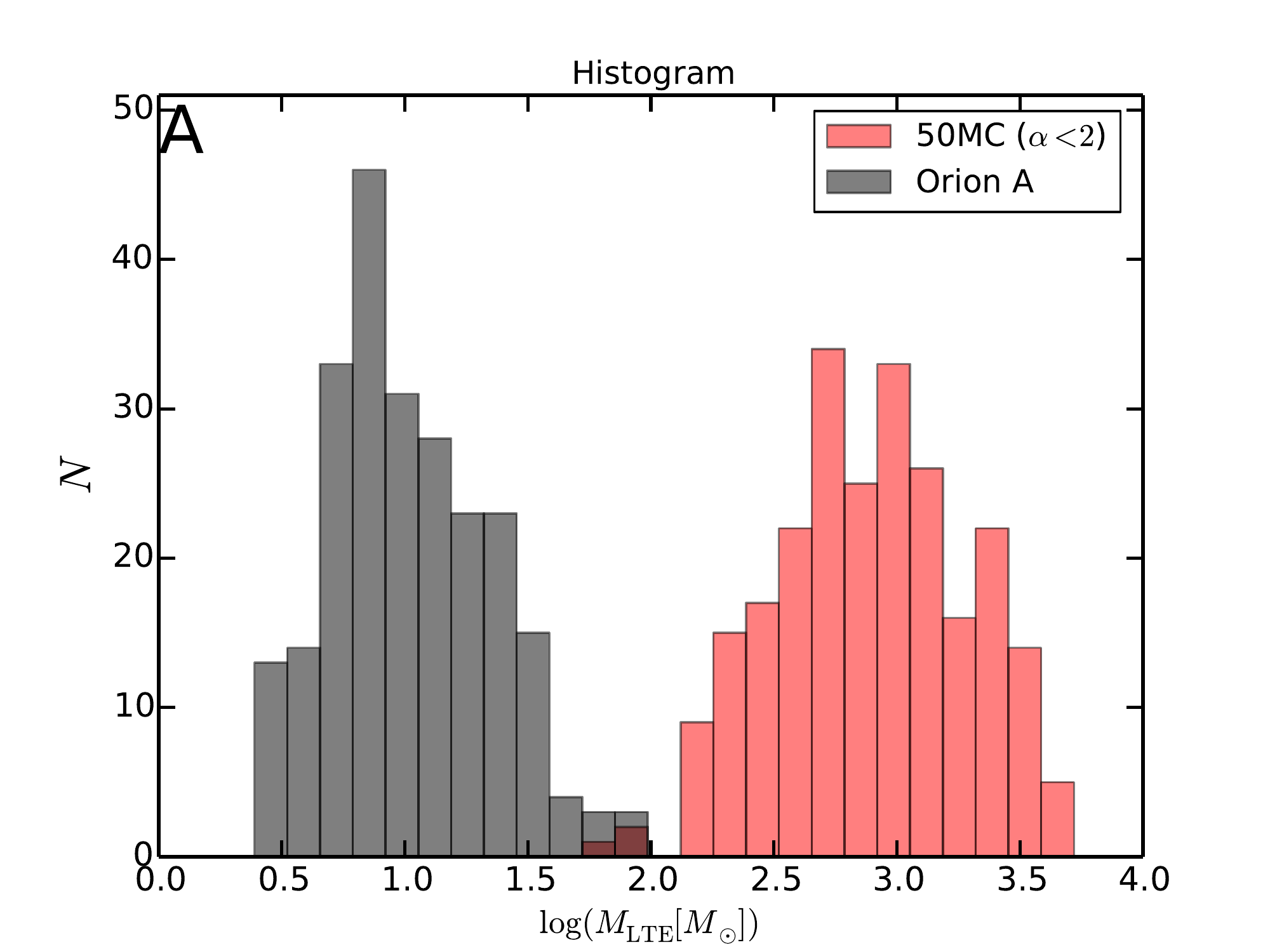}{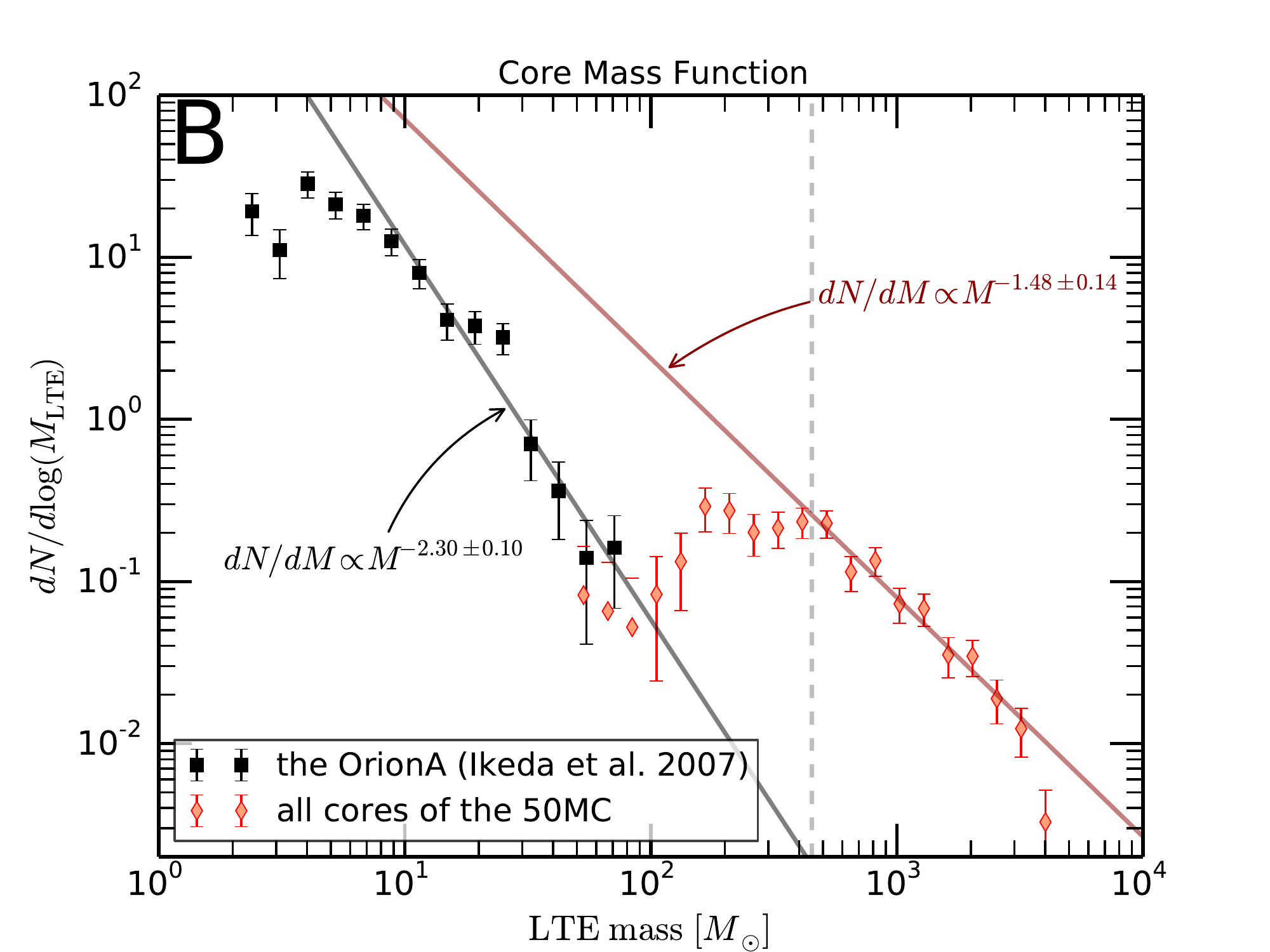}
    \caption{(A) Histogram of the bound ${\rm H^{13}CO^+}$ core mass in the 50MC (blue bars) and that of the Orion A cloud (black bars). 
    (B) Core mass function of the 50MC bound cores (red circles) with the best-fitted single power-law line (red solid line). 
    The power-law index is $1.48\pm0.14$. 
    The vertical gray dashed line shows $M_{\rm LTE}=450M_\odot$ and we analyze the CMF of the 50MC in the range of $>450M_\odot$. 
    The core mass function of the Orion A cloud is shown by the black squares \citep{ikeda2007} with the best-fitted single power-law line (black solid line). 
}\label{fig:cmf}
\end{figure}

\begin{figure}
        \plottwo{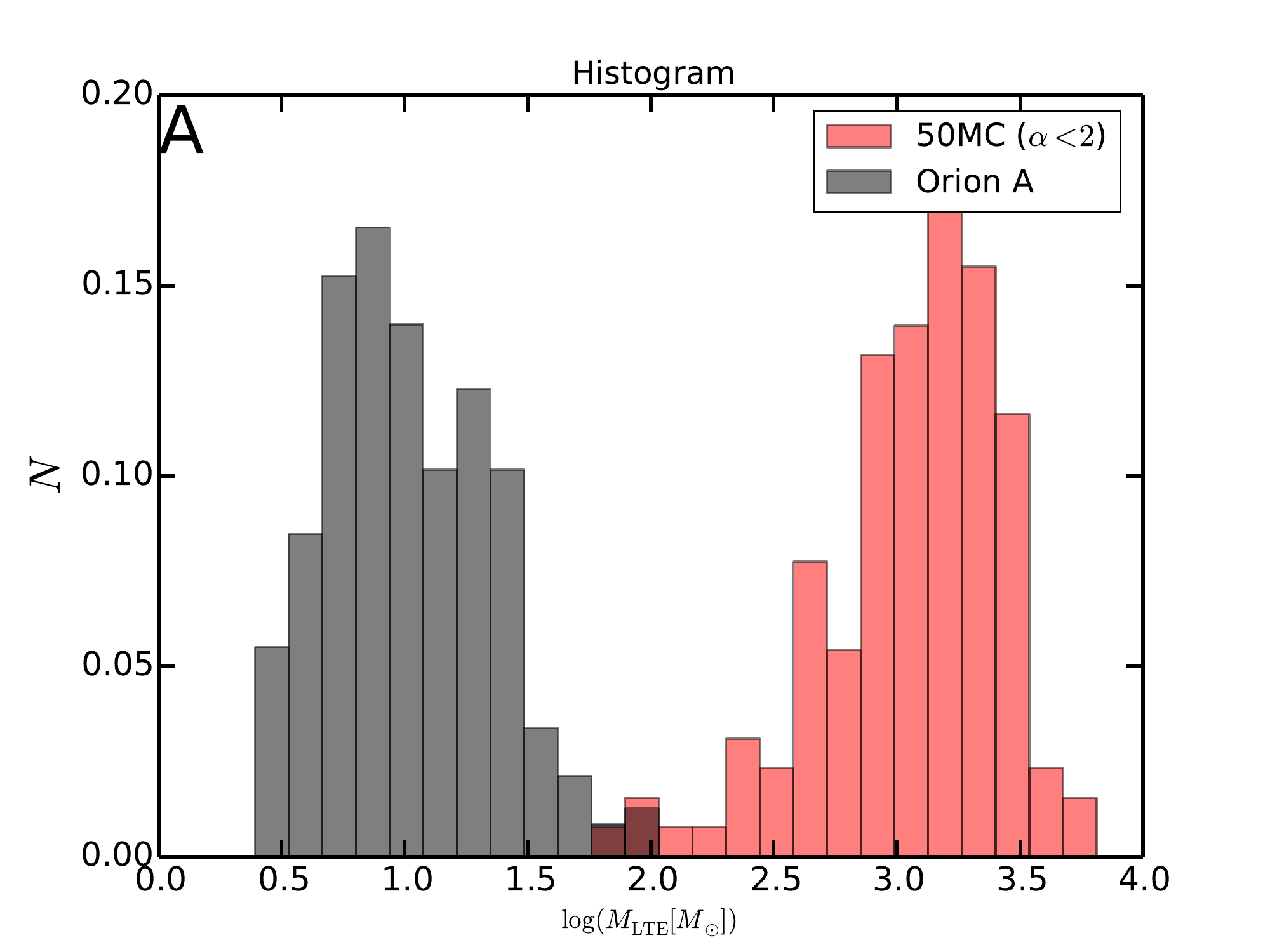}{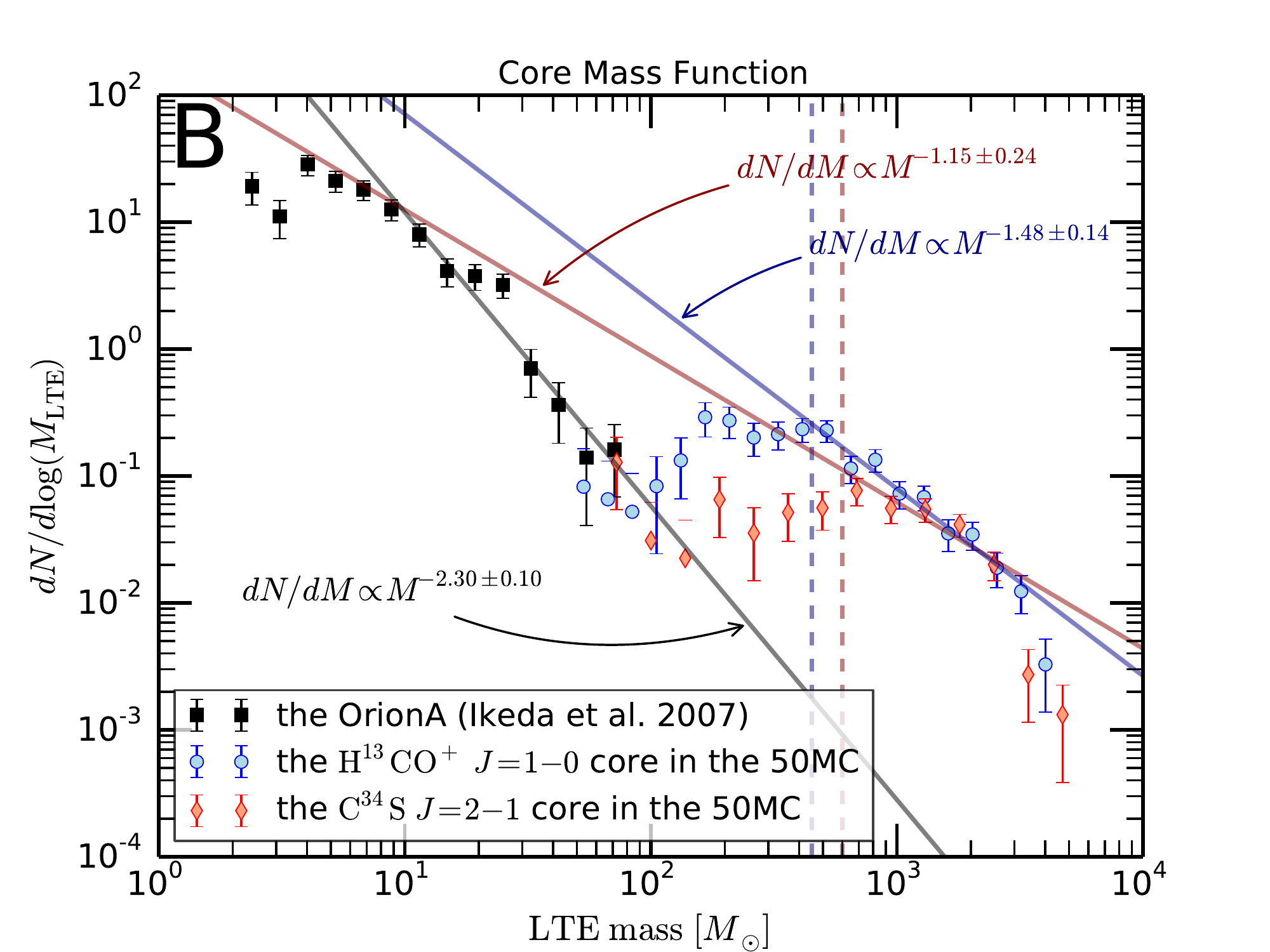}
    \caption{(A) Histogram of the bound ${\rm C^{34}S}$ core mass in the 50MC (blue bars) and that of the Orion A cloud (black bars). 
    (B) Core mass function of the bound ${\rm C^{34}S}$ cores in the 50MC (red circles) with the best-fitted single power-law line (red solid line). 
    The power-law index is $1.15\pm0.24$. 
    The vertical red dashed line shows $M_{\rm LTE}=600M_\odot$ and we analyze the CMF of the bound ${\rm C^{34}S}$ cores in the range of $>600M_\odot$. 
    The blue circles indicate the CMF of the bound ${\rm H^{13}CO^+}$ cores. 
    The blue circle and red solid line are the CMF and best-fitted single power-law line of the bound ${\rm H^{13}CO^+}$ cores, respectively. 
    The CMF of the Orion A cloud is shown by the black squares \citep{ikeda2007} with the best-fitted single power-law line (black solid line). 
}\label{fig:cmf-c34s}
\end{figure}

We also make the CMF of the bound ${\rm C^{34}S}$ cores with $\alpha<2$. 
The mean mass of the bound cores is $1300\pm890~M_\odot$ which is $\sim100$ times larger than that in the Orion A (see Figure \ref{fig:cmf-c34s}-A). 
Figure \ref{fig:cmf-c34s}-B shows the CMFs of the bound ${\rm C^{34}S}$ and ${\rm H^{13}CO^+}$ cores for comparison. 
We applied a single power-law function to the CMF of the bound ${\rm C^{34}S}$ cores in the mass range from $600~M_\odot$ to $2000~M_{\odot}$. 
The best-fit $\alpha_{\rm cmf}$ value is $1.15\pm0.24$. 
The power-law index of the bound ${\rm C^{34}S}$ cores is consistent with that of the bound ${\rm H^{13}CO^+}$ cores within the uncertainties. 

Therefore, we conclude that the bound cores in the 50MC have a top-heavy mass distribution compared with those in the Orion A. 
}

\subsection{The bound cores in the CCC region}\label{sec:bccc}
{ In this section, we discuss the influence of the CCC on the bound cores in terms of massive star formation. 
}

\subsubsection{The CCC in the 50MC}\label{sec:ccc}
{  \cite{2015PASJ...67..109T} found the half-shell-like shock structure with the brightness temperature ratio $R_T=T_{\rm B}({\rm SiO}~J=2-1)/T_{\rm B}({\rm H^{13}CO^+}~J=1-0)$ higher than 4 in the $l-b-v$ space observed by the NRO45. 
The $R_T$ is used as a shock tracer because the abundance of SiO molecules is increased by C-shock in molecular clouds, while the H$^{13}$CO$^+$ molecules are not affected by the shock \citep[e.g.][]{amo2011}. 
This shock structure is consistent with simulations of the CCC. 
Additionally, the 44GHz class I methanol masers \citep{pihlstrom2011} are located intensively around the northeastern boundary of the half-shell-like structure, although the class II methanol and $\rm H_{2}O$ maser has not been detected in this cloud yet. 
\cite{2015PASJ...67..109T} considered that the northeastern boundary is likely the front of the propagating shock wave at present. 
}

Figure \ref{fig:ratio}-A shows the distribution of the brightness temperature ratio $R_T=T_{\rm B}({\rm SiO}~J=2-1)/T_{\rm B}({\rm H^{13}CO^+}~J=1-0)$ in the range of $V_{\rm LSR} = 30-40\rm~km~s^{-1}$ smoothed to the resolution equal to the beam size of the NRO45 ($=26''$), which is made from our data observed by ALMA. 
We confirm the half-shell-like structure as shown in \cite{2015PASJ...67..109T}. 
The black filled circles show the positions of the 44GHz class I methanol maser \citep{2016ApJ...832..129M} which is another shock tracer. 
The half-shell-like structure would depict the shape of the shock front.

Figure \ref{fig:ratio}-B shows the first moment map in the ${\rm H^{13}CO^+}~J=1-0$ emission line. 
The 44GHz class I methanol masers are plotted with white filled circles in the inset panel. 
The $30-40\rm~km~s^{-1}$ cloud component (in red) has a filamentary structure and is located on a line where the HII regions A to C line up, whereas the $50-65\rm~km~s^{-1}$ component is widespread across the half-shell-like structure and the methanol masers. 
These facts can be interpreted as the CCC between clouds with different sizes, masses, and velocities of $30-40\rm~km~s^{-1}$ and $50-65\rm~km~s^{-1}$, hereafter referred to as ”cloud 1” and ”cloud 2”, respectively. 
The $40-50\rm~km~s^{-1}$ component is the velocity bridge feature connecting the two clouds that indicate the shocked gas created by the collision. 
Additionally, in the position-velocity diagram, we confirm a v-shaped gas structure in Figure \ref{fig:pv-ccc}. 
The methanol masers (black open circles) are associated with the v-shaped gas structure. 
This structure in the position-velocity diagram is an observational signature of the CCC which indicates the collision  between large and small clouds \citep{2015MNRAS.450...10H}. 
These results and the previous work strongly suggest that the two clouds collide with the different radial velocities of $V_{\rm LSR}\sim35\rm~km~s^{-1}$ and $\sim55\rm~km~s^{-1}$ and that the collision point propagates southwest to northeast. 

{
The line-of-sight collision velocity, $V_{\rm CCC}$, of $\sim20~ {\rm km~s^{-1}}$ between the two clouds can be attributed to the velocity vector difference between the orbital motions of the two clouds around Sgr A*. 
Because the projected distance is $\sim5-10\rm~pc$ between Sgr A$^{*}$ and the 50MC and the supermassive black hole (SMBH) associated with Sgr A$^{*}$ has a mass of $4\times10^6~M_{\odot}$ \citep{ghez2003,ghez2005}, the orbital velocities of the two clouds around Sgr A* are estimated to be $\sim50~{\rm km~s^{-1}}$ by the SMBH gravitational potential. 
Here we assume that the projected distance is comparable to the real distance between the 50MC and Sgr A*. 
Additionally, we consider that the cloud 2 moves along the direction parallel to the line of sight with an orbital velocity of $\sim50~ {\rm km~s^{-1}}$ and the cloud 1 moves along the direction inclined at an angle of $\sim50$ degrees from the line of sight. 
In this case, the two clouds collide with each other with the line-of-sight collision velocity $V_{\rm CCC}$ of $\sim20~ {\rm km~s^{-1}}$. 
Therefore, it is possible that the large fraction of the radial velocity difference between the two clouds originates from the orbital motion around Sgr A*. 
}

To discuss the influence of the CCC on the core properties, we define the region with the high brightness temperature ratio, $R_T>2.5$, as the CCC region because this region can cover the maser distribution. 
The last column of Table \ref{tab:core} indicates whether each core candidate is within the CCC or non-CCC region. 

\begin{figure}
        \plottwo{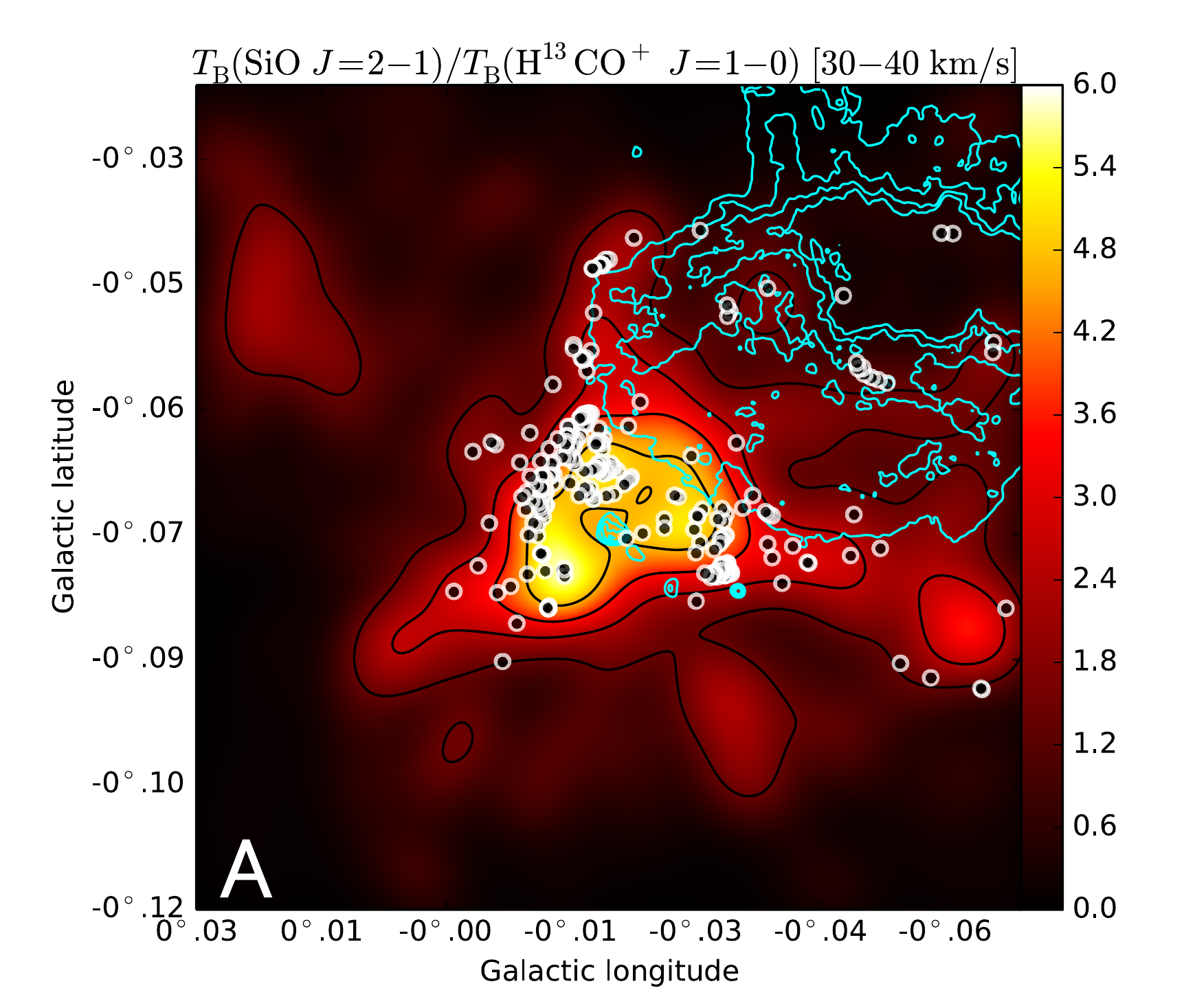}{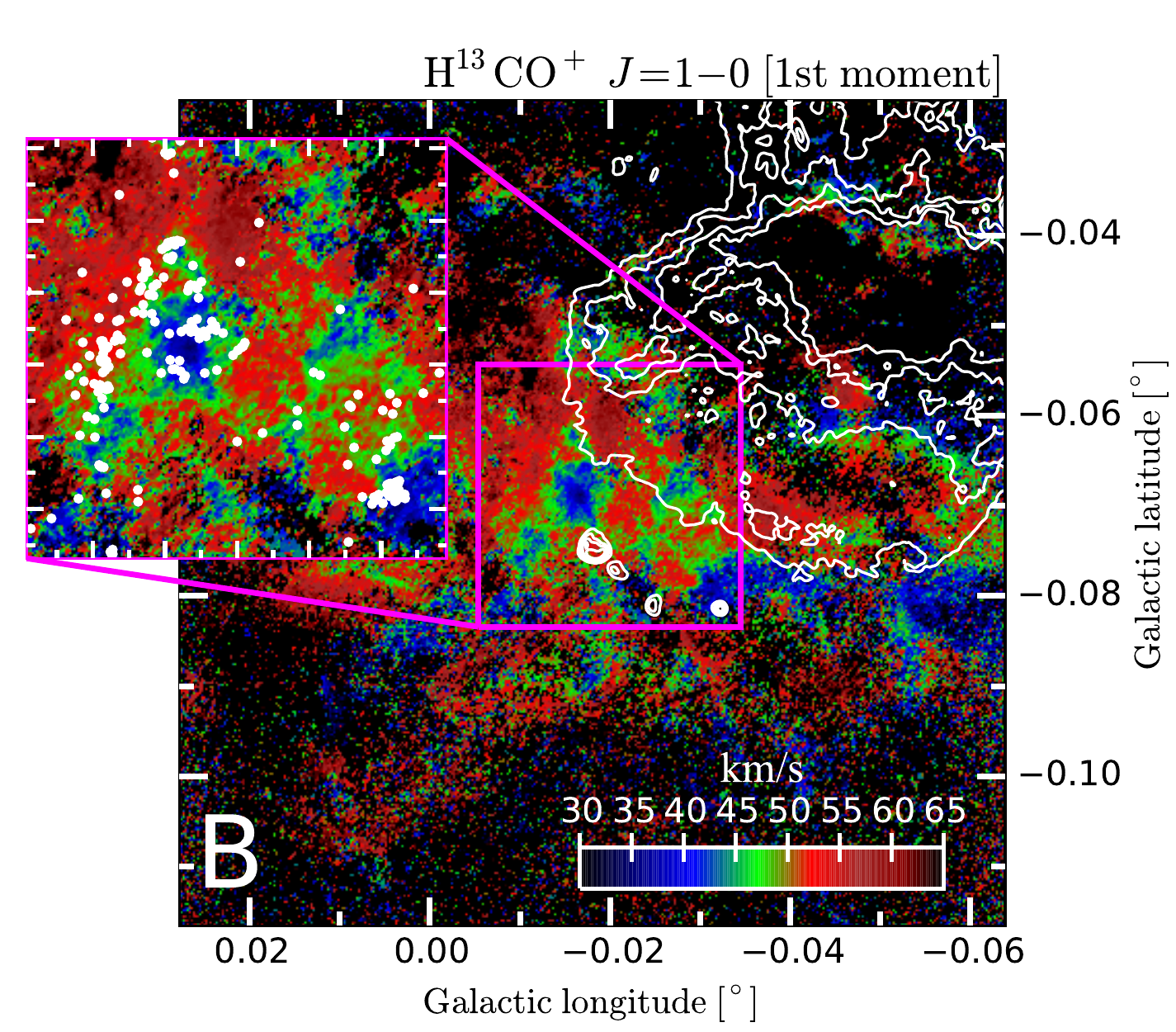}
    \caption{[A] Color map with contours (black solid line) of $R_T=T_{\rm B}({\rm SiO}~J=2-1)/T_{\rm B}({\rm H^{13}CO^+}~J=1-0)$ smoothed to $26''$ resolution. The ratio scale is shown in the color bar on the right-hand side of the panel, and the contour levels are 1.4,2.5,3.6,4.7 and 5.8. 
    The black filled circles show the positions of the 44GHz methanol masers. 
    The 6cm continuum emission is shown by the cyan contours, and the contour levels are $\rm 0.01,0.02,0.03$ and $0.04\rm~Jy/beam$. 
    The 6cm continuum emission indicates the compact HII regions A-D and the Sgr A east. 
    [B] First moment map in the ${\rm H^{13}CO^+}~J=1-0$ emission line with the 6cm continuum emission indicated by white contours. 
    The inset map of panel B shows the first moment map with the positions of the 44GHz methanol masers (white filled circles). 
    {We supply the RGB movie (\url{http://www.vsop.isas.jaxa.jp/~nakahara/figure_uehara/201812/map_h13co+_6km_rgb.mp4})}
    }\label{fig:ratio}
\end{figure}

\begin{figure}
        \plotone{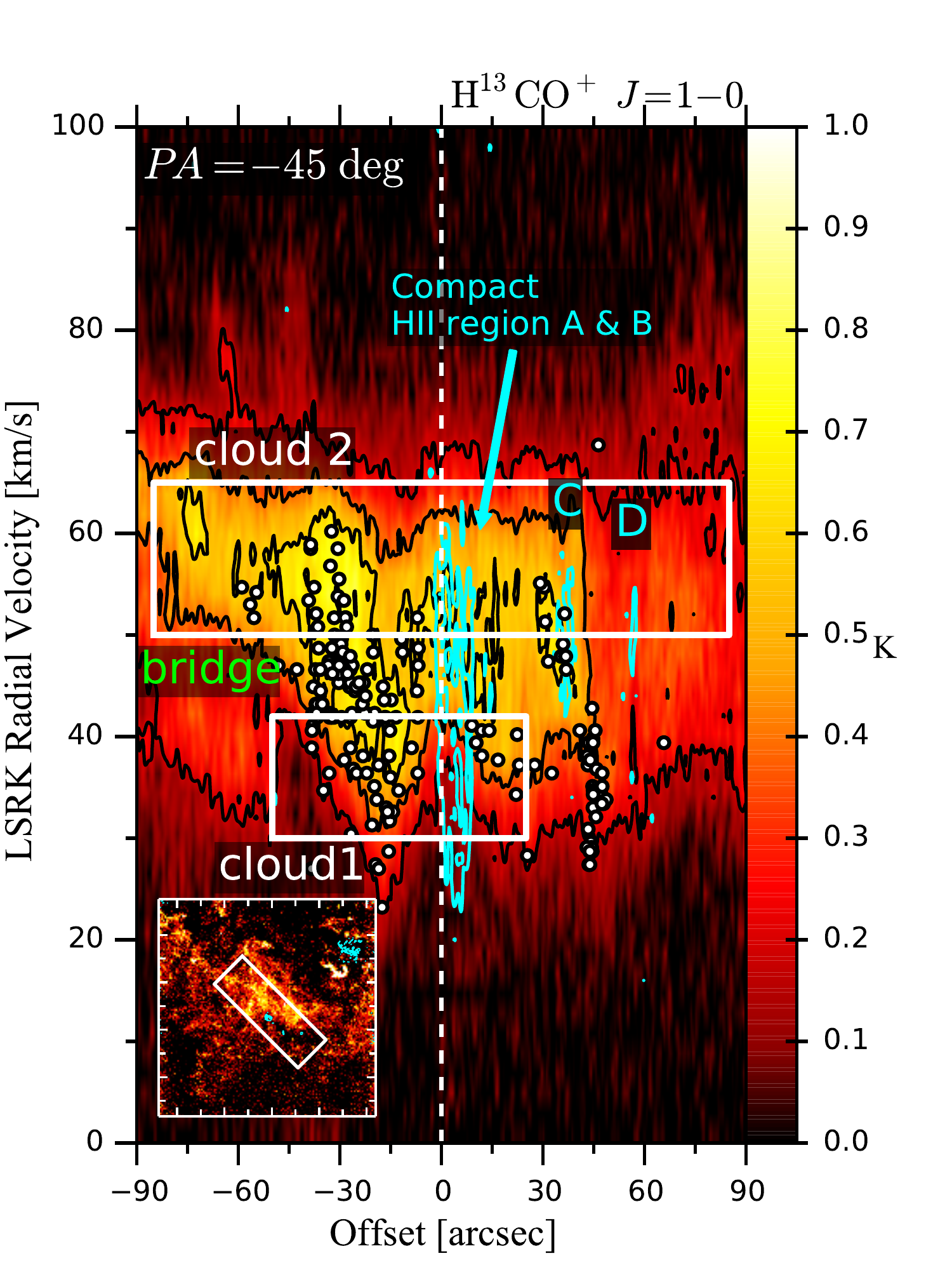}
    \caption{Psition-velocity diagram along four HII regions with a position angle of $-45^{\circ}$ in the ${\rm H^{13}CO^+}$ emission line with black contours. 
    The vertical white dashed lines show offset $=0\rm~arcsec$. 
    The cyan contours show the $\rm H_{42}\alpha$ recombination line of our ALMA data. 
    Open black circles indicate the positions of the 44 GHz class I methanol maser. 
    The inset map shows the $T_{\rm peak}$ maps of the ${\rm H^{13}CO^+}~J=1-0$ emission line. 
    The white open rectangles on the inset maps indicate the cutting direction of each position-velocity diagram with the cutting widths of $1\rm~arcmin$. 
	The velocity resolution is $2\rm~km~s^{-1}$. 
    }\label{fig:pv-ccc}
\end{figure}

\begin{figure}
    \plottwo{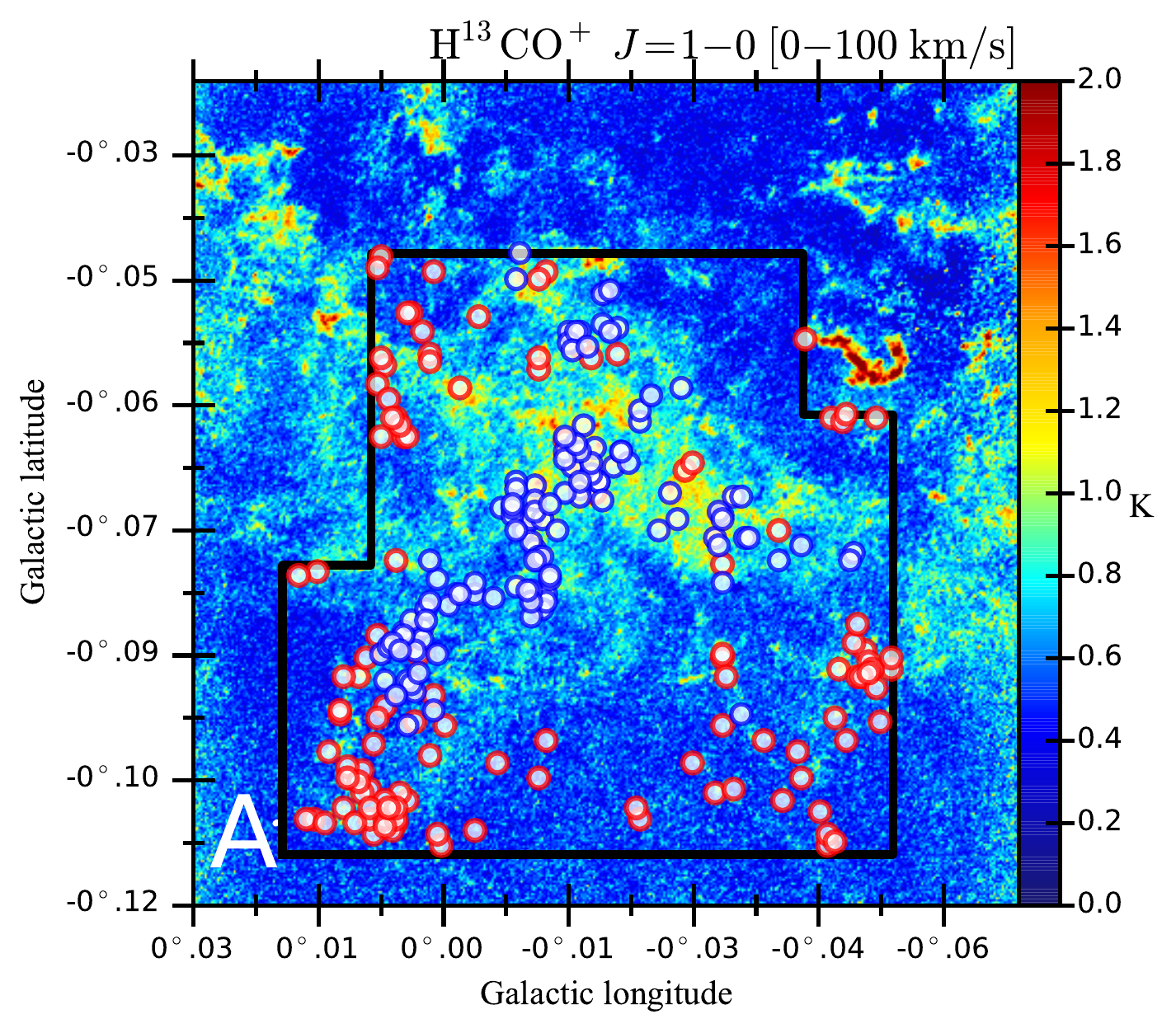}{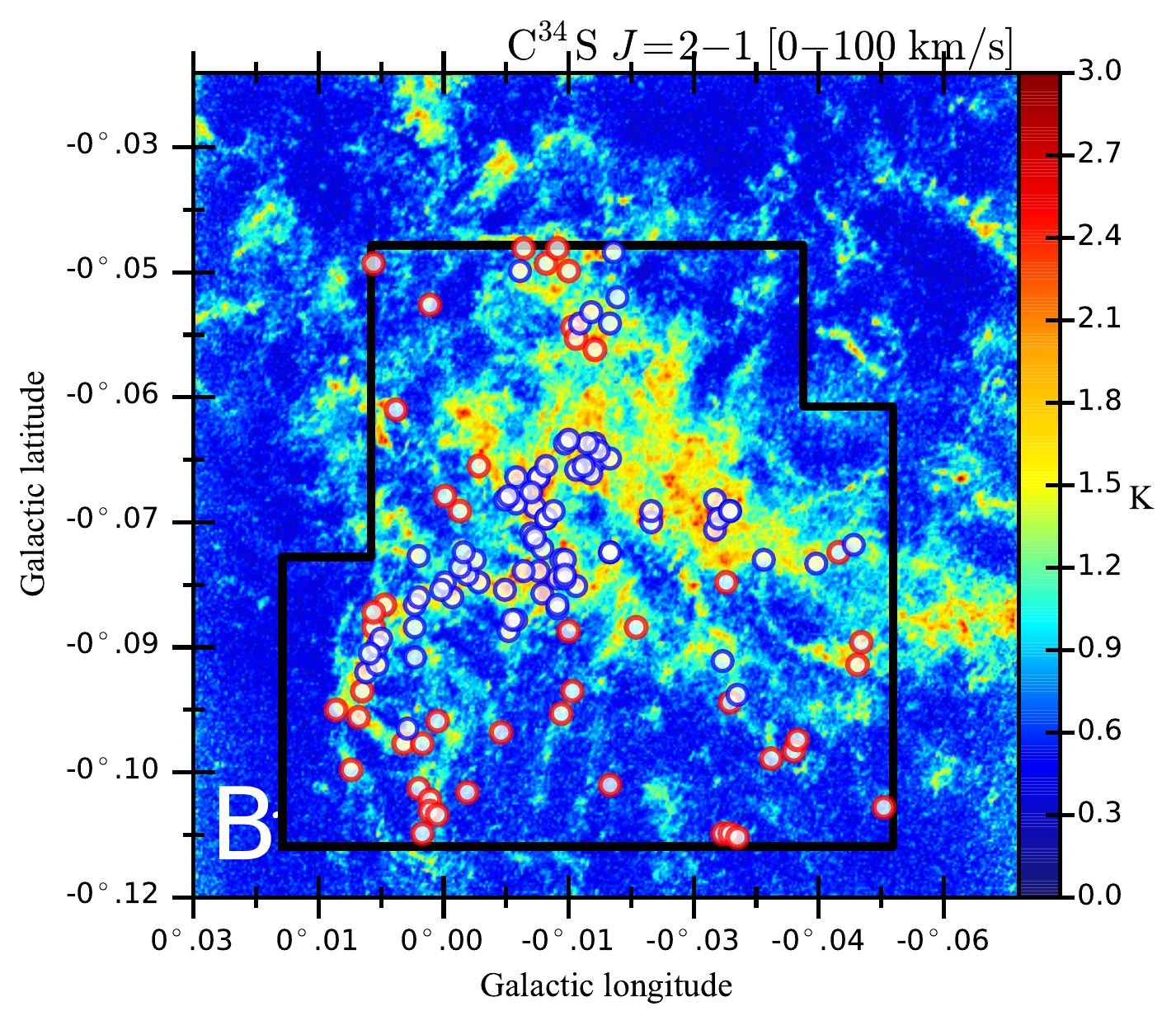}
    \caption{[A] Positions of the identified bound cores with $\alpha<2$ on the $T_{\rm peak}$ map of the ${\rm H^{13}CO^+}~J=1-0$ emission line in the velocity range of $V_{\rm LSR}=0-100\rm~km~s^{-1}$. The intensity scale is in brightness temperature units of K and is shown in the color bar on the right-hand side of the panel. The blue circles indicate the bound cores in the CCC region and the red circles indicate the bound cores in the non-CCC region. The cores outside the polygon by the black thick line are rejected. 
    [B] Positions of the identified bound cores on the $T_{\rm peak}$ map of the ${\rm C^{34}S}$ emission line in the velocity range of $V_{\rm LSR}=0-100\rm~km~s^{-1}$. }\label{fig:ccc}
\end{figure}

\subsubsection{The comparison of the densities of the bound cores between the CCC and Non-CCC regions}\label{sec:N}
We analyze the bound cores statistically in order to estimate the influence of the CCC. 
A total of 119 bound ${\rm H^{13}CO^+}$ cores are located in the CCC region, while 122 bound cores are located
in the non-CCC region (see  Figure \ref{fig:ccc}-A). 
The percentage of the bound cores in the CCC is $49\%(=119/241)$, while that in the non-CCC region $51\%~(=122/241)$. 
The physical parameters of the bound ${\rm H^{13}CO^+}$ cores in the CCC and non-CCC regions are summarized in Table \ref{tab:ccc}. 
{ Additionally, the pixel number ratio of the CCC region to the non-CCC region is $14\%~(=\rm 5329920~pixels/37077957~pixels)$ in the core identified region. 
Thus, the area of the CCC region is much smaller than that of the non-CCC region, but the numbers of the bound cores in the CCC and non-CCC regions are similar to each other. 
The surface density of the cores in the CCC region is an order of magnitude larger than that in the non-CCC region. }
On the other hand, the physical parameters of the bound ${\rm C^{34}S}$ cores in the CCC and non-CCC regions are summarized in Table \ref{tab:cccc34s}. 
The positions of these cores are shown in Figure \ref{fig:ccc}-B. 
The number ratio of the bound ${\rm C^{34}S}$ cores in the CCC region to all the cores is $64\%~(=82/129)$, while that in the non-CCC region $37\%~(=48/129)$. 

{ 
Figure \ref{fig:nccc}-A shows the CDFs for the number densities of the bound ${\rm H^{13}CO^+}$ cores in the CCC (black line) and non-CCC (red line) regions. 
The average and range of the number densities are indicated in Figure \ref{fig:nccc}-A. 
The distribution of the core number densities in the CCC region seems to be biased to a larger density than that in the non-CCC region. 
Similarly, the distribution of the column densities toward the bound cores in the CCC region is biased to a larger column density than that in the non-CCC region in Figure \ref{fig:colccc}-A. 
Especially, there exit 26 cores with more than $8.1\times10^{23}\rm~cm^{-2}$ only in the CCC region. 

The bound ${\rm C^{34}S}$ cores in the CCC region also have larger number and column densities than those in the non-CCC region as shown in Figure \ref{fig:nccc}-B and \ref{fig:colccc}-B. 
There are 27 cores with more than $5.6\times10^{23}\rm~cm^{-2}$ only in the CCC region. 
Consequently, it is most likely that the CCC compresses the molecular gas and increases the number of the bound cores with high densities. 
}

\subsubsection{The comparison of the masses of the bound cores in the CCC and non-CCC regions}

Figure \ref{fig:hist-ccc}-A shows the histograms of the LTE masses of the bound ${\rm H^{13}CO^+}$ cores in the CCC (blue bar) and non-CCC (red bar) regions. 
The average and range of the masses of the bound ${\rm H^{13}CO^+}$ cores in the CCC region are $1300\pm970~M_{\odot}$ and $190-4500~M_{\odot}$, respectively. 
Meanwhile, the average and range of the masses in the non-CCC region are $610\pm510~M_{\odot}$ and $48-2400~M_{\odot}$, respectively. 
The mean mass of the bound ${\rm H^{13}CO^+}$ cores in the CCC region seems larger than that in the non-CCC regions. 

{ 
The mass distribution peak of the cores in the CCC region is at $\log_{10}\left(M_{\rm LTE}~[M_{\odot}]\right)=3$ derived by the Gaussian fitting, whereas that in the non-CCC region is at $\log_{10}\left(M_{\rm LTE}~[M_{\odot}]\right)=2.7$. 
Note that the massive bound cores with masses of $2400~M_{\odot}$ or more exist only in the CCC region as well as the dense cores with more than $8.1\times10^{23}\rm~cm^{-2}$ exist only in the CCC region. 
The total bound core mass in the CCC region is $68~\%(=1.6\times10^{5}~M_{\sun}/2.3\times10^{5}~M_{\sun})$ of the total bound core mass in the whole 50MC.  
It is likely that the CCC efficiently formed the massive bound ${\rm H^{13}CO^+}$ cores by compressing the molecular gas. 

Additionally, the total mass of the bound ${\rm C^{34}S}$ core in the CCC region is $76\%(=1.3\times10^{5}M_{\odot}/1.7\times10^{5}M_{\odot})$ of the total bound core mass in the whole 50MC. 
The mass distribution peak of the cores in the CCC region is at $\log_{10}\left(M_{\rm LTE}~[M_{\odot}]\right)=3.1$ derived by the Gaussian fitting, whereas that in the non-CCC region is at $\log_{10}\left(M_{\rm LTE}~[M_{\odot}]\right)=2.8$ in Figure \ref{fig:hist-ccc}-B. 
Note that the massive bound cores with masses of $3000~M_{\odot}$ or more exist only in the CCC region as well as the bound cores with high column densities exist only in the CCC region. 
It is also likely that the CCC efficiently formed the massive bound ${\rm C^{34}S}$ cores by compressing the molecular gas. 
}

\begin{figure}
    \plottwo{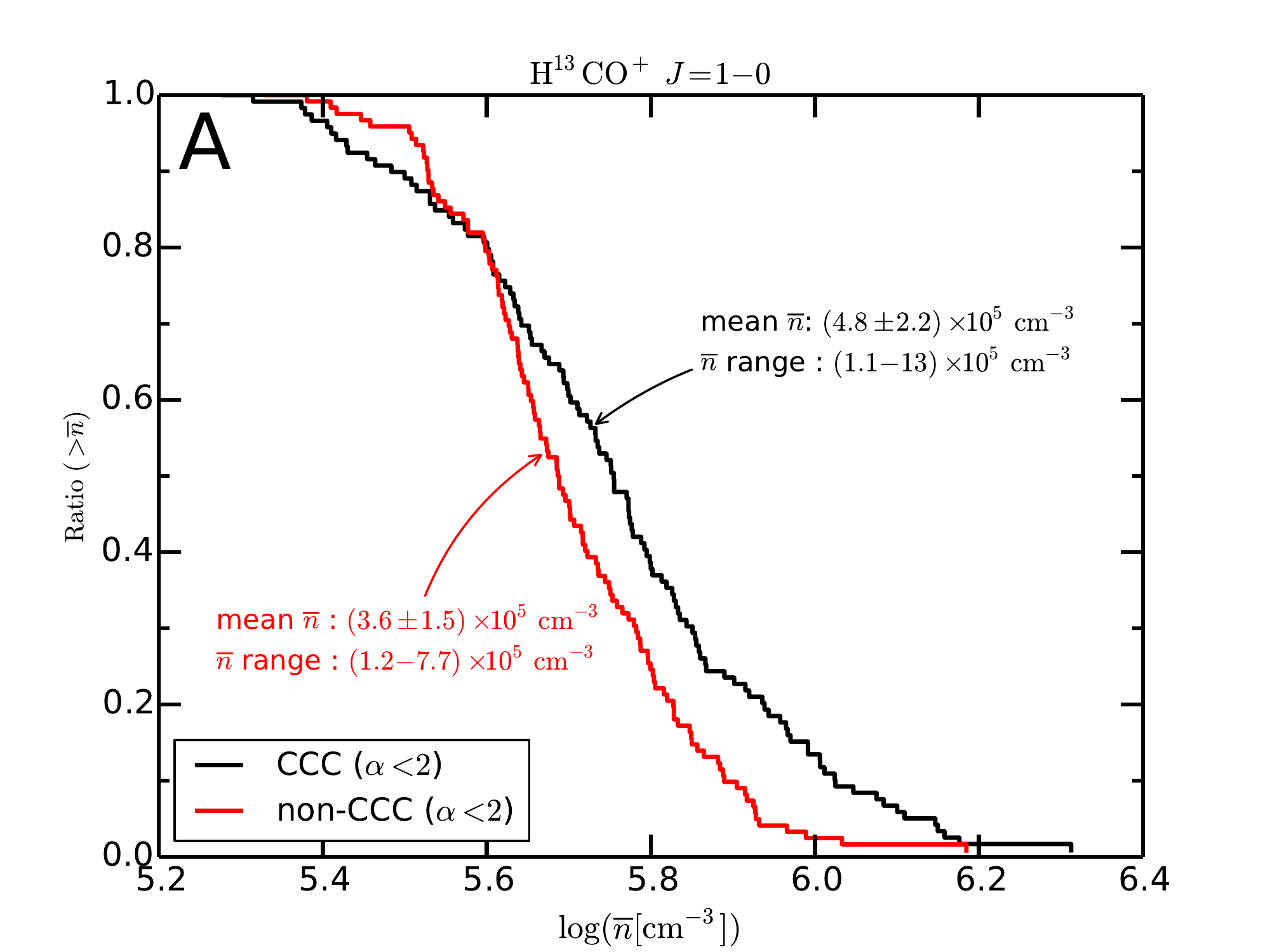}{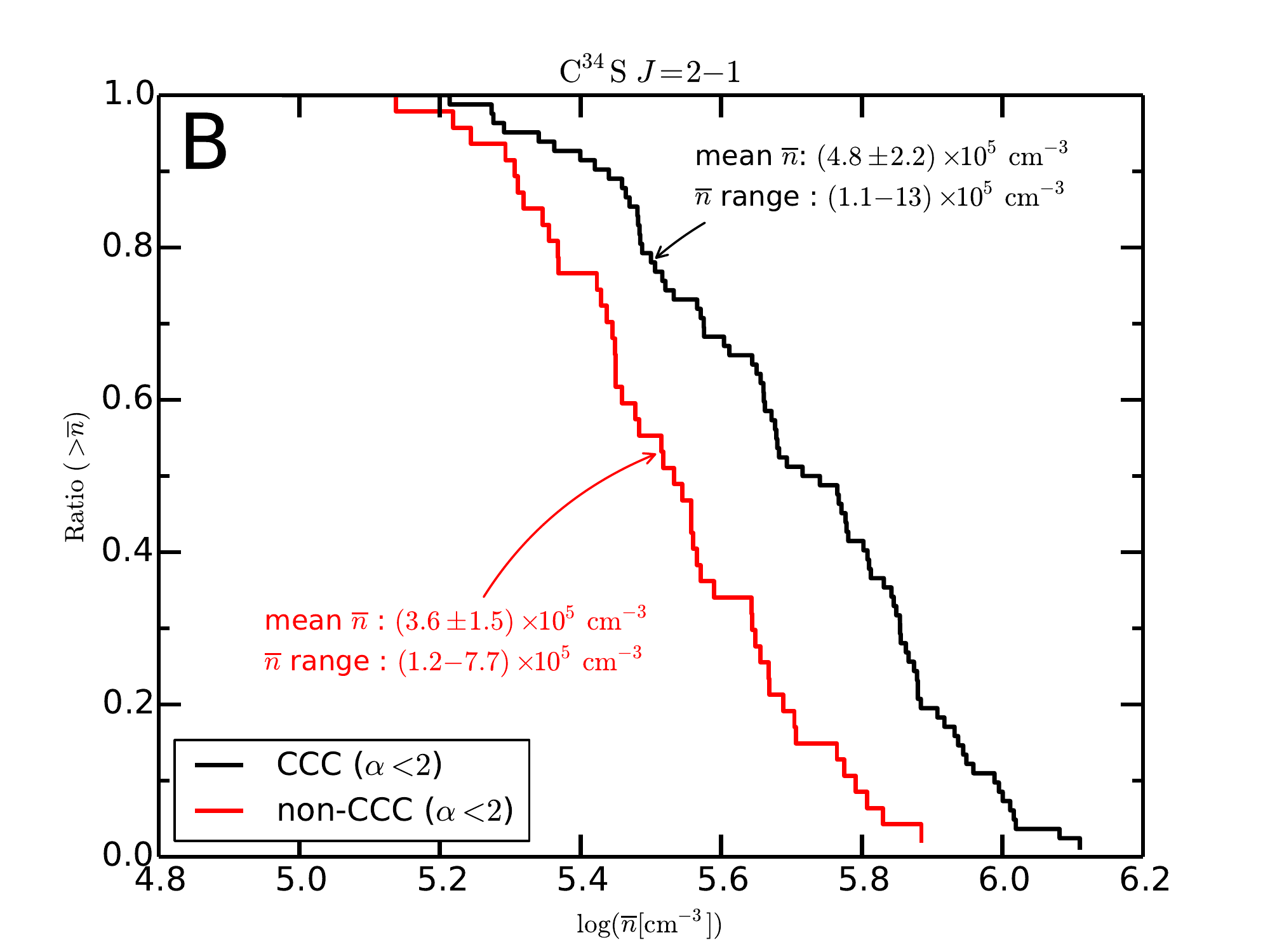}
    \caption{[A] CDF of the bound ${\rm H^{13}CO^+}~J=1-0$ core number density. The black thick line shows the distribution of the bound cores in the CCC region, while the red thick line shows the distribution of the bound cores in the non-CCC region. [B] CDF of the ${\rm C^{34}S}~J=2-1$ core number density. 
     }\label{fig:nccc}
\end{figure}

\begin{figure}
    \plottwo{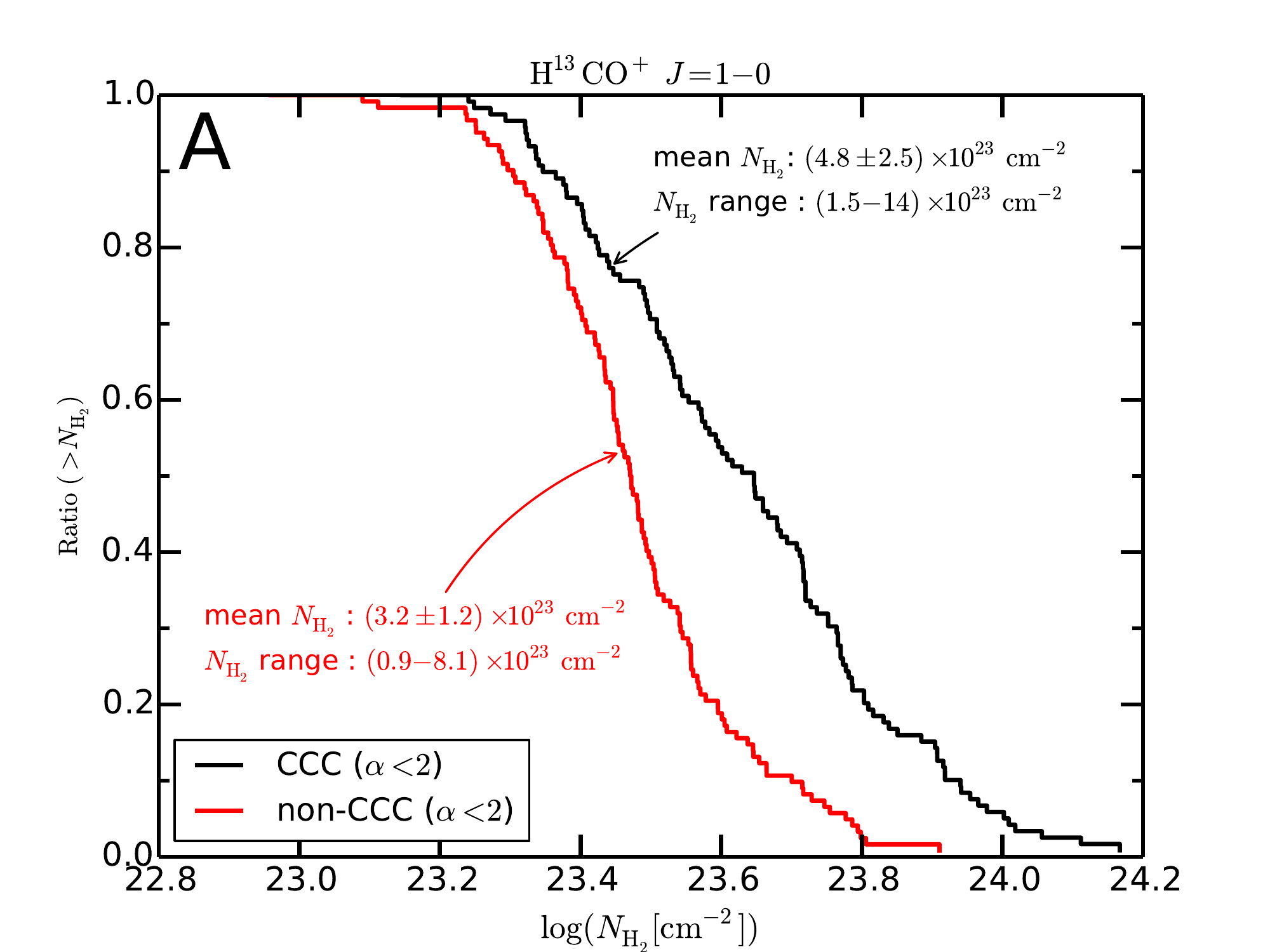}{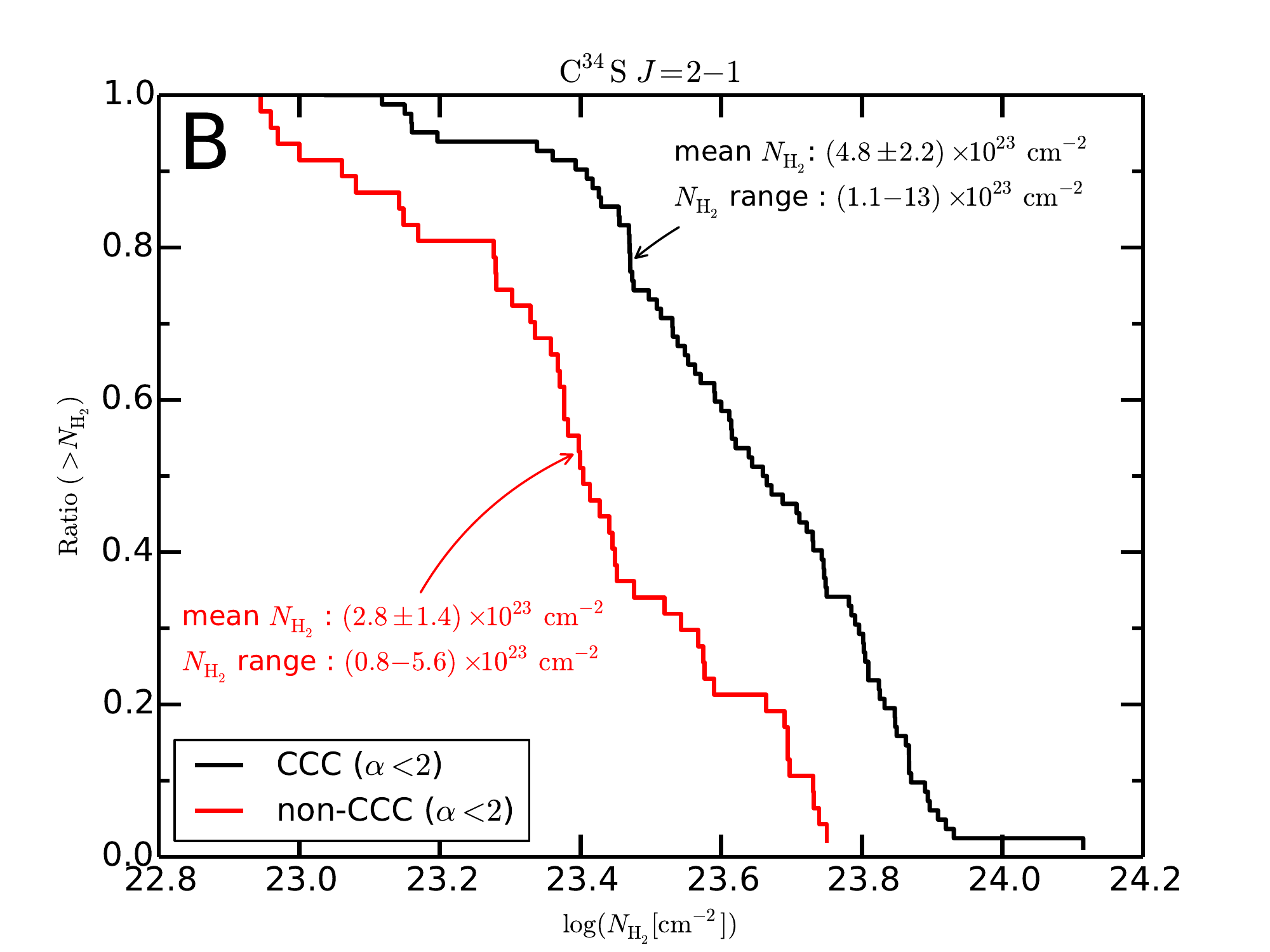}
    \caption{[A] CDF of the bound ${\rm H^{13}CO^+}~J=1-0$ core column density. The black thick line shows the distribution of the bound cores in the CCC region, while the red thick line shows the distribution of the bound cores in the non-CCC region. [B] CDF of the ${\rm C^{34}S}~J=2-1$ core column density. 
     }\label{fig:colccc}
\end{figure}

{ 
Figure \ref{fig:mccc}-A shows the CDFs for the LTE masses of the bound ${\rm H^{13}CO^+}$ cores in the CCC (black line) and non-CCC (red line) regions. 
The distribution of the core LTE masses in the CCC region is biased to a larger mass than that in the non-CCC region. 
The CMFs of the bound ${\rm H^{13}CO^+}$ cores within the CCC and non-CCC regions in the 50MC are shown by the blue circles and the red diamonds in the Figure \ref{fig:cmf-ccc}-A, respectively. 
We apply single power-law functions of Eq. (\ref{eq:spl}) to the CMFs in the CCC and non-CCC regions in the mass range of $>600~M_{\odot}$ and $>200~M_{\odot}$, respectively. 
The best-fit values of $\alpha_{\rm cmf}$ in the CCC and non-CCC regions are $1.38\pm0.20$ and $1.37\pm0.13$, respectively. 
The CMFs in the CCC and non-CCC regions show top-heavy distributions compared with that in the Orion A \citep{ikeda2007} and in the previous work \citep{2015PASJ...67..109T}, and the CMF index in the CCC region is consistent with that in the non-CCC region within the uncertainties. 
We conclude that the slope of the CMF was not changed so much, but the compression by the CCC efficiently formed the massive bound ${\rm H^{13}CO^+}$ cores, especially the bound cores with masses of $2400~M_{\odot}$ or more. 

On the other hand, the CMF index of the bound ${\rm C^{34}S}$ cores in the CCC region is consistent with that in the non-CCC region within the uncertainties (see Figure \ref{fig:cmf-ccc}). 
Figure \ref{fig:mccc}-B shows the CDFs for the LTE masses of the bound ${\rm C^{34}S}$ cores in the CCC (black line) and non-CCC (red line) regions. 
The distribution of the core LTE masses in the CCC region is biased to a larger mass than that in the non-CCC region. 
The analysis of the bound ${\rm C^{34}S}$ cores gives the same conclusion as in the analysis of the bound ${\rm H^{13}CO^+}$ cores; the slope of the CMF is not changed so much in the mass range of $100~M_{\odot}$ to $3000~M_{\odot}$ but the CCC efficiently formed the massive bound cores.

\begin{figure}
        \plottwo{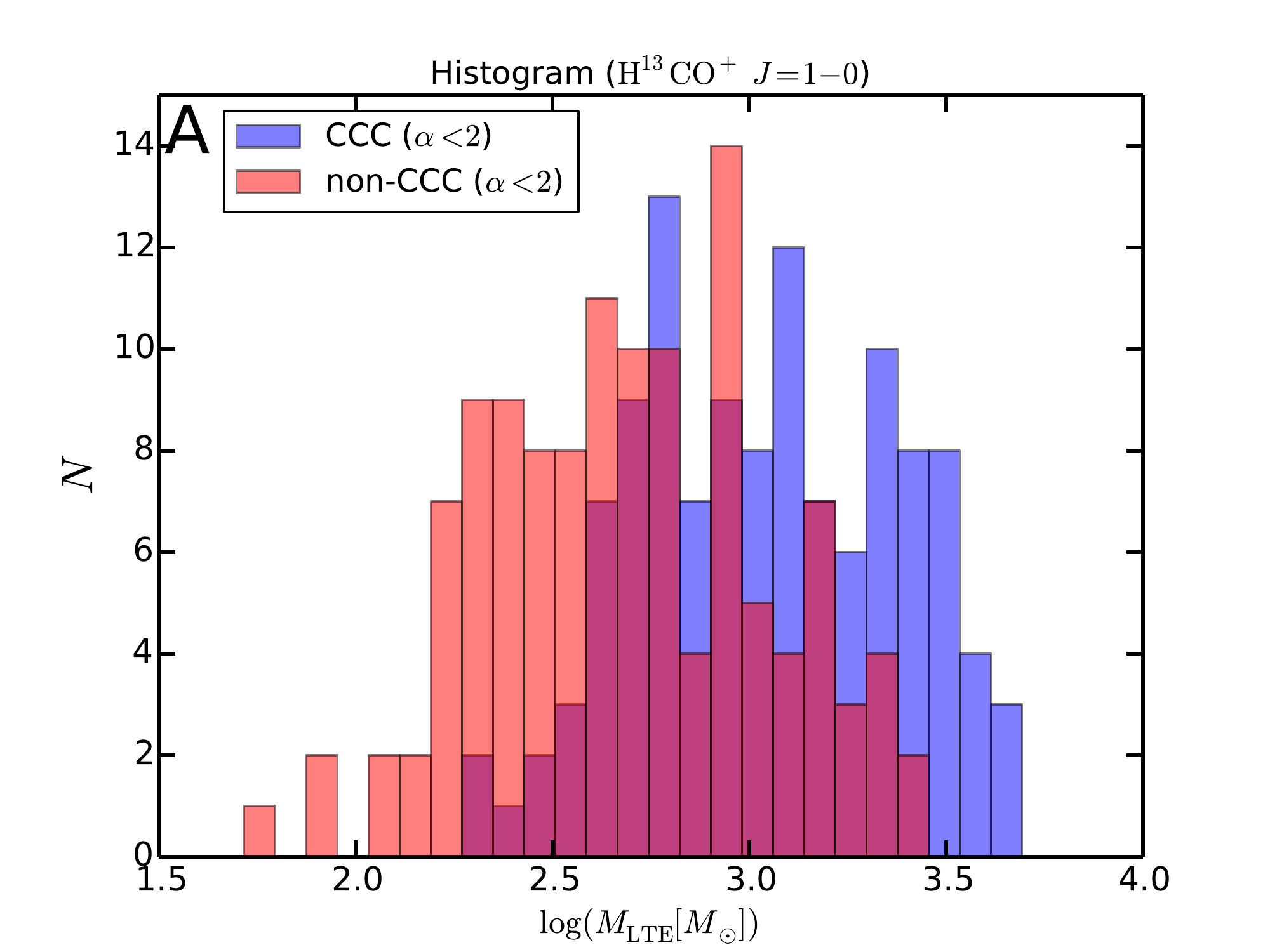}{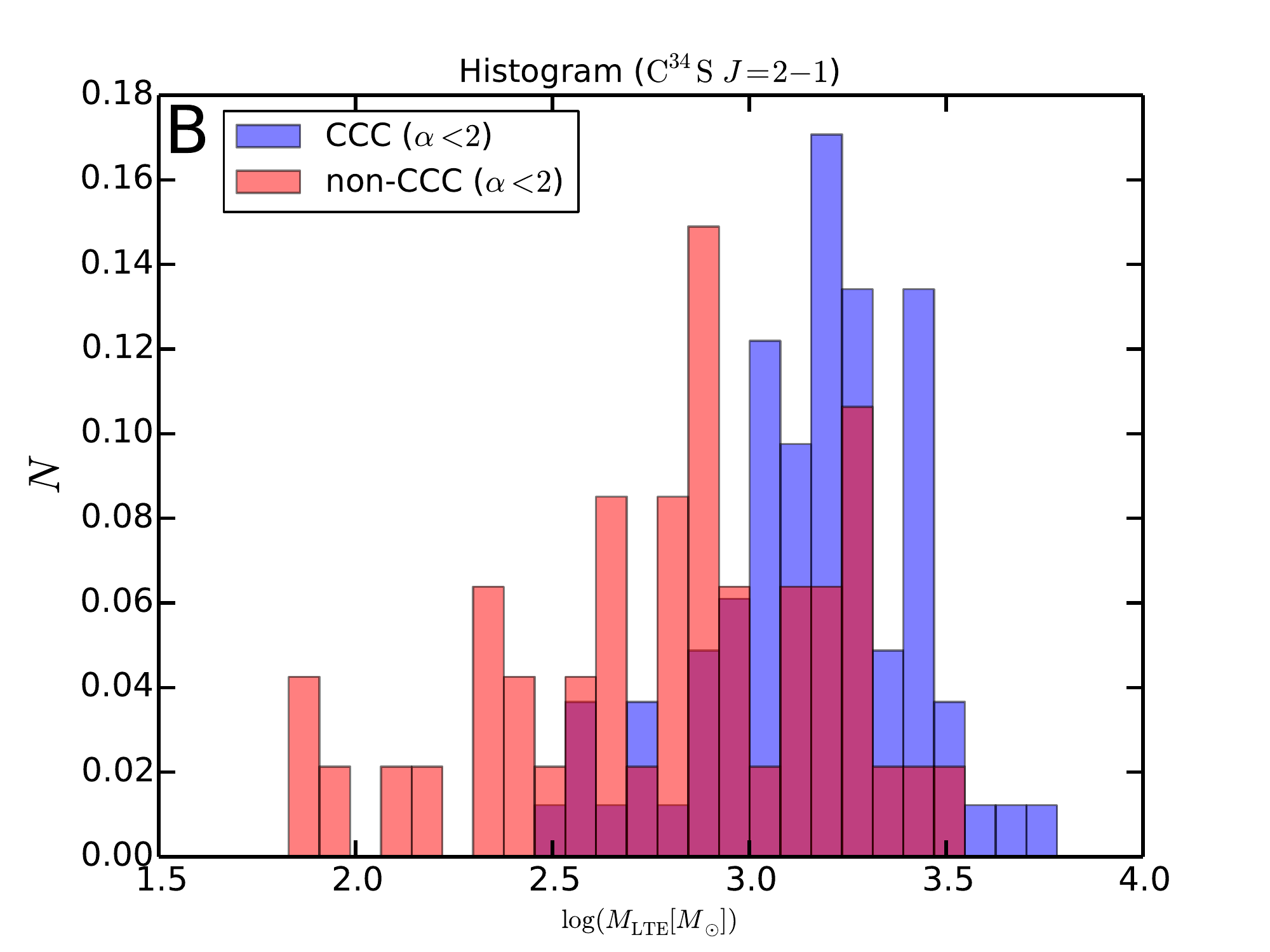}
    \caption{ (A) Histogram of the masses of the bound ${\rm H^{13}CO^+}$ cores in the CCC region (blue bars) and that in the non-CCC region (red bars). 
    (B) Histogram of the masses of the bound ${\rm C^{34}S}$ cores in the CCC region (blue bars) and that in the non-CCC region (red bars). 
   }\label{fig:hist-ccc}
\end{figure}

\begin{figure}
    \plottwo{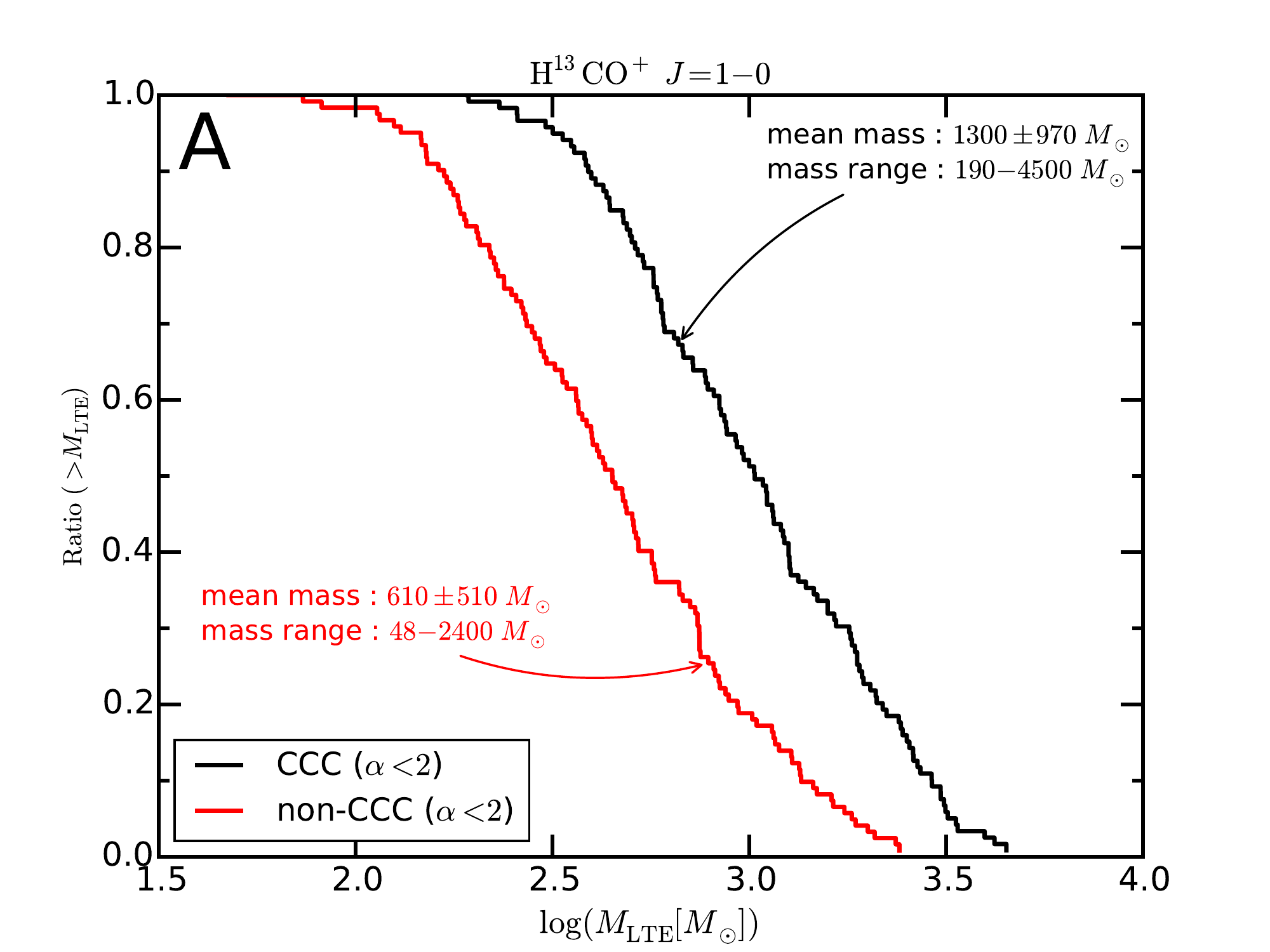}{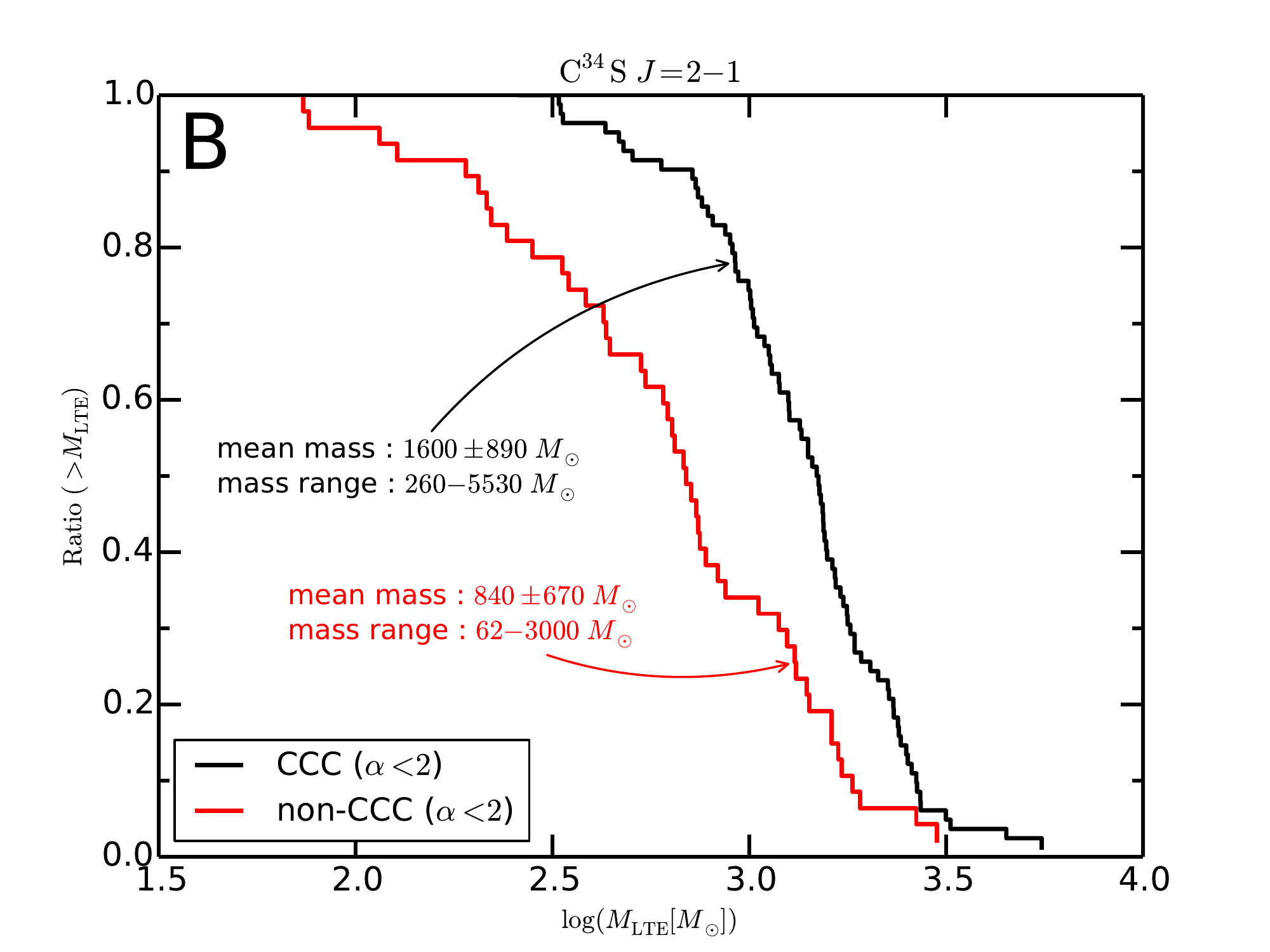}
    \caption{[A] CDF of the bound ${\rm H^{13}CO^+}~J=1-0$ core LTE mass. The black thick line shows the distribution of the bound cores in the CCC region, while the red thick line shows the distribution of the bound cores in the non-CCC region. [B] CDF of the ${\rm C^{34}S}~J=2-1$ core LTE mass. 
     }\label{fig:mccc}
\end{figure}

Additionally, the 50MC interacts with the Sgr A east \citep{1985ApJ...288..575H} (e.g. Figure \ref{fig:guide}-C in this paper). 
We discuss the effect of the interaction with the Sgr A east on the bound cores because it is possible that the 50MC has been compressed by the Sgr A east. 
A region of $l<-0\fdg016$ which is the right half region of the core identified region, is defined as the interaction region with the Sgr A east, whereas a region of $l>-0\fdg016$ is defined as the non-interaction region. 
Among the bound ${\rm H^{13}CO^+}$ cores, 83 cores are located in the interaction region, while 158 cores are located in the non-interaction region. 
The average and range of the LTE masses are shown in Figure \ref{fig:mcum_int}-A and the average and range in the interaction region are similar to those in the non-interaction region. 
The distributions are also consistent with each other. 
Additionally, we applied the same analysis to the cores only in the CCC region and obtained the same results. 
The bound core formation might be affected by the Sgr A east, but it is likely that the influence of the Sgr A east on the bound core distribution is small, suggesting that the compression of the 50MC by the Sgr A east does not have a significant influence on the bound core formation.

}

We conclude that the molecular gas compression by the CCC efficiently formed the massive bound cores, especially the bound cores with high masses of $\sim2500-3000~M_{\odot}$ or more, even if the slope of the CMF is not changed so much by the CCC.

\begin{figure}
        \plottwo{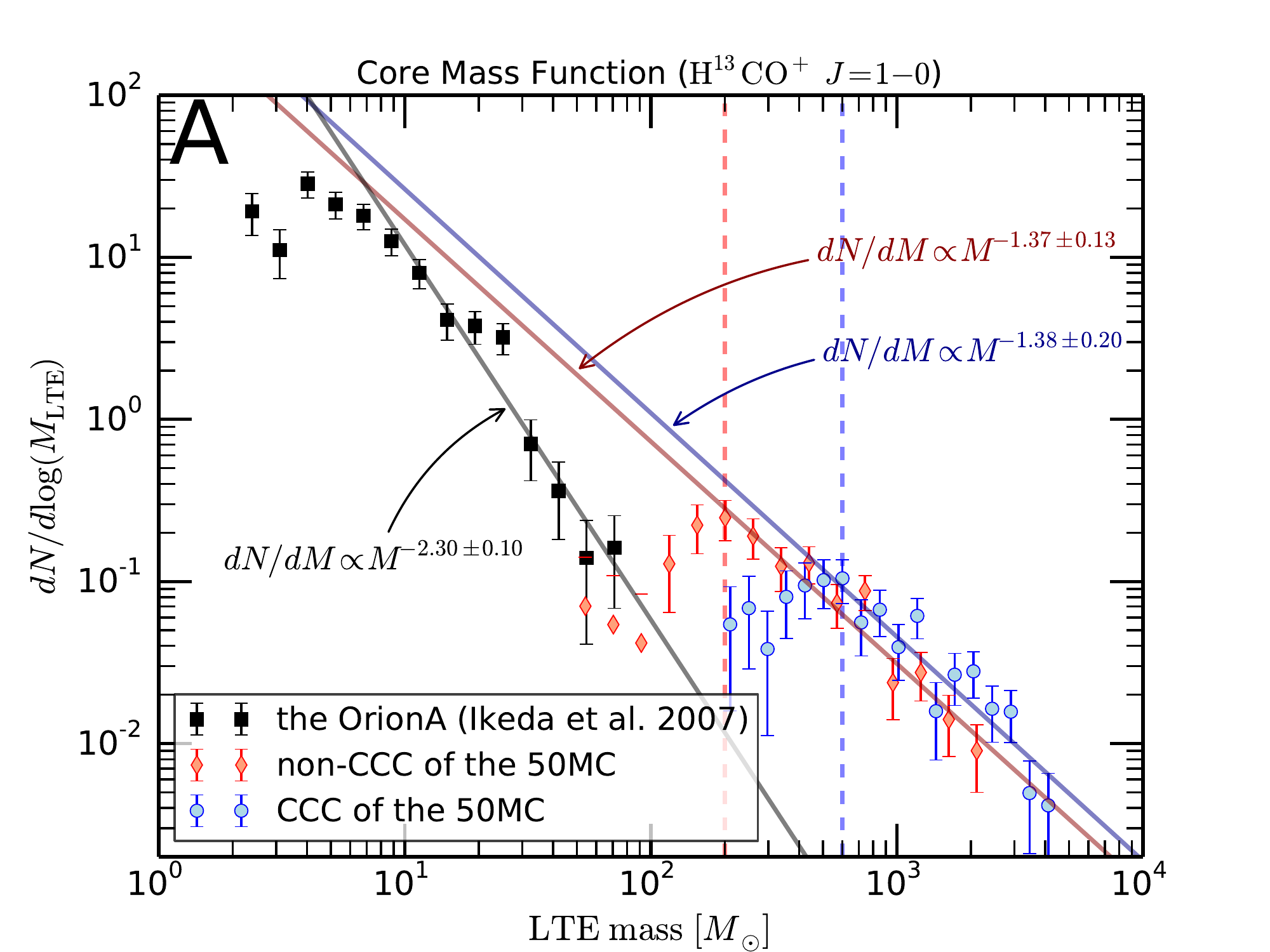}{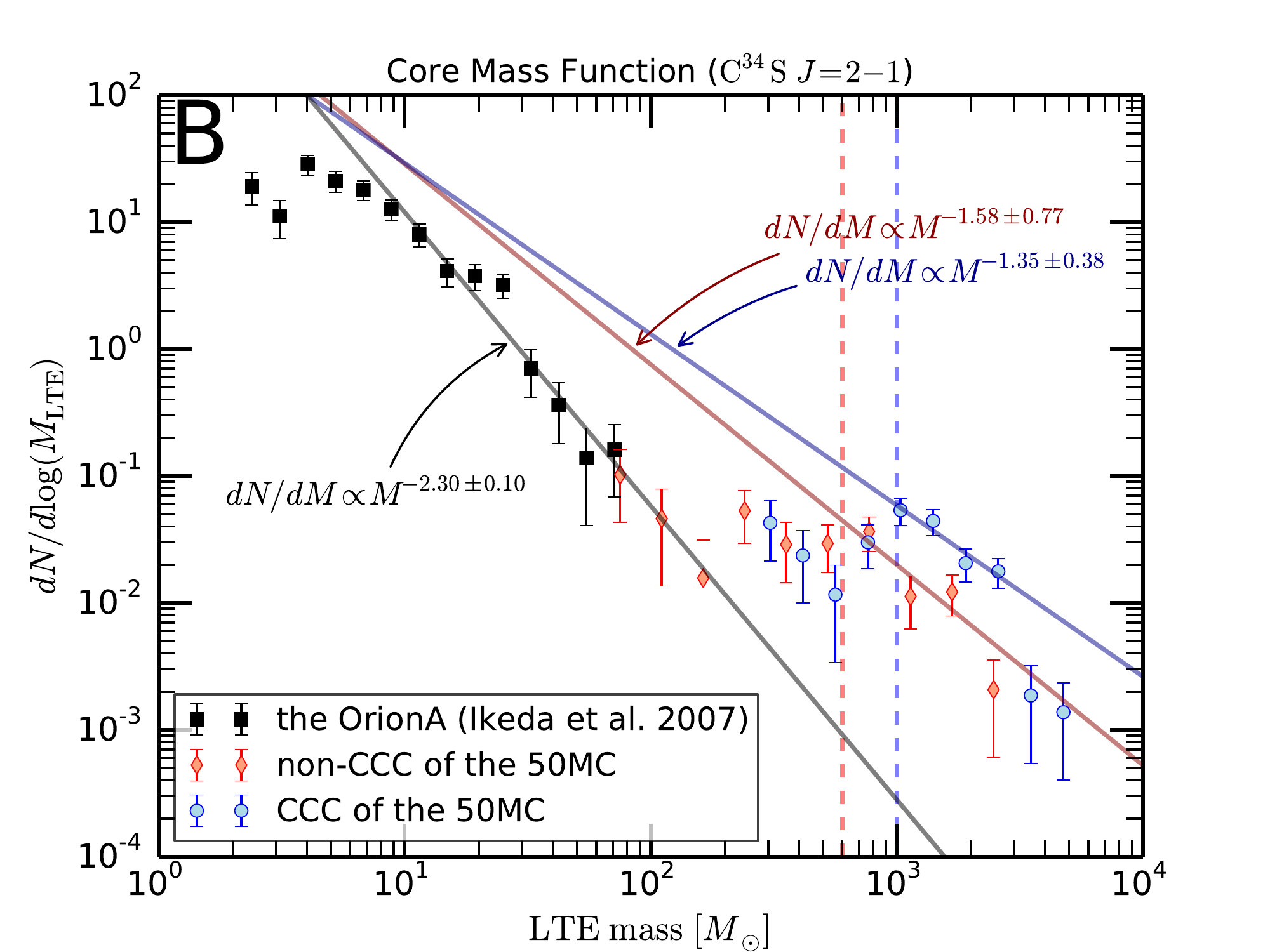}
    \caption{(A) Core mass function of the bound ${\rm H^{13}CO^+}$ cores in the CCC region (blue circles) with the best-fitted single power-law line (blue solid line). The power-law index is $1.38\pm0.20$. The vertical blue dashed line shows $M_{\rm LTE}=600M_\odot$ and we analyze the CMF in the range of $>600M_\odot$. 
    The core mass function in the non-CCC region is shown by the red diamonds with the best-fitted single power-law line (red dashed line). The power-law index is $1.37\pm0.13$. The vertical red dashed line shows $M_{\rm LTE}=200M_\odot$ and we analyze the CMF in the range of $>200M_\odot$. 
    The core mass function of the Orion A cloud is shown by the black squares \citep{ikeda2007}. 
    (B) Core mass function of the bound ${\rm C^{34}S}$ cores in the CCC region (blue circles) with the best-fitted single power-law line (blue solid line). The power-law index is $1.35\pm0.38$. The vertical blue dashed line shows $M_{\rm LTE}=1000M_\odot$ and we analyze the CMF in the range of $>1000M_\odot$. 
    The core mass function in the non-CCC region is shown by the red diamonds with the best-fitted single power-law line (red dashed line). The power-law index is $1.58\pm0.77$. The vertical red dashed line shows $M_{\rm LTE}=600M_\odot$ and we analyze the CMF in the range of $>600M_\odot$. 
   }\label{fig:cmf-ccc}
\end{figure}

\begin{figure}
    \plottwo{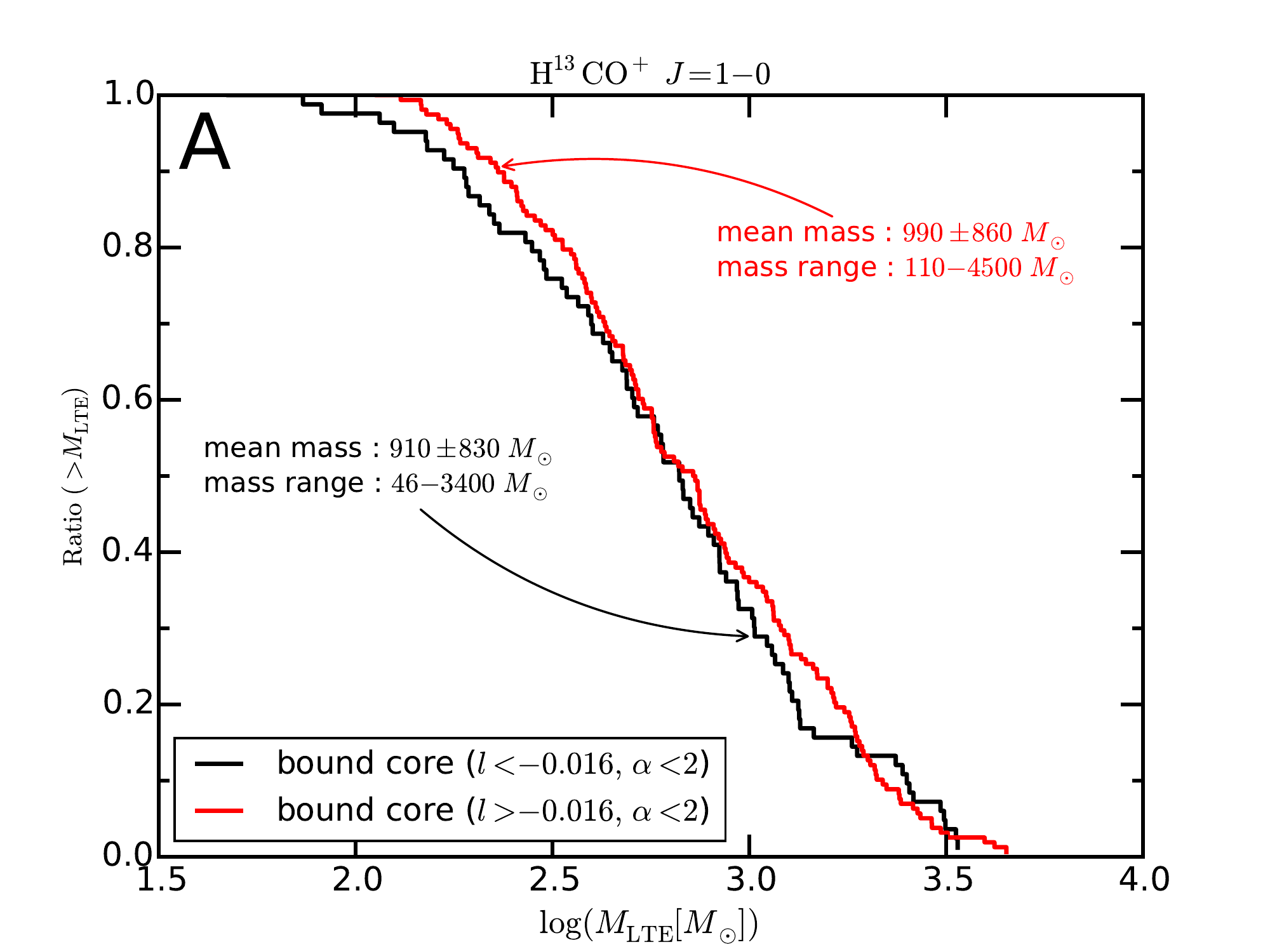}{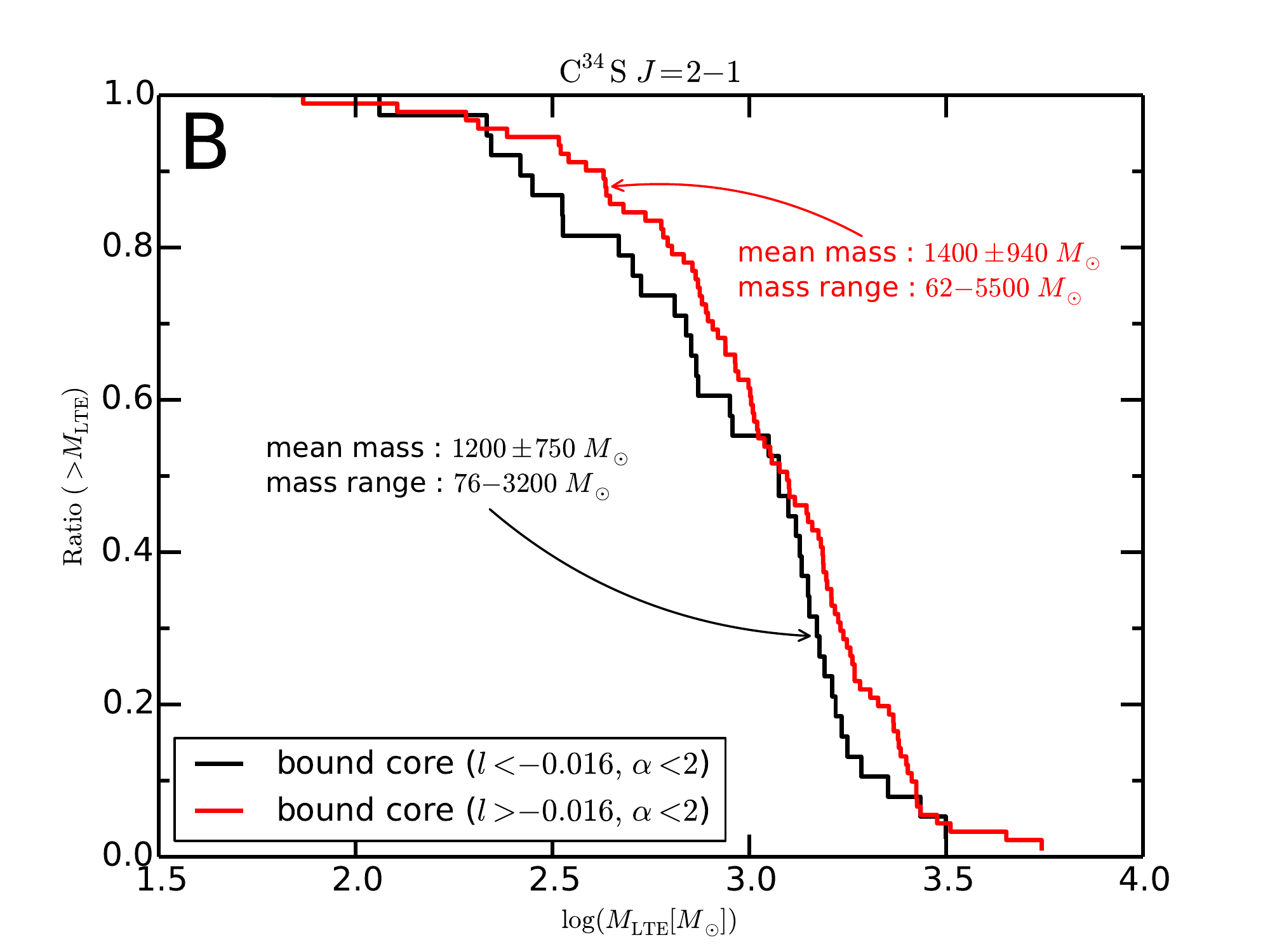}
    \caption{[A] CDF of the LTE mass of the bound ${\rm H^{13}CO^+}~J=1-0$ core. 
    The black thick line shows the distribution in the interaction region of $l<-0\fdg016$, while the red thick line shows the distribution in the interaction region of $l>-0\fdg016$. 
    [B] CDF of the LTE mass of the bound ${\rm C^{34}S}~J=2-1$ core. }\label{fig:mcum_int}
\end{figure}

\begin{table}
	\caption{The best-fit parameters of a single power-law for the CMFs. }\label{tab:fitpara}
	\begin{center}
 	\begin{tabular}{l|ccc}
		\hline\hline
		Parameter	&The whole of the 50MC	&CCC	&Non-CCC \\
		$({\rm H^{13}CO^+}~J=1-0)$&&\\
		\hline
		$\alpha_{\rm cmf}$	&$1.48\pm0.14$		&$1.38\pm0.20$	
							&$1.37\pm0.13$\\
		$\chi^2/{\rm dof}$	&$0.66~(4.63/7)$		&$1.11~(9.95/9)$	
							&$0.83~(6.67/8)$\\
		\hline\hline
		Parameter	&The whole of the 50MC	&CCC	&Non-CCC \\
		$({\rm C^{34}S}~J=2-1)$ &&\\
		\hline
		$\alpha_{\rm cmf}$	&$1.15\pm0.24$		&$1.35\pm0.38$	
							&$1.58\pm0.77$\\
		$\chi^2/{\rm dof}$	&$1.46~(4.36/3)$		&$0.53~(1.07/2)$	
							&$1.96~(1.96/1)$\\
		\hline\hline
    	\end{tabular}
	\end{center}
\end{table}

Finally, Figure \ref{fig:smcore} shows the smallest $\alpha$ core (ID 1706) and the most massive core (ID 1530) in the bound ${\rm H^{13}CO^+}$ cores. 
These bound cores are located on a line where the $\rm H_{II}$ regions A to C line up and at the region considered as the shock front discussed in \S\ref{sec:ccc}. 
The radial velocities of the HII regions are $40-48\rm~km~s^{-1}$ as shown in Figure \ref{fig:pv-ccc} and the smallest alpha core and the most massive core have the radial velocities of $40$ and $38\rm~km~s^{-1}$, respectively. 
Additionally, in \S\ref{sec:ccc}, we argued that the clouds with $V_{\rm LSR}\sim35$ and $\sim55\rm~km~s^{-1}$ collide and that the shock front propagates southwest to northeast. 
{
It is likely that the HII regions A-C in the southwest part of the line formed first in the CCC process. 
This is because the HII regions A-C have ages of $\sim10^{4-5}$ years \citep{yusef-zadeh2010,2011ApJ...735...84M}, which are smaller shorter than the dynamical time scale of the CCC. 
The dynamical time scale of the CCC is $L/V_{\rm CCC}\sim5\times10^5$ years, assuming that the size of the 50MC, $L$, is $\sim10\rm~pc$ and the collision velocity, $V_{\rm CCC}$, is $55~{\rm km~s^{-1}}-35~{\rm km~s^{-1}}=20~{\rm km~s^{-1}}$. 
}
Considering the lining up of the bound cores and the compact $\rm HII$ regions of A, B, and C {and that the ages of the HII regions are smaller than the CCC time scale}, the star formation might have occurred sequentially from the compact $\rm HII$ region C to A owing to the CCC, and the bound cores would produce massive stars to evolve into new compact $\rm HII$ regions. 

We will discuss the possibility of the massive star formation in the bound cores in the 50MC in a future paper.

\begin{figure}
        \plotone{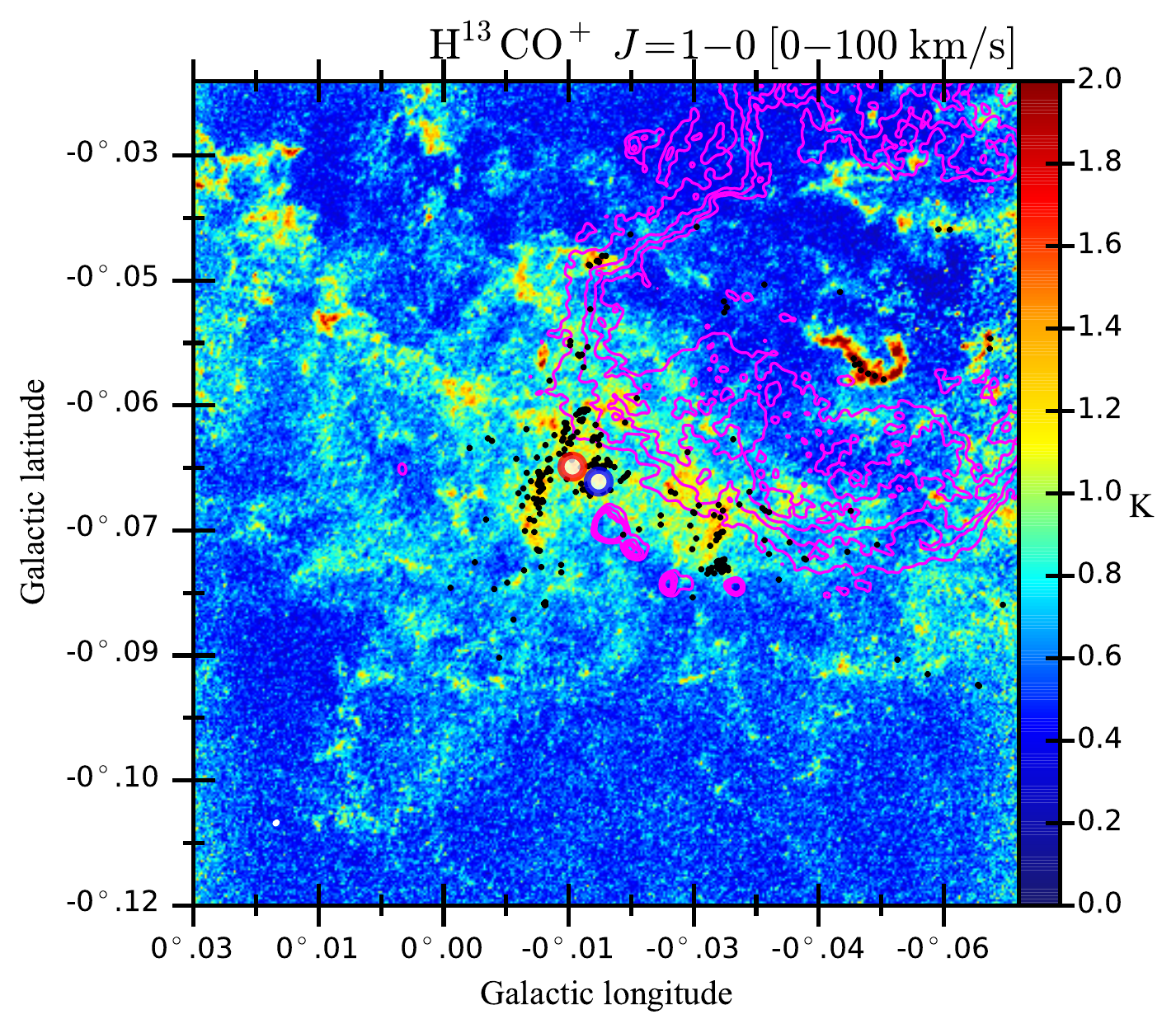}
    \caption{The bound ${\rm H^{13}CO^+}$ core with smallest $\alpha$ of 0.4 is plotted on the $T_{\rm peak}$ map of the ${\rm H^{13}CO^+}~J=1-0$ emission line in the velocity range of $V_{\rm LSR}=0-100\rm~km~s^{-1}$ with the blue open circle. 
    The red open circle shows the most massive bound ${\rm H^{13}CO^+}$ core of $4500~M_{\odot}$. 
    The 6cm continuum emission is shown by the magenta contours. 
    The positions of the 44 GHz methanol masers are shown by the black filled circles. 
   }\label{fig:smcore}
\end{figure}

\section{Summary}
We observed the whole of the 50MC by ALMA in the H$^{13}$CO$^+~J=1-0$ and ${\rm C^{34}S}~J=2-1$ emission lines with the high sensitivity of $\sigma=0.16\rm~K$ and with the high angular resolution of $2\farcs04\times1\farcs41$ and $2\farcs00\times1\farcs35$, respectively. 
Our results and conclusions are summarized as follows:
\begin{itemize}
\item
We identified 3293 and 3192 molecular cloud { core candidates} in the ${\rm H^{13}CO^+}~J=1-0$ and ${\rm C^{34}S}~J=2-1$ maps, respectively, and the number of the identified cores in these data is 100 times larger than that in the previous work in the CS $J=1-0$ emission line maps and is enough for good statistics. 
The mean mass of the identified dense cores is $\sim15$ times larger than that in the Orion A cloud, although the radii of the { core candidates} in the 50MC are similar to those in the Orion A. 

\item
The bound ${\rm H^{13}CO^+}$ cores with a virial parameter of less than 2 are $7\%~(= 241/3293)$ of all the identified ${\rm H^{13}CO^+}$ { core candidates}. 
The bound ${\rm C^{34}S}$ cores are $\sim4\%~(=129/3192)$ of all the identified ${\rm C^{34}S}$ { core candidates}. 

\item
The mean masses of the ${\rm H^{13}CO^+}$ and ${\rm C^{34}S}$ bound cores are $960\pm850M_\odot$ and $1300\pm890~M_{\odot}$, respectively. 
The mass ratio of the total bound ${\rm H^{13}CO^+}$ core mass to the total gas mass is $18~\%(=2.3\times10^{5}~M_{\sun}/1.3\times10^{6}~M_{\sun})$. 
On the other hand, the mass ratio of the total bound ${\rm C^{34}S}$ core mass is $23~\%(=1.7\times10^{5}~M_{\sun}/7.3\times10^{5}~M_{\sun})$. 
These low mass ratios are consistent with the low star formation efficiency in the CMZ. 

\item
The velocity widths of the cores in the 50MC are 10 times larger than those of the cores in the Orion A cloud, but the slopes of the $R$-$\Delta V$ relations in the CCC and non-CCC regions of the 50MC are agree with those in the Orion A. 
The bound cores in the 50MC and the Orion A cloud have similar radii, but the LTE masses and the virial parameters in the 50MC are one to two orders of magnitude larger than those in the Orion A. 
The virial parameters in the 50MC are $\sim10-100$ times larger than those in the Orion A. 
Most of the { core candidates} are not bound by self-gravity because of the large virial parameters. 
The CMFs of the bound cores in the 50MC have top-heavy distributions (${\rm H^{13}CO^+}:1.48\pm0.14$; ${\rm C^{34}S}:1.15\pm0.24$) to that of the Orion A cloud in the high-mass parts.

\item
The number ratio of the bound ${\rm H^{13}CO^+}$ cores to all the bound cores in the CCC region ($49\%=119/241$) is as large as that in the non-CCC region ($51\%=122/241$). 
The distribution of the core number and column densities in the CCC region seems to be biased to a larger density than those in the non-CCC region. 
Especially, 26 cores with more than $8.1\times10^{23}\rm~cm^{-2}$ exist only in the CCC region. 
The bound ${\rm C^{34}S}$ cores also have a density distribution biased toward the dense side. 
These results indicate that the CCC compresses the molecular gas and increases the number of the bound cores with high densities.

\item
The mean mass in the CCC region ($1300\pm970~M_{\odot}$) is also $\sim2$ times larger than that in the non-CCC region ($610\pm510~M_{\odot}$). 
The total bound core mass ratio in the CCC region to the all regions is $68~\%(=1.6\times10^{5}~M_{\sun}/2.3\times10^{5}~M_{\sun})$. 
The mass distribution peak of the cores in the CCC region is also positioned on the larger mass side than that in the non-CCC region. 
In addition, the massive bound cores with masses of $2400~M_{\odot}$ or more exist only in the CCC region. 
Additionally, the area of the CCC region is much smaller than that of the non-CCC region, but the numbers of the bound cores in the CCC and non-CCC regions are similar to each other. 
However, the slopes of the CMFs for the bound ${\rm H^{13}CO^+}$ cores in the CCC and non-CCC regions are $1.38\pm0.20$ and $1.37\pm0.13$ in the mass range of $>600~M_{\odot}$, respectively. 
Thus, the slope of the CMF is not changed so much in the mass range of $>600~M_{\odot}$, but the compression by the CCC efficiently formed the massive bound ${\rm H^{13}CO^+}$ cores. 

\item
The number ratio of the bound ${\rm C^{34}S}$ cores to all the cores in the CCC region ($64\%=82/129$) is larger than that in the non-CCC region ($37\%=48/129$). 
The mean mass in the CCC region ($1600\pm890~M_{\odot}$) is also $\sim2$ times larger than that in the non-CCC region ($840\pm670~M_{\odot}$). 
The total bound core mass ratio in the CCC region to the all regions is $76\%(=1.3\times10^{5}M_{\odot}/1.7\times10^{5}M_{\odot})$. 
The mass distribution peak of the cores in the CCC region is positioned on the larger mass side than that in the non-CCC region. 
Additionally, the massive bound cores with masses of $3000~M_{\odot}$ or more exist only in the CCC region. 
The slopes of the CMFs for the bound ${\rm H^{13}CO^+}$ cores in the CCC and non-CCC regions are $1.35\pm0.38$ and $1.58\pm0.77$ in the mass range of $1000~M_{\odot}$ to $3000~M_{\odot}$, respectively. 
\end{itemize}

We conclude that the compression by the CCC efficiently formed the massive bound cores, especially the bound cores in high-mass end of $\sim 2500-3000$ or more, even if the CMF slope is not changed so much by the CCC. 

\acknowledgments
This work is supported in part by the Grant-in-Aid from the Ministry of Eduction, Sports, Science and Technology (MEXT) of Japan, No.16K05308.
The National Radio Astronomy Observatory is a facility of the National Science Foundation operated under cooperative agreement by Associated Universities, Inc. USA. This paper makes use of the following ALMA data:
ADS/JAO.ALMA\#2012.1.00080.S. ALMA is a partnership of ESO (representing its member states), NSF (USA) and NINS (Japan), together with NRC(Canada), NSC and ASIAA (Taiwan), and KASI (Republic of Korea), in cooperation with the Republic of Chile. The Joint ALMA Observatory is operated by ESO, AUI/NRAO and NAOJ. 


\begin{table}[!h]
  \caption{The property of identified molecular cloud core candidates in the ${\rm H^{13}CO^+}~J=1-0$ emission line.  
The columns from 2 to 4 are the three-dimensional positions of the most intense pixels in the {core candidate}s with $T_{\rm B.peak}$ in the column 7 and those from 5 to 12 are the estimated parameters. 
The last column indicates whether each {core candidate} is located in the CCC region or not. 
In the end of the tables, we summarize the mean, standard deviation, maximum and minimum values of each parameter. }
\footnotesize
  \begin{tabular}{l|ccccccccccc|l}
\hline\hline
{No.	}&{$l$ \tablenotemark{a}}&{$b$ \tablenotemark{a}	}&{$V_{\rm LSR}$ \tablenotemark{a}	}&{$R$}&{$\Delta V$	}&{$T_{\rm B, peak}$	}&{$\frac{M_{\rm vir}}{10^{2}}$	}&{$\frac{M_{\rm LTE}}{10^{2}}$	}&{$\alpha$	}&{$\frac{\overline{n}}{10^{5}}$	}&{$\frac{p_{\rm ex}/k_{\rm B}}{10^{8}}$ }&{CCC/}\\
	{}&{[deg]	}&{[deg]}&{[km/s]}&{[pc]	}&{[km/s]	}&{[K]	}&{$[M_\odot]$	}&{$[M_\odot]$	}&{}&{$[\rm cm^{-3}]$	}&{$[\rm K~cm^{-3}]$}&{non-CCC}\\
\hline
1	&0.0175 	&-0.1042	&56	&0.156	&6.5	&0.69	&14	&3.2	&4.4	&2.9	&5.9	&non-CCC	\\
2	&0.0175 	&-0.0938	&42	&0.311	&11	&0.75	&82	&1.3	&62	&0.15	&1.2	&non-CCC	\\
3	&0.0175 	&-0.0892	&58	&0.285	&12	&0.74	&79	&2.4	&33	&0.36	&2.8	&non-CCC	\\
4	&0.0171 	&-0.1047	&68	&0.120	&7.4	&0.65	&14	&0.12	&120	&0.24	&0.78	&non-CCC	\\
5	&0.0171 	&-0.1001	&14	&0.141	&5.7	&0.55	&9.7	&0.17	&56	&0.21	&0.42	&non-CCC	\\
6	&0.0171 	&-0.0826	&58	&0.163	&6.9	&0.78	&16	&1.8	&8.7	&1.5	&3.8	&non-CCC	\\
7	&0.0171 	&-0.0805	&62	&0.160	&8.5	&0.68	&24	&0.79	&31	&0.67	&2.9	&non-CCC	\\
8	&0.0171 	&-0.0805	&46	&0.212	&7.5	&0.67	&25	&0.71	&36	&0.26	&0.87	&non-CCC	\\
9	&0.0171 	&-0.0797	&54	&0.143	&5.3	&0.68	&8.4	&3.4	&2.5	&4.0	&4.0	&non-CCC	\\
10	&0.0171 	&-0.0793	&64	&0.163	&10	&0.78	&35	&0.60	&59	&0.48	&3.0	&non-CCC	\\
11	&0.0167 	&-0.1088	&68	&0.120	&6.6	&0.64	&11	&2.6	&4.2	&5.2	&11	&non-CCC	\\
12	&0.0167 	&-0.1072	&54	&0.069	&5.0	&0.66	&3.6	&0.87	&4.1	&9.1	&10	&non-CCC	\\
13	&0.0167 	&-0.1063	&46	&0.110	&2.6	&0.68	&1.5	&0.098	&16	&0.26	&0.097	&non-CCC	\\
14	&0.0167 	&-0.1026	&94	&0.144	&7.0	&0.56	&15	&2.2	&6.7	&2.5	&6.4	&non-CCC	\\
15	&0.0167 	&-0.0876	&68	&0.155	&7.2	&0.73	&17	&0.49	&35	&0.45	&1.4	&non-CCC	\\
16	&0.0167 	&-0.0867	&68	&0.170	&7.3	&0.70	&19	&1.2	&15	&0.87	&2.7	&non-CCC	\\
17	&0.0167 	&-0.0818	&54	&0.160	&6.7	&0.67	&15	&5.0	&3.0	&4.2	&7.8	&non-CCC	\\
18	&0.0163 	&-0.1105	&30	&0.147	&7.3	&0.73	&17	&3.0	&5.6	&3.2	&8.7	&non-CCC	\\
19	&0.0163 	&-0.1076	&66	&0.092	&8.4	&0.69	&14	&0.48	&28	&2.2	&9.0	&non-CCC	\\
20	&0.0163 	&-0.1063	&34	&0.098	&6.9	&0.68	&9.7	&1.6	&6.2	&5.8	&14	&non-CCC	\\
21	&0.0163 	&-0.1055	&40	&0.174	&12	&0.72	&54	&0.53	&100	&0.35	&3.1	&non-CCC	\\
22	&0.0163 	&-0.1042	&26	&0.084	&9.1	&0.73	&14	&1.2	&12	&7.1	&33	&non-CCC	\\
23	&0.0163 	&-0.1042	&46	&0.151	&6.1	&0.71	&12	&3.0	&3.8	&3.0	&5.0	&non-CCC	\\
24	&0.0163 	&-0.1038	&68	&0.091	&9.2	&0.77	&16	&0.084	&190	&0.39	&2.0	&non-CCC	\\
25$^{\rm b}$	&0.0163 	&-0.0801	&56	&0.162	&7.8	&0.97	&21	&10	&2.0	&8.6	&16	&non-CCC	\\
26	&0.0163 	&-0.0797	&64	&0.155	&9.3	&0.72	&28	&0.71	&40	&0.65	&3.3	&non-CCC	\\
27	&0.0155 	&-0.1092	&54	&0.114	&5.8	&0.81	&8.0	&1.7	&4.8	&4.0	&6.5	&non-CCC	\\
28$^{\rm b}$	&0.0155 	&-0.1072	&64	&0.119	&4.4	&0.76	&4.8	&2.7	&1.8	&5.4	&2.9	&non-CCC	\\
29	&0.0155 	&-0.1072	&52	&0.084	&6.8	&0.74	&8.2	&1.6	&5.2	&9.2	&21	&non-CCC	\\
30	&0.0155 	&-0.1047	&64	&0.104	&4.2	&0.67	&3.9	&0.14	&27	&0.44	&0.46	&non-CCC	\\
\hline\hline
Ave.	&	&	&	&0.164	&6.7	&0.80	&18	&2.3	&20	&1.7	&3.5	&	\\
Std.	&	&	&	&0.040	&2.1	&0.20	&15	&3.7	&25	&2.0	&5.5	&	\\
Max.	&	&	&	&0.330	&22	&2.1	&230	&45	&450	&25	&120	&	\\
Min.	&	&	&	&0.042	&2.1	&0.50	&0.81	&0.043	&0.40	&0.10	&-11	&	\\
\hline\hline
\end{tabular}\label{tab:core}
\tablecomments{Table \ref{tab:core} is published in its entirety in the machine-readable format (\url{http://www.vsop.isas.jaxa.jp/~nakahara/figure_uehara/d-ron/tab4.txt}). 
A portion is shown here for guidance regarding its form and content.}
\tablenotetext{1}{The position and $V_{\rm LSR}$ of each core are those of the most intense pixel with $T_{\rm B,peak}$ in the core. }
\tablenotetext{2}{The bound cores with $\alpha\le2$. }
\tablenotetext{3}{The bound cores identified in both the ${\rm H^{13}CO^+}~J=1-0$ and ${\rm C^{34}S}~J=2-1$ emission lines. }
\end{table}

\begin{table}[!h]
\caption[The property of identified molecular cloud core candidates in the ${\rm C^{34}S}~J=2-1$ emission line. ]{The property of identified molecular cloud core candidates in the ${\rm C^{34}S}~J=2-1$ emission line. 
The columns from 2 to 4 are the three-dimensional positions of the most intense pixels in the {core candidate}s with $T_{\rm B.peak}$ in the column 7 and those from 5 to 12 are the estimated parameters. 
The last column indicates whether each {core candidate} is located in the CCC region or not. 
In the end of the tables, we summarize the mean, standard deviation, maximum and minimum values of each parameter. 
}
\footnotesize
  \begin{tabular}{l|ccccccccccc|l}
\hline\hline
{No.	}&{$l$ \tablenotemark{a}}&{$b$ \tablenotemark{a}	}&{$V_{\rm LSR}$ \tablenotemark{a}	}&{$R$}&{$\Delta V$	}&{$T_{\rm B, peak}$	}&{$\frac{M_{\rm vir}}{10^{2}}$	}&{$\frac{M_{\rm LTE}}{10^{2}}$	}&{$\alpha$	}&{$\frac{\overline{n}}{10^{5}}$	}&{$\frac{p_{\rm ex}/k_{\rm B}}{10^{8}}$ }&{CCC/}\\
	{}&{[deg]	}&{[deg]}&{[km/s]}&{[pc]	}&{[km/s]	}&{[K]	}&{$[M_\odot]$	}&{$[M_\odot]$	}&{}&{$[\rm cm^{-3}]$	}&{$[\rm K~cm^{-3}]$}&{non-CCC}\\
\hline
1	&0.0175 	&-0.0947	&56	&0.171	&11	&0.65	&40	&0.16	&250	&0.11	&0.75	&non-CCC	\\
2	&0.0175 	&-0.0788	&58	&0.114	&6.3	&0.79	&9.6	&0.27	&36	&0.62	&1.5	&non-CCC	\\
3	&0.0171 	&-0.1063	&58	&0.099	&4.0	&0.61	&3.3	&0.46	&7.3	&1.6	&1.4	&non-CCC	\\
4	&0.0171 	&-0.1047	&70	&0.230	&9.7	&0.78	&46	&0.29	&160	&0.081	&0.47	&non-CCC	\\
5	&0.0171 	&-0.1017	&72	&0.135	&6.3	&0.70	&11	&0.13	&87	&0.18	&0.44	&non-CCC	\\
6	&0.0171 	&-0.1013	&48	&0.171	&12	&0.65	&56	&0.13	&430	&0.091	&0.86	&non-CCC	\\
7	&0.0171 	&-0.0851	&66	&0.197	&10	&0.68	&44	&0.46	&97	&0.21	&1.3	&non-CCC	\\
8	&0.0171 	&-0.0805	&58	&0.159	&4.9	&0.78	&8.0	&0.51	&16	&0.44	&0.60	&non-CCC	\\
9	&0.0171 	&-0.0805	&62	&0.185	&7.8	&0.67	&24	&0.27	&88	&0.15	&0.55	&non-CCC	\\
10	&0.0167 	&-0.1084	&78	&0.207	&12	&0.61	&66	&0.48	&140	&0.19	&1.7	&non-CCC	\\
11	&0.0167 	&-0.0917	&28	&0.114	&5.8	&0.85	&7.9	&0.11	&71	&0.26	&0.53	&non-CCC	\\
12	&0.0167 	&-0.0897	&54	&0.172	&11	&0.56	&41	&0.19	&210	&0.13	&0.90	&non-CCC	\\
13	&0.0167 	&-0.0872	&30	&0.191	&15	&0.67	&92	&0.19	&480	&0.095	&1.3	&non-CCC	\\
14	&0.0167 	&-0.0826	&64	&0.204	&10	&0.65	&47	&0.54	&86	&0.22	&1.5	&non-CCC	\\
15	&0.0163 	&-0.0951	&66	&0.131	&7.6	&0.66	&16	&0.075	&210	&0.12	&0.41	&non-CCC	\\
16	&0.0159 	&-0.1097	&80	&0.121	&3.5	&0.65	&3.1	&0.055	&56	&0.11	&0.08	&non-CCC	\\
17	&0.0159 	&-0.1059	&34	&0.131	&8.9	&0.69	&22	&0.76	&28	&1.2	&5.5	&non-CCC	\\
18	&0.0159 	&-0.0801	&56	&0.126	&5.3	&0.68	&7.5	&1.1	&6.6	&2.0	&2.9	&non-CCC	\\
19	&0.0155 	&-0.1105	&32	&0.143	&5.3	&0.69	&8.6	&1.1	&7.9	&1.3	&1.9	&non-CCC	\\
20	&0.0155 	&-0.1105	&62	&0.164	&8.2	&0.66	&23	&1.4	&17	&1.1	&4.1	&non-CCC	\\
21	&0.0155 	&-0.1080	&30	&0.133	&6.7	&0.94	&13	&1.4	&9.3	&2.0	&4.9	&non-CCC	\\
22	&0.0155 	&-0.1042	&48	&0.147	&5.8	&0.81	&10	&0.20	&51	&0.22	&0.45	&non-CCC	\\
23	&0.0155 	&-0.0930	&54	&0.290	&12	&0.65	&87	&0.57	&150	&0.081	&0.70	&non-CCC	\\
24	&0.0155 	&-0.0813	&54	&0.167	&5.9	&0.66	&12	&1.6	&7.7	&1.2	&2.2	&non-CCC	\\
25	&0.0155 	&-0.0809	&58	&0.134	&3.6	&0.77	&3.7	&0.32	&12	&0.45	&0.33	&non-CCC	\\
26	&0.0155 	&-0.0793	&58	&0.135	&10	&0.74	&29	&0.31	&92	&0.44	&2.7	&non-CCC	\\
27	&0.0150 	&-0.1038	&14	&0.226	&13	&0.60	&80	&2.7	&30	&0.80	&7.9	&non-CCC	\\
28	&0.0150 	&-0.1026	&52	&0.104	&12	&0.67	&31	&0.57	&55	&1.7	&15	&non-CCC	\\
29	&0.0150 	&-0.0934	&24	&0.110	&4.4	&0.69	&4.4	&0.085	&52	&0.22	&0.25	&non-CCC	\\
30	&0.0146 	&-0.1059	&30	&0.160	&16	&0.68	&83	&1.5	&56	&1.3	&19	&non-CCC	\\
\hline\hline
Ave.	&	&	&	&0.171	&6.8	&1.0	&19	&1.9	&39	&1.0	&2.4	&	\\
Std.	&	&	&	&0.042	&2.0	&0.40	&14	&3.6	&52	&1.4	&3.5	&	\\
Max.	&	&	&	&0.319	&20	&3.9	&140	&55	&800	&15	&53	&	\\
Min.	&	&	&	&0.050	&2.0	&0.50	&0.76	&0.024	&0.80	&0.056	&-5.3	\\
\hline\hline
\end{tabular}\label{tab:core34}
\tablecomments{Table \ref{tab:core34} is published in its entirety in the machine-readable format (\url{http://www.vsop.isas.jaxa.jp/~nakahara/figure_uehara/d-ron/tab5.txt}). 
A portion is shown here for guidance regarding its form and content. }
\tablenotetext{1}{The position and $V_{\rm LSR}$ of each core are those of the most intense pixel with $T_{\rm B,peak}$ in the core. }
\tablenotetext{2}{The bound cores with $\alpha\le2$. }
\tablenotetext{3}{The bound cores identified in both the ${\rm H^{13}CO^+}~J=1-0$ and ${\rm C^{34}S}~J=2-1$ emission lines. }
\end{table}

\begin{table}[!h]
	\begin{center}
  \caption{The physical parameter of the bound ${\rm H^{13}CO^+}$ and ${\rm C^{34}S}$ cores. }
  \begin{tabular}{l|cccccc}
\hline\hline
${\rm H^{13}CO^+}$ 	&$R$[pc]	&$\Delta V[\rm km~s^{-1}]$	&$\frac{M_{\rm vir}}{10^{2}}[M_\odot]$	&$\frac{M_{\rm LTE}}{10^{2}}[M_\odot]$	&$\alpha(=M_{\rm vir}/M_{\rm LTE})$	&$\frac{n}{10^{5}}[\rm cm^{-3}]$	\\
\hline
Ave.	&0.170	&5.4	&13	&9.6	&1.4	&5.8	\\
Std.		&0.045	&1.8	&11	&8.5	&0.37	&2.7	\\
Max.	&0.302	&11	&75	&45	&2.0	&21	\\
Min.		&0.067	&2.1	&0.81	&0.48	&0.40	&1.9	\\
\hline\hline
${\rm C^{34}S}$ 	&$R$[pc]	&$\Delta V[\rm km~s^{-1}]$	&$\frac{M_{\rm vir}}{10^{2}}[M_\odot]$	&$\frac{M_{\rm LTE}}{10^{2}}[M_\odot]$	&$\alpha(=M_{\rm vir}/M_{\rm LTE})$	&$\frac{n}{10^{5}}[\rm cm^{-3}]$	\\
\hline
Ave.	&0.204	&6.3	&19	&13	&1.5	&4.9	\\
Std.		&0.040	&1.8	&12	&8.9	&0.29	&2.5	\\
Max.	&0.305	&10	&52	&55	&2.0	&13	\\
Min.		&0.097	&2.0	&0.83	&0.62	&0.80	&0.95	\\
\hline\hline
\end{tabular}\label{tab:bound}
	\end{center}
\end{table}

\begin{table}[!h]
	\begin{center}
  \caption{The physical parameter of the bound ${\rm H^{13}CO^+}$ cores in the CCC and non-CCC regions. }
  \begin{tabular}{l|cccccc}
\hline\hline
CCC	&$R$[pc]	&$\Delta V[\rm km~s^{-1}]$	&$\frac{M_{\rm vir}}{10^{2}}[M_\odot]$	&$\frac{M_{\rm LTE}}{10^{2}}[M_\odot]$	&$\alpha(=M_{\rm vir}/M_{\rm LTE})$	&$\frac{n}{10^{5}}[\rm cm^{-3}]$	\\
\hline
Ave.		&0.188	&6.2		&18	&13	&1.4	&6.3	\\
Std.			&0.040	&1.8		&13	&9.7	&0.36	&3.2	\\
Max.		&0.302	&11		&75	&45	&2.0	&21	\\
Min.			&0.100	&2.8		&2.6	&1.9	&0.40	&1.9	\\
\hline\hline
Non-CCC	&$R$[pc]	&$\Delta V[\rm km~s^{-1}]$	&$\frac{M_{\rm vir}}{10^{2}}[M_\odot]$	&$\frac{M_{\rm LTE}}{10^{2}}[M_\odot]$	&$\alpha(=M_{\rm vir}/M_{\rm LTE})$	&$\frac{n}{10^{5}}[\rm cm^{-3}]$	\\
\hline
Ave.		&0.151	&4.6		&8.3	&6.1	&1.4	&5.3	\\
Std.			&0.043	&1.4		&6.9	&5.1	&0.39	&1.9	\\
Max.		&0.245	&8.2		&31	&24	&2.0	&15	\\
Min.			&0.067	&2.1		&0.81	&0.48	&0.49	&2.1	\\
\hline\hline
\end{tabular}\label{tab:ccc}
	\end{center}
\end{table}

\begin{table}[!h]
	\begin{center}
  \caption{The physical parameter of the bound ${\rm C^{34}S}$ cores in the CCC region. }
  \begin{tabular}{l|ccccccc}
\hline\hline
CCC	&$R$[pc]	&$\Delta V[\rm km~s^{-1}]$	&$\frac{M_{\rm vir}}{10^{2}}[M_\odot]$	&$\frac{M_{\rm LTE}}{10^{2}}[M_\odot]$	&$\alpha(=M_{\rm vir}/M_{\rm LTE})$	&$\frac{n}{10^{5}}[\rm cm^{-3}]$	\\
\hline
Ave.		&0.212	&6.8		&22	&16	&1.5	&5.6	\\
Std.			&0.033	&1.6		&11	&8.9	&0.31	&2.6	\\
Max.		&0.296	&10		&52	&55	&2.0	&13	\\
Min.			&0.146	&3.0		&3.4	&2.6	&0.80	&0.95	\\
\hline\hline
Non-CCC	&$R$[pc]	&$\Delta V[\rm km~s^{-1}]$	&$\frac{M_{\rm vir}}{10^{2}}[M_\odot]$	&$\frac{M_{\rm LTE}}{10^{2}}[M_\odot]$	&$\alpha(=M_{\rm vir}/M_{\rm LTE})$	&$\frac{n}{10^{5}}[\rm cm^{-3}]$	\\
\hline
Ave.		&0.189	&5.3		&14	&8.4	&1.6	&3.6	\\
Std.			&0.047	&1.8		&11	&6.7	&0.23	&1.5	\\
Max.		&0.305	&9.7		&52	&30	&2.0	&7.7	\\
Min.			&0.097	&2.0		&0.83	&0.62	&1.1	&1.3	\\
\hline\hline
\end{tabular}\label{tab:cccc34s}
	\end{center}
\end{table}

\bibliographystyle{apj}
\bibliography{apj-jour,refer}

\end{document}